  \documentclass[a4paper,11pt]{article}
  \pdfoutput=1
  \usepackage{silence}
  \usepackage{jheppub} 
  
    \makeatletter
    \gdef\@fpheader{Prepared for submission to \textbf{Machine Learning: Science and Technology}}
    \gdef\@journal{mlst}
    \makeatother
    
\newcommand{\eref}[1]{\eqref{#1}} 
\newcommand{\erefs}{} 
\newcommand{\Eref}[1]{\eqref{#1}} 
\newcommand{\fref}[1]{figure~\ref{#1}}
\newcommand{\Fref}[1]{Figure~\ref{#1}}
\newcommand{\sref}[1]{section~\ref{#1}}
\newcommand{\Sref}[1]{Section~\ref{#1}}
\newcommand{\tref}[1]{table~\ref{#1}}

\newcommand{\aref}[1]{appendix~\ref{#1}}

\usepackage[utf8]{inputenc}
\usepackage{amsthm}
\usepackage[T1]{fontenc}
\usepackage{mathtools}
\usepackage{bm}
\usepackage[usenames,dvipsnames]{xcolor}
\usepackage[normalem]{ulem}

\usepackage[ruled,vlined,linesnumbered]{algorithm2e}
\newcommand{\code}[1]{\texttt{#1}}

\usepackage{refcount}
\newcommand{\footref}[1]{\hyperref[#1]{\footnotemark[\getrefnumber{#1}]}}

\allowdisplaybreaks 

  \usepackage{longtable}
  \usepackage{array}
  \newcolumntype{P}[1]{>{\centering\arraybackslash}p{#1}}
  \newcolumntype{M}[1]{>{\centering\arraybackslash}m{#1}}
  \newcolumntype{B}[1]{>{\centering\arraybackslash}b{#1}}
  \usepackage{dcolumn}
  \newcolumntype{.}{D{.}{.}{-1}}
  \usepackage{booktabs}
  \newcommand{\cellcenter}[1]{\multicolumn{1}{c}{#1}}
\DeclareMathOperator*{\mean}{\mathbb{MEAN}}
\DeclareMathOperator{\E}{\mathbb{E}}
\DeclareMathOperator*{\argmax}{arg\,max}
\DeclareMathOperator{\erf}{erf}
\DeclareMathOperator{\logit}{logit}

\newcommand{\myS}{\mathcal{S}}
\newcommand{\Hn}{\mathbb{H}_b}
\newcommand{\Ha}{\mathbb{H}_{b+s}}
\newcommand{\e}{{\bm{e}}}
\newcommand{\x}{{\bm{x}}}
\newcommand{\cond}{\,;\,}
\newcommand{\dneym}{D_{\mathrm{Neym}\chi^2}}
\newcommand{\dpear}{D_{\mathrm{Pear}\chi^2}}
\newcommand{\dkl}{D_{\mathrm{KL}}}
\newcommand{\drevkl}{D_{\mathrm{revKL}}}
\newcommand{\djeff}{D_{\mathrm{J}}}
\newcommand{\db}{D_{\mathrm{B}}}
\newcommand{\dstat}{D_{\mathrm{stat}}}

\newcommand{\tdneym}{\tilde{D}_{\mathrm{Neym}\chi^2}}
\newcommand{\tdpear}{\tilde{D}_{\mathrm{Pear}\chi^2}}
\newcommand{\tdkl}{\tilde{D}_{\mathrm{KL}}}
\newcommand{\tdrevkl}{\tilde{D}_{\mathrm{revKL}}}
\newcommand{\tdjeff}{\tilde{D}_{\mathrm{J}}}
\newcommand{\tdb}{\tilde{D}_{\mathrm{B}}}
\newcommand{\tdstat}{\tilde{D}_{\mathrm{stat}}}

\newcommand{\edneym}{\Lambda_{\mathrm{Neym}\chi^2}}
\newcommand{\edpear}{\Lambda_{\mathrm{Pear}\chi^2}}
\newcommand{\edkl}{\Lambda_{\mathrm{KL}}}
\newcommand{\edrevkl}{\Lambda_{\mathrm{revKL}}}
\newcommand{\edjeff}{\Lambda_{\mathrm{J}}}
\newcommand{\edb}{\Lambda_{\mathrm{B}}}
\newcommand{\edstat}{\Lambda_{\mathrm{stat}}}

\newcommand{\thneym}{p_{\mathrm{Neym}\chi^2}^\text{thresh}}
\newcommand{\thpear}{p_{\mathrm{Pear}\chi^2}^\text{thresh}}
\newcommand{\thkl}{p_\mathrm{KL}^\text{thresh}}
\newcommand{\threvkl}{p_\mathrm{revKL}^\text{thresh}}
\newcommand{\thjeff}{p_\mathrm{J}^\text{thresh}}
\newcommand{\thb}{p_\mathrm{B}^\text{thresh}}
\newcommand{\thstat}{p_\mathrm{stat}^\text{thresh}}

\usepackage{anyfontsize}
\DeclareSymbolFont{symbols}{OMS}{cmsy}{m}{n}
\DeclareMathSymbol{\myprec}{\mathrel}{symbols}{"1E}
\DeclareMathSymbol{\mypreceq}{\mathrel}{symbols}{"16}

\makeatletter
\newcommand{\vast}{\bBigg@{3}}
\makeatother

\newcommand{\thickbrickurl}{https://prasanthcakewalk.gitlab.io/thickbrick/}

\begin{document}


\title{Optimal event selection and categorization in high energy physics, Part 1: Signal discovery}
\author{Konstantin T. Matchev,}
\author{Prasanth Shyamsundar}
\affiliation{Institute for Fundamental Theory, Physics Department, University of Florida, Gainesville, FL 32611, USA}

\emailAdd{matchev@ufl.edu}
\emailAdd{prasanths@ufl.edu}


\abstract{We provide a prescription to train optimal machine-learning-based event selectors and categorizers that maximize the statistical significance of a potential signal excess in high energy physics (HEP) experiments, as quantified by any of six different performance measures. For analyses where the signal search is performed in the distribution of some event variables, our prescription ensures that only the information complementary to those event variables is used in event selection and categorization. This eliminates a major misalignment with the physics goals of the analysis (maximizing the significance of an excess) that exists in the training of typical ML-based event selectors and categorizers. In addition, this decorrelation of event selectors from the relevant event variables prevents the background distribution from becoming peaked in the signal region as a result of event selection, thereby ameliorating the challenges imposed on signal searches by systematic uncertainties. Our event selectors (categorizers) use the output of machine-learning-based classifiers as input and apply optimal selection cutoffs (categorization thresholds) that depend on the event variables being analyzed, as opposed to flat cutoffs (thresholds). These optimal cutoffs and thresholds are learned iteratively, using a novel approach with connections to Lloyd's k-means clustering algorithm. We provide a public, Python 3 implementation of our prescription called \href{\thickbrickurl}{ThickBrick}, along with usage examples.
}

\maketitle
\flushbottom

\section{Introduction} \label{sec:intro}

The increase in complexity and scale of high energy physics (HEP) experiments over the years has been accompanied by an 
increased sophistication of the techniques employed in the data analysis. Among the more recent additions to the medley of analysis techniques used in experimental HEP are machine learning algorithms. Machine learning techniques have found applications in various aspects of the analysis \cite{Guest:2018yhq, Albertsson:2018maf, Bourilkov:2019MLreview}, and are an active area of research, 
both from the standpoint of finding new applications as well as refining their current usage.

An important step in any collider\footnote{In what follows, for brevity we shall refer to ``collider'' events such as those collected at the Large Hadron Collider (LHC), but our treatment is equally applicable to other areas of HEP as well, and maybe even outside physics.} data analysis is event selection and/or categorization. Event selection is the process of selecting a subset of collider events for further analysis. Event categorization is the process of grouping the selected events into disjoint categories, each of which will be individually analyzed.

Event selection serves two purposes. First, it helps bring down the volume of recorded data within the computational limits of the experiment (storage and processing power). This is done at the trigger level \cite{Matsushita:2017yzh, MontejoBerlingen:2019ukc} and works by making rapid decisions on whether or not an event is ``interesting enough'' to warrant further processing, given the goals of the experiment and its computational limitations. Secondly, event selection is used in individual analyses to reduce the amount of ``background'' in the event sample and make it more ``signal''-rich, with signal and background events defined appropriately. This is the type of event selection we will focus on in this paper. 

A sample of collider events can often be thought of as being composed of several subsamples, although real collider events do not necessarily carry a label indicating which subsample they are from. These \emph{mixture models} are used in the Monte Carlo simulation of events, with simulated events carrying a label to identify the subsample they are from. These subsamples can be labeled as signal and background (or bg1, bg2, etc), as appropriate for the particular analysis at hand. The event selection process can then be thought of as selectively including signal-like events (and rejecting background-like events) to ``improve'' the analysis. Let us call this the ``signal is better than background'' heuristic.

The goal of event selection in an analysis searching for a signal (due to new physics or an interesting Standard Model process), is to increase the expected significance of a potential signal excess, or improve the expected upper-limit on the signal cross-section (in the case of a null result). It is easy to see how the heuristic mentioned above can help improve such an analysis. On the other hand, the goal in a parameter measurement analysis (say a top quark or a $W$-boson mass measurement) is to increase the precision of the measurement. In such analyses, subsamples whose distributions are independent of the parameter(s) being measured can be labeled as background, and subsamples whose distributions do depend on the parameter(s) being measured can be labeled as signal. Again, we can see how the ``signal is better than background'' heuristic can improve the precision of measurement by making the distribution of the selected event sample more sensitive to the parameter(s) being measured.

The ``signal is better than background'' heuristic allows us to rephrase the event selection problem as an event classification problem. Over the last two decades, machine learning methods like artificial neural networks \cite{Rosenblatt:1958perceptron} and boosted decision trees \cite{Breiman:1984tree, Freund:1996adaboost} originally developed to solve classification problems have been successfully employed in collider physics analyses \cite{Guest:2018yhq, Albertsson:2018maf, Bourilkov:2019MLreview}. These algorithms are trained on simulated data samples to learn features that distinguish signal and background events. They can then provide a measure of how signal-like an event in the real data is, and the event selection can be done based on this measure.

But despite the rationale behind the ``signal is better than background'' heuristic, it is not perfectly aligned with the physics goals of improving the search sensitivity and measurement precision. In this work, we construct optimal event selection and categorization prescriptions for physics analyses within the supervised machine learning paradigm, with cost functions defined on a per-event basis. Along the way, we will point out exactly how our methods realign the goals of event selection and categorization with the physics goals mentioned above. To facilitate clarity of presentation and improve accessibility, we will present our work in three parts (tentative).

\begin{itemize}
\item In part 1 (this paper), we will focus on event selection and categorization for signal discovery (and upper-limit setting). We will provide a prescription to train event selectors and categorizers, which are optimal for an exactly specified signal distribution.
\begin{itemize}
\item We will not account for unspecified signal parameter(s), like the mass of the dark photon in a dark photon search. In other words, in this paper we will train our categorizers to be optimal for a given set of values of signal parameters.
We will also not concern ourselves with systematic uncertainties in the model specification. Both of these issues will be addressed in part 3.
\end{itemize}
\item In part 2 \cite{Matchev:ML2}, we will provide event selection and categorization prescriptions for reducing the expected \textbf{statistical} uncertainty of the measurement in parameter measurement experiments. 
\begin{itemize}
 \item As in part 1, we will be optimizing the measurement precision only for a given underlying value of the parameter being measured, and we will not be accounting for model (systematic) uncertainties---both of those issues will be addressed in part 3.
\end{itemize}
\item In part 3 \cite{Matchev:ML3}, we will provide prescriptions to train optimal event selectors and categorizers accounting for unspecified signal parameters and systematic uncertainties. In other words,
\begin{itemize}
 \item We will provide prescriptions to train event selectors and categorizers that improve sensitivity over a range of unspecified signal parameter values for signal discovery, and over a range of underlying parameter values for parameter measurement.
 \item We will also provide prescriptions to maximize the significance of a signal excess in the presence of model uncertainties, and minimize the combined statistical and systematic uncertainty in parameter measurement experiments.
\end{itemize}
\end{itemize}

\subsection{Dimensionality reduction and complementary information} \label{intro:dimreduce}
Data recorded in collider experiments is typically high dimensional. For example, in LHC experiments, a typical recorded event leaves thousands of hits in the detector. Even at the parton level, depending on the number of reconstructed final state particles, events can have $\sim 10$ or more attributes. Ideally one would want to draw statistical inferences on the underlying physics by analyzing the distribution of events in these high dimensional spaces.

But analyzing the distribution of collider data in a high dimensional space (using multi-dimensional histograms, likelihood-ratio tests, or the Matrix Element Method) comes with several challenges---the computational power needed to systematically scan the phase-space grows exponentially with the number of dimensions; the volume of data (real and simulated) needed to populate the space grows exponentially with the number of dimensions; Monte Carlo validation in high dimensional phase-space can be challenging; directly evaluating the likelihood-ratio using the Matrix Element Method is computationally expensive \cite{Gainer:2013iya}, etc.

One way to deal with the curse of dimensionality is to reduce the dimensionality---by constructing event variables like pseudorapidities, transverse momenta, invariant or transverse masses of sets of final state particles, etc. After reducing the dimensionality of data to a more tractable number of event variables, we can limit the analyses to studying the distribution of these event variables.

But this dimensionality reduction is accompanied by information loss---typically the dimensionality-reduced data contains less information than the original data, be it for the purpose of signal search or parameter measurement. In the case of signal search analyses, the information loss can be understood as follows: All the information contained in a collider event for the purpose of testing for the presence of a signal over a background can be captured by the likelihood-ratio of the event under the signal and background distributions (a consequence of the Neyman--Pearson lemma). Reducing the dimensionality of the data mixes up regions of phase-space with different values of this likelihood-ratio, but the same value of the event variable(s). To see how such a mixing causes information loss, consider the simple scenario with two regions of phase-space where the expected number of signal and background events are, respectively, $(S_1, B_1)$ and $(S_2, B_2)$, all non-zero. Let the events in the two regions be mixed and analyzed together, and let the combined (expected) signal and background counts be $(S_\text{tot}, B_\text{tot}) \equiv (S_1+S_2, B_1+B_2)$. Note that
\begin{equation}
B_1~\frac{S_1}{B_1} + B_2~\frac{S_2}{B_2} = (B_1+B_2)~\frac{S_\text{tot}}{B_\text{tot}}~~.
\end{equation}
This implies, by Jensen's inequality, that for any convex function $f_\text{convex}$,
\begin{equation}
B_1~f_\text{convex}\left(\frac{S_1}{B_1}\right) + B_2~f_\text{convex}\left(\frac{S_2}{B_2}\right) \geq B_\text{tot}~f_\text{convex}\left(\frac{S_\text{tot}}{B_\text{tot}}\right)~~,
\end{equation}
with equality occurring iff $S_1/B_1 = S_2/B_2$. By choosing appropriate forms of $f_\text{convex}$, it can be shown that
\begin{subequations}\label{eqn:inf_loss}
\begin{align}
\frac{S_1^2}{B_1} + \frac{S_2^2}{B_2} &\geq \frac{S_\text{tot}^2}{B_\text{tot}} \label{eqn:inf_loss1}~~,\\
\frac{S_1^2}{N_1} + \frac{S_2^2}{N_2} &\geq \frac{S_\text{tot}^2}{N_\text{tot}} \label{eqn:inf_loss2}~~,\\
\left[-S_1 + N_1 \ln\left(\frac{N_1}{B_1}\right)\right] + \left[-S_2 + N_2 \ln\left(\frac{N_2}{B_2}\right)\right] &\geq -S_\text{tot} + N_\text{tot} \ln\left(\frac{N_\text{tot}}{B_\text{tot}}\right)~~,
\end{align}
\end{subequations}
where 
\begin{equation*}
N_1 \equiv S_1 + B_1~, \qquad N_2 \equiv S_2 + B_2~, \qquad N_\text{tot} \equiv S_\text{tot} + B_\text{tot}~.
\end{equation*}
\Eref{eqn:inf_loss} illustrates the loss of sensitivity of the analysis to the presence of a signal caused by phase-space mixing or dimensionality reduction, with sensitivity quantified by several measures, including the familiar ``$S/\sqrt{B}$ or $S/\sqrt{N}$ added in quadrature''.

Many commonly used analysis techniques, including event selection and categorization, can be thought of as attempts to mitigate the information loss caused by dimensionality reduction. Event categorization helps by separating regions of phase-space, \textbf{that would otherwise be mixed together}, into disjoint categories. Similarly, event selection would help when not including certain parts of phase-space leads to better sensitivity than mixing them with regions of higher purity---in our example, if $S_1^2/B_1 > S_\text{tot}^2/B_\text{tot}$, then rejecting region 2 is better than mixing the two regions together.

\begin{figure}[htp]
\centering
 \includegraphics[width=.5\columnwidth]{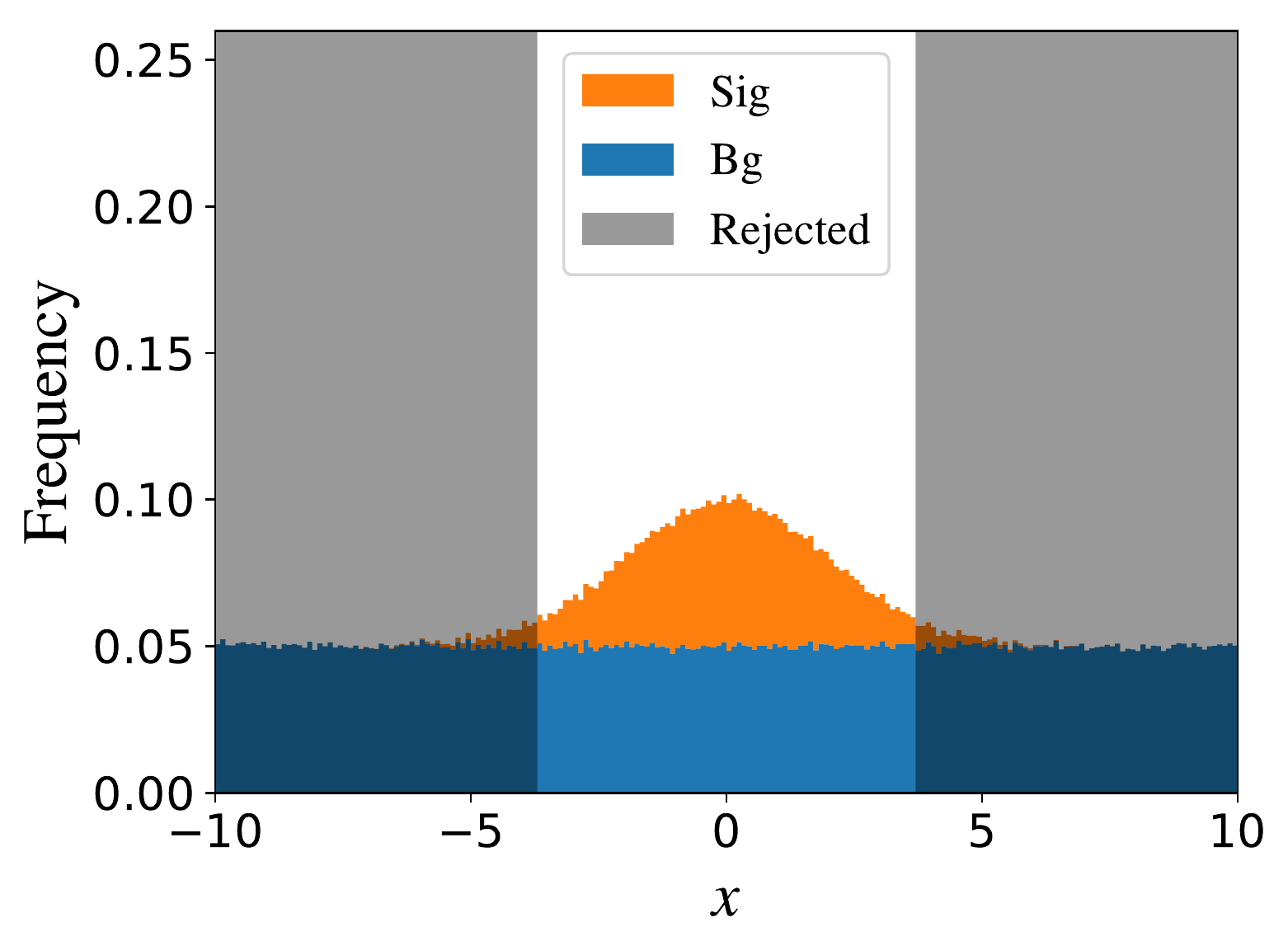}
 \caption{\label{fig:complementary} An illustration of event selection based solely on the event variable to be analyzed after the selection process. The dark shaded region represents the rejected region of phase-space. This kind of event selection does not improve the sensitivity of the analysis. To improve sensitivity, background needs to be selectively removed from each bin as opposed to removing entire bins.}
\end{figure}

Let $\e$ represent all the observable information pertaining to a single collider event. Let $\x$ be an event variable constructed out of $\e$ whose distribution will be analyzed to search for the presence of a signal. In order to mitigate the loss of information caused by dimensionality reduction $\e \rightarrow \x$, event selection or categorization needs to be done based on the information lost during dimensionality reduction, i.e., information complementary to $\x$. Event selection or categorization based on the information contained in $\x$ itself cannot improve the sensitivity of the experiment, and can potentially be detrimental to it.\footnote{Even in the absence of dimensionality reduction, event selection and categorization can still be useful in computationally expensive analyses (like the Matrix Element Method) to reduce the size of the dataset, and hence the computational cost, with an acceptably low loss of sensitivity.} This is illustrated in \fref{fig:complementary}, which shows event selection performed based only on $\x$. This only removes some bins from the analysis, hence removing some non-negative, potentially positive, terms from the sum $\sum\limits_{i\in \x\text{\,bins}} s_i^2/b_i$ (or other measures of sensitivity). 

Typically event selection based on machine learning (ML) classifiers (trained on the full event information $\e$) is done by choosing a working point or selection threshold on the ML output. The problem here is that the ML classifier learns to distinguish between signal and background based on the full information in $\e$, and not just the complementary information. As a result, this selection procedure has to find a compromise between the gain in sensitivity from tapping into complementary information and the loss in sensitivity resulting from using non-complementary information. This problem is particularly severe when $\x$ is a ``good event variable'' in terms of separating signal and background, since such a variable will be strongly correlated with the classifier output.

This is the source of ``misalignment'' addressed in this paper. As we will demonstrate, such a compromise is not necessary, and it is possible to force an event selector to learn from only information complementary to $\x$. This can be done by using different selection thresholds at different values of $\x$. Intuitively this can be understood as follows: To maximize $\sum\limits_{i\in \x\text{\,bins}} s_i^2/b_i$ using an event selector, one needs to maximize $s_i^2/b_i$ in each $\x$-bin, by choosing an appropriate, bin dependent, threshold on the ML classifier output. On the other hand, even the best bin independent threshold will only be a compromise value.

In the case of event categorizers, although using information contained in the event variable $\x$ is not detrimental to the sensitivity, it still does not improve it. So a similar compromise between using complementary and non-complementary information exists in typical ML classifier based categorization as well (flat thresholds between categories), which can also be avoided using $\x$ dependent thresholds between categories.

\begin{figure}
\centering
 \includegraphics[width=.3\columnwidth]{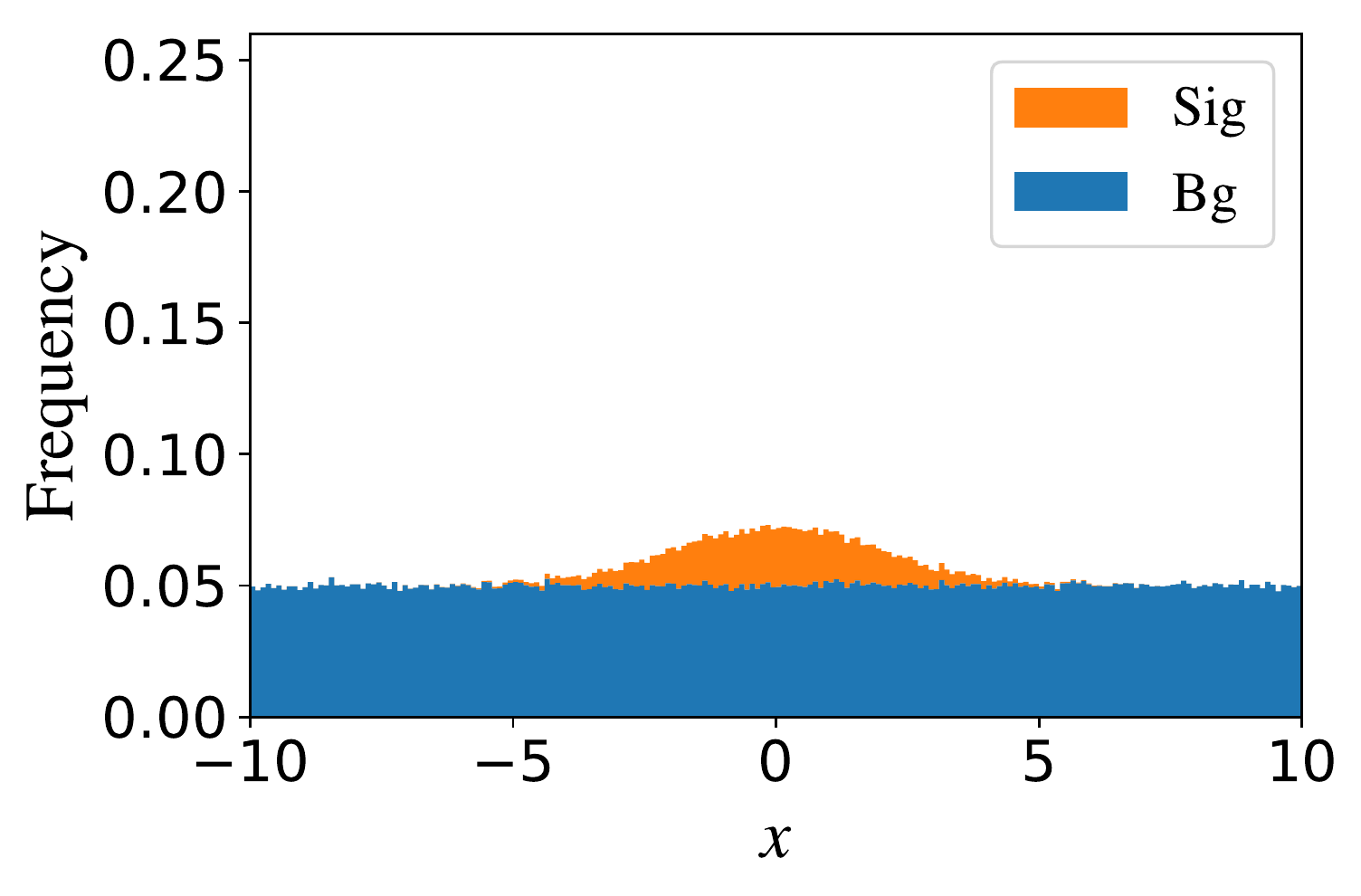}
 \hskip 5mm
 \includegraphics[width=.3\columnwidth]{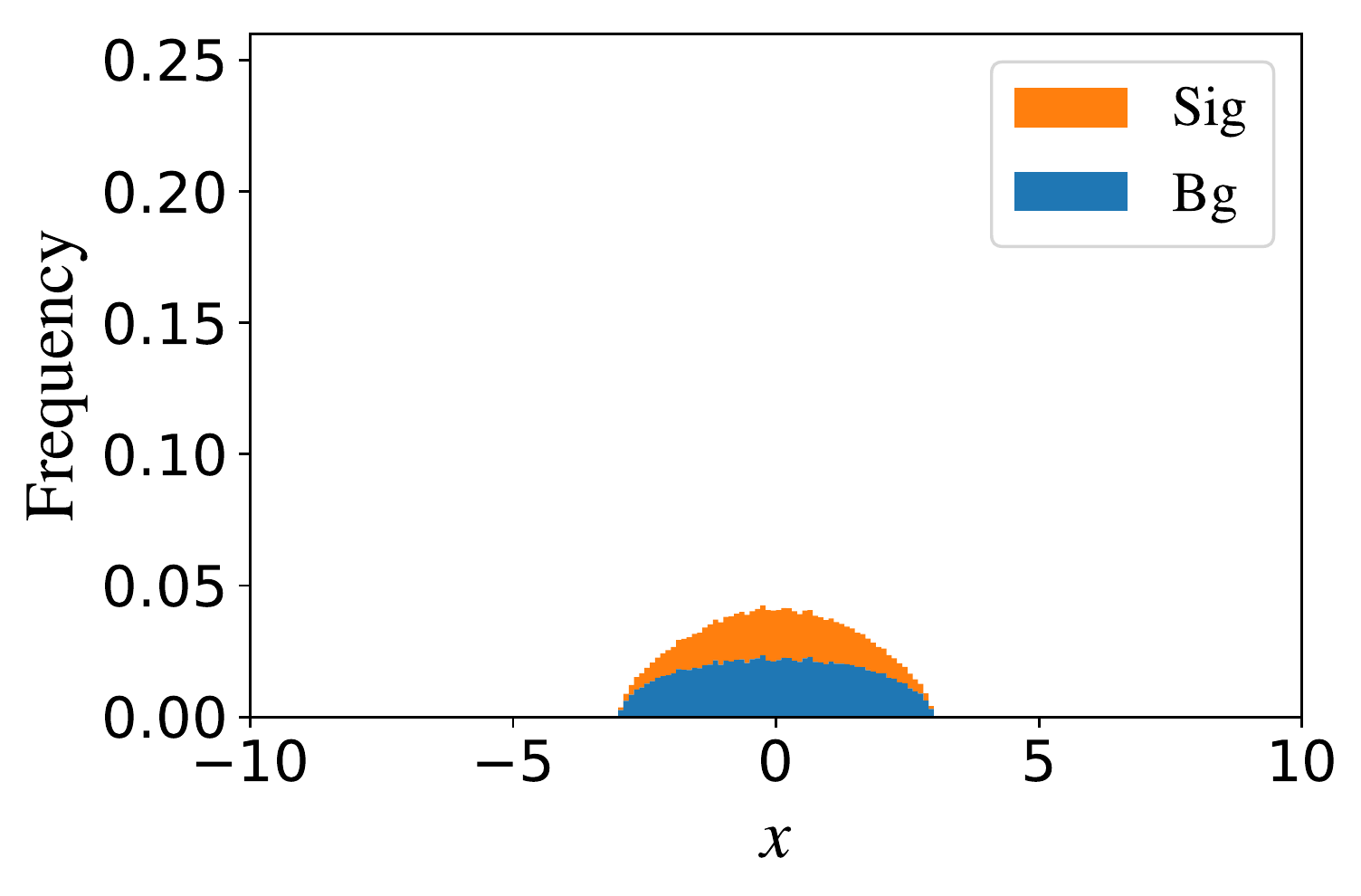}
 \hskip 5mm
 \includegraphics[width=.3\columnwidth]{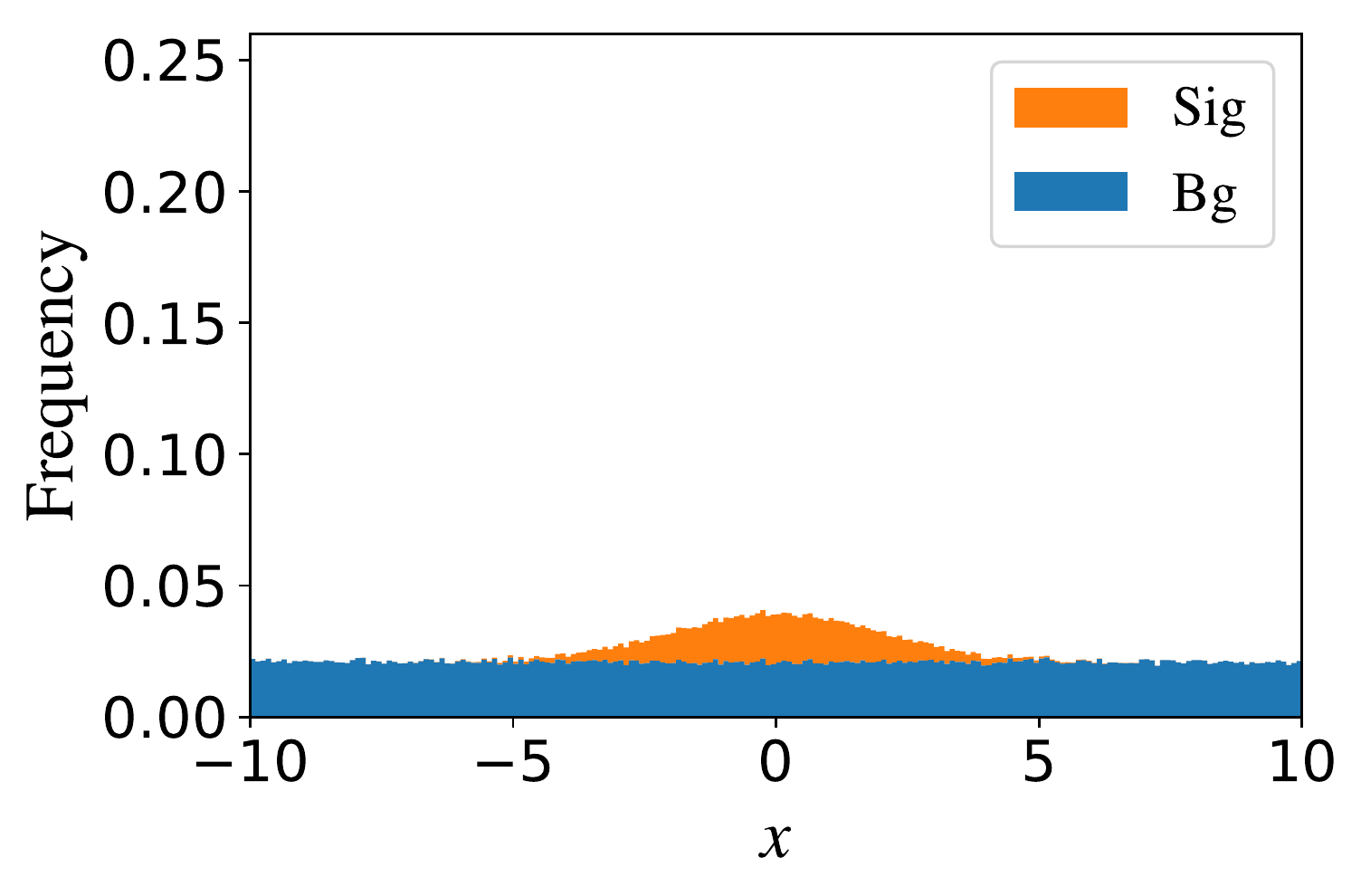}
 \caption{\label{fig:complementary2} Plots from the toy example considered in \sref{subsec:toy}.
The left panel shows the distribution of signal and background before event selection, with the 
background normalized to 1 and a signal-to-background ratio of 1:9. 
The middle panel shows the result from a selection based on full event information, including the event variable $\x$. 
The right panel shows the result from the procedure outlined in \sref{sec:prescription} below, which uses only information complementary to the event variable $\x$. }
\end{figure}

In addition to leading to sub-optimal event selectors and categorizers in terms of statistical sensitivity,  
non-complementary information used in event selection often causes the background distribution (post-selection) to be peaked in the same region as the signal. This amplifies the challenge posed to signal searches by systematic uncertainties in the background specification. Although this paper is not geared towards reducing systematic uncertainties (which we will do in part 3 \cite{Matchev:ML3}), event selectors decorrelated from the event variables $\x$ can make it easier to estimate the background distribution from data itself, thereby reducing the impact of systematic uncertainties. This is illustrated in \fref{fig:complementary2} using plots from a toy example discussed in \sref{subsec:toy}. The left panel shows the distributions of signal and background before selection. The remaining two panels show their distribution post-selection using event selectors based on the full information $\e$ (middle panel) and complementary information only (right panel). We observe that in the middle panel, the background distribution becomes sculpted and peaks in the same region as the signal, while in the right panel, with the procedure which we advocate below, the background distribution remains flat.

\subsection{The landscape of related methods}

It was pointed out in \cite{Whiteson:2007iaai, Whiteson:2009eaai} that the performance of event selectors in terms of classification accuracy is not perfectly correlated with the sensitivity and precision of the physics analysis that uses them---neural networks trained to directly optimize for sensitivity (as in a Higgs search) or precision (as in a top mass measurement) were demonstrated to outperform event selectors trained for maximal classification accuracy (with thresholds or working points chosen to optimize precision). It is precisely this misalignment that we seek to explain and alleviate in this paper and the subsequent\footnote{The findings in \cite{Whiteson:2007iaai, Whiteson:2009eaai} relate more to the forthcoming parts 2 and 3 than the current paper.} work \cite{Matchev:ML2, Matchev:ML3}.

In \cite{Whiteson:2007iaai, Whiteson:2009eaai}, an evolutionary method for training neural networks called NeuroEvolution of Augmenting Topologies (NEAT) \cite{Stanley:2002neat} was used in the direct optimization of event selectors. Another approach to directly optimizing neural networks for discovery significance was introduced in \cite{Elwood:2018qsr}, using cost functions defined over batches of training events. Our prescription, on the other hand, relies on simply using the output of typical ML-based classifiers in an optimal manner.

At the heart of this paper lies the following observation: The task of finding the event selector/categorizer that maximizes the statistical sensitivity of a search analysis (performed in the distribution of an event variable) is intrinsically related to the space of event selectors/categorizers over which the maximization is performed. We claim that selectors/categorizers based only on classifier output (flat thresholds) lack the flexibility needed to achieve global optimality (i.e., in the space of \textbf{all} selectors/categorizers), and expanding the space of categorizers under consideration can potentially lead to improved results. This is in line with the findings of \cite{Bourilkov:2019als}, where, in addition to using thresholds on a BDT classifier output, the flexibility of cutting on another carefully chosen variable was added to the categorization process, leading to a better performing categorizer. We show in \sref{sec:groundwork} that, from an information theory standpoint, global optimality can be attained with our prescription to use the ML classifier output as well as the event variable $\x$, and impose $\x$-dependent thresholds. 

Several approaches exist in the literature for decorrelating the output of a classifier from a particular event variable. One possibility is to use only features decorrelated from the event variable as inputs to the ML classifier \cite{Dolen:2016kst, Aguilar-Saavedra:2017rzt, Chang:2017kvc}. Another way is to force decorrelation during training with appropriate penalties \cite{Stevens:2013dya, Shimmin:2017mfk, Bradshaw:2019ipy}. In our case, the decorrelation is achieved naturally as a byproduct of maximizing performance measures more in line with the sensitivity of the physics analysis than, say, classification accuracy.
\vskip 5mm
The rest of this paper is organized as follows: In \sref{sec:groundwork}, we will introduce several performance measures for event selectors/categorizers, and lay the groundwork for a training prescription to maximize them. We will also explain in more detail how our methods realign the goals of event selection and categorization with the physics goals. \Sref{sec:prescription} will be a description of our training algorithm, intended to serve as an implementation and usage guide. To this end, we will keep the material in \sref{sec:prescription} mostly self-contained, with minimal dependence on \sref{sec:groundwork}. In \sref{sec:thickbrick}, we will demonstrate the working of our algorithm using a toy example analyzed using \href{\thickbrickurl}{ThickBrick} \cite{TB}, our publicly available implementation of the algorithm. Finally, we will provide some outlook on the future of this work and possible extensions in sections~\ref{sec:conclusions} and \ref{sec:epilogue}.

\section{Groundwork} \label{sec:groundwork}
\subsection{Setup and notation}
In this subsection we will establish the setup and notation for a generic signal search procedure considered in this paper. Throughout this paper, we will work under the frequentist approach to statistical inference, so no notion of probability will be associated with the presence or absence of a signal.

Let $\myS$ be a collection of collider events. We will treat a given instance of $\myS$ as one of the possible outcomes of the experiment performed to produce it. The events in $\myS$ are assumed to be independent and identically distributed. The total number of events in $\myS$ is assumed to be a Poisson distributed random variable whose mean could possibly depend on the underlying theory. In practice, $\myS$ can be a collection of events recorded at the collider that pass some event pre-selection criteria.

Let $\e$ be a random vector which contains all the observable information (that is available to an analysis) pertaining to a single collider event. In practice, $\e$ can include \textbf{estimated} number and ids of parton-level final state particles in the event, their \textbf{reconstructed} energies and momenta, etc.

Let $\x$ be an event variable (also a random vector) constructed out of $\e$ whose distribution is to be analyzed to search for the presence of a signal. For example, $\x$ could be the 4-lepton invariant mass for a Higgs search in the 4-lepton channel.

In general, the vectors $\e$ and $\x$ can have continuous (real valued) and discrete components.\footnote{For categorical/discrete valued components of $\e$, like the number and ids of reconstructed final state particles, it is a good idea in practice to group events based on those features and analyze each group separately, although our formalism does not assume this to have been done.} Let $\Omega_\e$ and $\Omega_\x$ represent, respectively, the sample spaces $\e$ and $\x$ belong to, with appropriate probability measure functions.\footnote{Using as $\sigma$-algebra, the product of Borel algebras for real valued components and powersets for discrete valued components.}
\vskip 2mm\noindent
We consider a generic signal search analysis performed in the following two steps:
\begin{enumerate}
 \item[Step 1:] \textbf{Event selection/categorization}
 
 From $\myS$, events are categorized based on their $\e$ values into $C$ categories ($C \geq 1$) numbered $1,\dots,C$. If we allow for event rejection, an event can also be not included in any of the $C$ categories (rejected) based on its $\e$ value. For now we will assume that the number of categories, $C$, is predetermined.
 
 Let $\myS_c$ be the collection of events in category $c$ for $c \in \{1,\dots,C\}$. 
 
 Note that if we use only one category and allow for event rejection, this behaves as a simple in-or-out event selection procedure. \textbf{Henceforth, ``event categorization'' will refer to both event selection and event categorization, unless a distinction is explicitly made.}
 
 \item[Step 2:] \textbf{Number density analysis}
 
 The event variable $\x$ is calculated for events in each of the $C$ categories.
 
 The number density distribution of $\x$ in each of the categories $\myS_c$ is then analyzed to test for the presence of a signal, say, using the likelihood-ratio test or the $\chi^2$ difference test.
 
\end{enumerate}

We will assume that there are two subpopulations of events, background and signal, each with its own exactly specified \textbf{number density} distribution of $\e$. The null hypothesis, $\Hn$ is that only the background subpopulation exists in the true event population, and the alternative hypothesis, $\Ha$, is that both signal and background events exist in the true event population.

The goal of this paper is to provide a prescription to perform the event categorization step (step 1) ``optimally''. To this end, we need a measure of how distinguishable the two hypotheses are after reducing each event to its $\x$ value and the category it belongs to, i.e., after event categorization. We lay the notational groundwork for that in \tref{tab:notation}.

\begin{longtable}{P{.1\textwidth} p{.8\textwidth}}
\caption{\label{tab:notation} Notation}\\

\toprule[1.5\heavyrulewidth]
Notation & \cellcenter{Definition/Interpretation}\\
\midrule[1.5\heavyrulewidth]
$\e$ & Random vector which contains all the observable information (that is available to an analysis) pertaining to a single collider event. Randomness of $\e$ refers to the randomness of collider events.\\
\cmidrule(lr){1-2}
$\x$ or $\x(\e)$ & Event variable (also a random vector) constructed out of $\e$ whose distribution is to be analyzed to search for the presence of a signal.\\
\cmidrule(lr){1-2}
$\Omega_\e$ & The sample space of event $\e$.\\
\cmidrule(lr){1-2}
$\Omega_\x$ & The sample space of event variable $\x$.\\
\cmidrule(lr){1-2}
$\Hn$ & Null hypothesis that only the background subpopulation exists in the true event population.\\
\cmidrule(lr){1-2}
$\Ha$ & Alternative hypothesis that both background and signal subpopulations exist in the true event population.\\
\cmidrule(lr){1-2}
$\eta(\e)$ & Represents the event categorizer used in the analysis. $\eta$ is a function of $\e$ and returns the category (integer) between $1$ and $C$ to place an event in. $\eta$ can also return $\code{None}$ if event rejection is allowed.\\
\cmidrule(lr){1-2}
$B$ & Expected total number of events in the collection $\myS$, under the null hypothesis $\Hn$. It is also the expected number of background events under $\Ha$.\\
\cmidrule(lr){1-2}
$N$ & Expected total number of events in the collection $\myS$, under the alternative hypothesis $\Ha$.\\
\cmidrule(lr){1-2}
$S$ & $S \equiv N - B$. Expected number of signal events under $\Ha$.\\
\cmidrule(lr){1-2}
$b(\e)$ & \textbf{Number density} distribution of $\e$ under the null hypothesis. It is proportional to the probability density function for $\e$ under $\Hn$, and is normalized to $B$.\\
\cmidrule(lr){1-2}
$n(\e)$ & Number density distribution of $\e$ under the alternative hypothesis $\Ha$ (normalized to $N$).\\
\cmidrule(lr){1-2}
$s(\e)$ & $s(\e) \equiv n(\e)-b(\e)$. Number density distribution of signal events under the alternative hypothesis. It is non-negative and normalized to $S$.\\
\cmidrule(lr){1-2}
$B_c$ & Expected number of events in category $c$ under $\Hn$. This also equals the expected number of background events in category $c$ under $\Ha$.\\
\cmidrule(lr){1-2}
$N_c$ & Expected number of events in category $c$ under $\Ha$.\\
\cmidrule(lr){1-2}
$S_c$ & $S_c \equiv N_c - B_c$. Expected number of signal events in category $c$ under $\Ha$.\\
\cmidrule(lr){1-2}
$b_c(\x)$ & Number density distribution of $\x$ in category $c$ under $\Hn$ (background distribution). It is normalized to $B_c$.\\
\cmidrule(lr){1-2}
$n_c(\x)$ & Number density distribution of $\x$ in category $c$ under $\Ha$. It is normalized to $N_c$.\\
\cmidrule(lr){1-2}
$s_c(\x)$ & $s_c(\x) \equiv n_c(\x) - b_c(\x)$. Number density distribution of $\x$ in category $c$ for signal events. It is normalized to $S_c$.\\
\cmidrule(lr){1-2}
$$p(\e)$$ & \begin{equation} p(\e) \equiv \frac{s(\e)}{n(\e)}\end{equation} The probability that an event sampled from the alternative hypothesis $\Ha$ is a signal event, conditional on its $\e$ value.\\
\cmidrule(lr){1-2}
$$p_c(\x)$$ & \begin{equation} p_c(\x) \equiv \frac{s_c(\x)}{n_c(\x)}\end{equation} The probability that an event sampled from the alternative hypothesis $\Ha$ is a signal event, conditional on its $\x$ value and the category $c$ it belongs to.

$p_c(\x)$ can also be thought of as the expected value of $p(\e)$ conditional on $(c, \x)$.\\
\bottomrule[1.5\heavyrulewidth]\\
\textbf{Note :}& \textbf{All quantities with the subscript $c$ implicitly depend on the categorizer $\eta$ used.}\\ \\
\end{longtable}

Note that both the null and the alternative hypotheses are point hypotheses, i.e., there are no free parameters in either $\Hn$ or $\Ha$, including the relative total abundance of signal and background, $S/B$.
The effect of scaling the signal and background cross-sections relative to each other\footnote{An overall scaling of signal and background number densities (keeping $S/B$ fixed) will have no effect on our event selection/categorization prescriptions.} will become apparent at the end of the construction.

\subsection{Statistical distances}
To formalize the problem of optimizing the event selection/categorization procedure, we need a measure of how distinguishable the null and the alternative hypotheses are after reducing each event to the category it belongs to and its $\x$ value. In \eref{eqn:d} we provide six such measures, referred to as \textbf{statistical distances} between the hypotheses. The prescription we develop in this paper is compatible with any of these distance measures; they represent a choice available when using our event categorization prescription.

\begin{subequations} \label{eqn:d}
\begin{align}
\dneym &= \sum\limits_{c=1}^C~\int\limits_{\Omega_\x} d\x~\frac{s_c^2(\x)}{n_c(\x)} \label{eqn:dneym}\\
\dpear &= \sum\limits_{c=1}^C~\int\limits_{\Omega_\x} d\x~\frac{s_c^2(\x)}{b_c(\x)} \label{eqn:dpear}\\
\dkl &= \sum\limits_{c=1}^C~\int\limits_{\Omega_\x} d\x \left[-s_c(\x) - n_c(\x) \ln\left[1-\frac{s_c(\x)}{n_c(\x)}\right]\right] \label{eqn:dkl}\\
\drevkl &= \sum\limits_{c=1}^C~\int\limits_{\Omega_\x} d\x \left[s_c(\x) + b_c(\x) \ln\left[1-\frac{s_c(\x)}{n_c(\x)}\right]\right] \label{eqn:drevkl}\\
\djeff &= \sum\limits_{c=1}^C~\int\limits_{\Omega_\x} d\x \left[-s_c(\x) \ln\left[1-\frac{s_c(\x)}{n_c(\x)}\right]\right] \label{eqn:djeff}\\
\db &= \sum\limits_{c=1}^C~\int\limits_{\Omega_\x} d\x \left[n_c(\x) - \frac{s_c(\x)}{2} - n_c(\x)\sqrt{1-\frac{s_c(\x)}{n_c(\x)}}~\right] \label{eqn:db}
\end{align}
\end{subequations}

We will only provide a brief description of these distances here, and relegate the derivation of these formulas to \aref{appendix:a}. We also provide a comparison between the distance measures in \aref{appendix:d}.
\begin{itemize}
 \item $\dneym$ is the asymptotic expected value of the Neyman-$\chi^2$ statistic \cite{Neyman:1949chi}, sometimes referred to as the modified $\chi^2$ statistic, in the fine binning limit under the alternative hypothesis. This may be familiar as ``$s/\sqrt{n}$ summed over bins in quadrature''.
  \item $\dpear$ is the asymptotic expected value of the Pearson-$\chi^2$ statistic \cite{Pearson:1900chi} in the fine binning limit under the alternative hypothesis. Again, this may be familiar as ``$s/\sqrt{b}$ summed over bins in quadrature''.
  \item $\dkl$ is the Kullback--Leibler divergence \cite{Kullback:1951kldiv} from the null hypothesis to the alternative hypothesis. Up to a factor of $1/2$, this form may be familiar as the (asymptotic) expected significance of excess under the alternative hypothesis, squared \cite{Cowan:2010js}. We provide an alternative interpretation of $\dkl$ in \aref{appendix:b}.
 \item $\drevkl$\footnote{$\drevkl$ is not standard notation. We intend for $\drevkl$ to be read as ``reverse Kullback--Leibler (divergence)'' or ``reverse KL (divergence)''.} is the Kullback--Leibler divergence \cite{Kullback:1951kldiv} from the alternative hypothesis to the null hypothesis. We provide an interpretation of $\drevkl$ in \aref{appendix:b}.
 \item $\djeff$ is Jeffreys divergence \cite{Jeffreys:1946jeff} between the null and alternative hypotheses. It is a symmetrized version of the Kullback--Leibler divergence defined as $\djeff \equiv \dkl + \drevkl$.
 \item $\db$ is the Bhattacharyya distance \cite{Kailath:1967bhatt} between the null and alternative hypotheses.
\end{itemize}
In the limit $s_c(\x) \ll b_c(\x)$ for all $c$ and almost all $\x$, we have
\begin{equation}
 \dneym \approx \dpear \approx 2\dkl \approx 2\drevkl \approx \djeff \approx 8\db
\end{equation}
up to leading order in $s_c(\x)/n_c(\x)$.
Generically we will refer to these distances as $\dstat$.
\begin{equation*}
\dstat \in \left\{\dneym,\dpear,\dkl,\drevkl,\djeff,\db\right\}~~.
\end{equation*}
Some key points about these distances relevant to our discussion are: 
\begin{enumerate}
 \item \textbf{\emph{They capture distinguishability of hypotheses.}} The measures given in \eref{eqn:d} do not depend on actual data. Instead they depend on the two hypotheses, $\Hn$ and $\Ha$, for the distribution of data after event categorization. The greater the value of these distances, the more distinguishable the hypotheses.

Although not explicit from their form, it can be shown that $\dstat \geq 0$, with equality occurring iff $b_c(\x) = n_c(\x)$ for all $c\in \{1,\dots, C\}$ and almost all $\x \in \Omega_{\x}$, i.e., iff the hypotheses $\Hn$ and $\Ha$ are experimentally indiscernible from one another.
 
 \item \textbf{\emph{They depend on the event categorization criteria.}} How events are categorized implicitly affects each of the distance measures. This allows us to phrase the problem of optimizing the event categorization procedure as maximizing $\dstat$ by varying the event categorization criteria.
 
 \item \textbf{\emph{They scale linearly with the integrated luminosity of the experiment.}} Scaling the number of signal and background events proportionally, without changing their distributions or relative abundance, will scale $\dstat$ by the same factor. As a result, we can write each of our statistical distances as $N$ times the expected value of a per-event contribution defined appropriately. We will use this in constructing a  
training prescription to maximize our distance measures.
 
 \item \textbf{\emph{They only depend on $\x$ and $p$ values of events}} that get placed in each category. In particular, they depend on $n_c(\x)$ and $p_c(\x)$---this is akin to the significance of excess depending on the number of events and signal-background ratio.   
Here $p_c(\x)$ is the expected value of $p(\e)$ for events in category $c$ with the given value of $\x(\e)$ (assuming events are sampled from $\Ha$).
Recall that $p(\e)$ and $p_c(\x)$ are the probabilities that an event sampled from $\Ha$ is a signal event, conditioned on $\e$ and $(c, \x)$, respectively.
 
 This means that there exists an optimal categorization criterion based only on the $\x$ and $p$ values of events. This is to be expected from the Neyman--Pearson lemma, a consequence of which is that $p(\e)$ contains all\footnote{Since the total number of events is itself a random variable with different distributions under $\Hn$ and $\Ha$, the mere occurrence/existence of an event in data also carries some information relevant to distinguishing between $\Hn$ and $\Ha$, in addition to $p(\e)$. But in the context of event categorization, this is a moot point.} the information in an event relevant to distinguishing between the hypotheses $\Hn$ and $\Ha$. The optimal categorization criterion will also take into account the $\x$ value of an event to make sure that only information complementary to $\x$ is used in the categorization.
\end{enumerate}

\subsection{Eventwise contribution to \texorpdfstring{$\dneym$}{DNeym}} \label{subsec:dneym}
Let us choose $\dneym$ as a demo distance measure to construct our prescription. For notational simplicity, wherever applicable, $\x$ will refer to $\x(\e)$ of the particular event $\e$ under consideration.

Let us begin with a (training) set of events sampled from $\Ha$, i.e., both signal and background events. We want to train an event categorizer on this event sample to maximize $\dneym$. We will assume that the values of $\x$ and $p(\e)$ are known for each event. $p(\e)$ can be estimated \textbf{separately} using machine learning techniques as shown in \cite{Cranmer:2015bka}. In \sref{subsec:learn_p}, we will discuss some practical aspects of learning $p(\e)$ for our event categorization purpose.

In the interest of providing some intuition for what follows, let us momentarily sacrifice mathematical rigor.\footnote{The following argument can be presented more rigorously by considering a probabilistic event categorizer that returns the probabilities of assigning an event to different categories, and using functional derivatives to capture \emph{small} changes to the categorizer. This, however, is unnecessary for our purposes here.\label{footnote:funcder}} Since we are looking to train an event categorizer, let us look at the effect of modifying a given event categorizer `slightly'. Let us modify an event categorizer so that a given event $\e$ moves from category $c_1$ to category $c_2$. This changes $\dneym$ by changing $s^2_{c_1}(\x)/n_{c_1}(\x)$ and $s^2_{c_2}(\x)/n_{c_2}(\x)$ for $\x = \x(\e)$. 
\begin{align}
 \Delta\left(\frac{s_c^2(\x)}{n_c(\x)}\right) &= -\frac{s^2_c(\x)}{n^2_c(\x)} \Delta n_c(\x) + 2 \frac{s_c(\x)}{n_c(\x)} \Delta s_c(\x)\\
 &= \Delta n_c(\x) \left[-p^2_c(\x) + 2 p_c(\x) p(\e) \right] \label{eqn:heuristic}
\end{align}
Here $\Delta n_c(\x)$ and $\Delta s_c(\x)$ represent small changes to $n_c(\x)$ and $s_c(\x)$ caused by the change in the event categorizer, with $\Delta s_c(\x) = p(\e)\Delta n_c(\x)$. Note that for our modification of moving event $\e$ from $c_1$ to $c_2$, $\Delta n_{c_2} = -\Delta n_{c_1}$. The first term captures the loss (gain) from increasing (decreasing) $n_c$, and the second term captures the gain (loss) from increasing (decreasing) $s_c$. Armed with this intuition, let us rewrite $\dneym$ as follows:

\begin{align}
\dneym &= \sum\limits_{c=1}^C~\int\limits_{\Omega_\x}d\x~n_c(\x)~\frac{s_c^2(\x)}{n_c^2(\x)} \\
&= \sum\limits_{c=1}^C~\int\limits_{\Omega_\e}d\e~n(\e)~\delta\big(c, \eta(\e)\big)~\frac{s_c^2(\x)}{n_c^2(\x)} \label{eqn:step1}\\
&= \int\limits_{\Omega_\e}d\e~n(\e)~\sum\limits_{c=1}^C~\delta\big(c, \eta(\e)\big)~\frac{s_c^2(\x)}{n_c^2(\x)} \label{eqn:step2}\\
&= \int\limits_{\Omega_\e}d\e~n(\e)~\sum\limits_{c=1}^C~\delta\big(c, \eta(\e)\big)~\left[-\frac{s_c^2(\x)}{n_c^2(\x)} + 2\frac{s_c(\x)}{n_c(\x)}\frac{s(\e)}{n(\e)}\right] \label{eqn:step3}\\
&= \int\limits_{\Omega_\e}d\e~n(\e)~\sum\limits_{c=1}^C~\delta\big(c, \eta(\e)\big)~\left[-p_c^2(\x) + 2p_c(\x)p(\e)\right] \label{eqn:step4}\\
&= N\times\mean_{\e\,\sim\,\Ha}~\left[\sum\limits_{c=1}^C~\delta\big(c, \eta(\e)\big)~\left[-p_c^2(\x) + 2p_c(\x)p(\e)\right]\right]~~,\label{eqn:step5}
\end{align}
where $\delta$ is the Kronecker delta, and $\displaystyle\mean_{\e\,\sim\,\Ha}\big[{\ldots}\big]$ represents the expected value computed over events $\e$ sampled from $\Ha$. As a reminder, in \erefs(\ref{eqn:step1}-\ref{eqn:step5}), $\x$ represents $\x(\e)$ of the event $\e$ being integrated over. Note the commonality between the right-hand sides of \erefs\eqref{eqn:heuristic} and \eqref{eqn:step5}.

We will first justify each step of the derivation, and then unpack the power of the last equation for our categorizer training purpose. \Eref{eqn:step1} follows from the fact that
\begin{equation}
n_c(\x') = \int\limits_{\Omega_\e}d(\e)~n(\e)~\delta(c, \eta(\e))~\delta_x(\x'-\x(\e)) ~~,
\end{equation}
where $\delta$ is the Kronecker delta function, and $\delta_x$ satisfies\footnote{$\delta_x$ can be thought of as a product of Kronecker deltas for the discrete components of $\x$, and Dirac delta functions for the real valued components.}
\begin{equation}
\int\limits_{\Omega_\x}d\x'~\delta_x(\x'-\x)~f(\x') = f(\x)~~.
\end{equation}
\Eref{eqn:step3} follows from the fact that $s_c(\x)/n_c(\x)$ is the expected value of $s(\e)/n(\e)$ over events sampled from $\Ha$, conditional on their $c$ and $\x$ values. More concretely, \eref{eqn:step3} can be shown to follow from
\begin{equation}
\frac{s_c(\x)}{n_c(\x)} = \frac{1}{n_c(\x)}~\int\limits_{\Omega_\e}d\e'~n(\e')~\delta(c, \eta(\e'))~\delta_x(\x-\x(\e'))~\frac{s(\e')}{n(\e')}~~. \label{eqn:sampleavg1}
\end{equation}
\Eref{eqn:step4} follows from the definitions of $p_c(\x)$ and $p(\e)$, and \eref{eqn:step5} follows from the definition of expected value of a random variable over events sampled from a certain distribution.

Based on the form of \eref{eqn:step5}, let us define a function $\edneym: [0, 1]\times[0, 1] \rightarrow \mathbb{R}$ as
\begin{equation}
 \edneym(U, V) \equiv -U^2 + 2UV~~. \label{eqn:edneym}
\end{equation}
$\edneym(U, V)$ represents the eventwise contribution to the statistical distance $\dneym$ of adding an event $\e$ with $p(\e)=V$ to a category $c$ with $p_c(\x(\e)) = U$. This lets us write \eref{eqn:step5} as
\begin{equation}
 \frac{\dneym}{N} = \mean_{\e\,\sim\,\Ha}~\bigg[\sum\limits_{c=1}^C~\delta(c, \eta(\e))~\edneym\big(p_c(\x), p(\e)\big)\bigg]~~.\label{eqn:eventwiseneym}
\end{equation}

To understand the meaning of the phrase ``eventwise contribution to the statistical distance'', let us see how \eref{eqn:eventwiseneym} can be used to estimate $\dneym\,/\,N$ for a given event categorizer $\eta$.
\vskip 2mm\noindent
Estimating $\dneym\,/\,N$ for a \textbf{given event categorizer} $\eta$:
\begin{enumerate}
 \item Use the event categorizer $\eta$ to categorize events sampled from $\Ha$.
 \item Estimate $p_c(\x)$ for all $c$ and $\x$ using the categorized events. This can be done either as the average of $p(\e)$ conditional on $(c, \x)$, or directly from the label carried by each training event indicating whether it is a signal or background event.
 \item Next for each event $\e$, estimate its contribution to $\dneym$ as $\edneym\big(p_c(\x), p(\e)\big)$, where $c$ is the category it belongs to \textbf{under the given event categorizer $\eta$}. If event rejection is allowed, the contribution from rejected events is just $0$. The average value of this contribution over training events is an estimate for $\dneym\,/\,N$.\footnote{Note that $N$ is the expected number of events in the real experiment under $\Ha$, not the number of training events.}
\end{enumerate}

In this way, the $\edneym$ term represents the eventwise contribution to $\dneym$. While this is admittedly a roundabout way of estimating $\dneym$ from a training sample, it leads to a method of training an optimal event categorizer iteratively, as we will see next.

\subsection{Iterative maximization of \texorpdfstring{$\dneym$}{DNeym}} \label{subsec:iterative}
The idea behind writing the statistical distance, $\dneym$, in terms of an eventwise contribution, $\edneym$, is that from event categorizer $\eta$, we can get a ``better'' event categorizer $\bar{\eta}$ by requiring that $\bar{\eta}$ places each event in the category that maximizes $\edneym\big(p_c(\x), p(\e)\big)$ (with $p_c(\x)$ corresponding to the categorizer $\eta$).
\vskip 2mm\noindent
This leads to the following iterative prescription to train an event categorizer:
\begin{enumerate}
 \item Let the current categorizer be $\eta$. Categorize events sampled from $\Ha$ as per $\eta$.
 \item Estimate $p_c(\x)$ for all $c$ and $\x$ (for the current $\eta$) using the categorized events.
 \item Construct a new categorizer $\bar{\eta}$ which places each event $\e$ in the category $c$ with the highest $\edneym\big(p_c(\x), p(\e)\big)$. If event rejection is allowed, then an event will be rejected by $\bar{\eta}$, iff $\edneym\big(p_c(\x), p(\e)\big) < 0$ for all categories $c$.
 \item Set $\eta = \bar{\eta}$. Repeat from step 1. Stop using some appropriate termination condition.
\end{enumerate}
We will postpone the detailed description of this training procedure to \sref{sec:prescription}, which is intended to serve as a usage guide independent from the build-up we are doing now.

\begin{figure}[htp]
\centering
 \includegraphics[width=.49\textwidth]{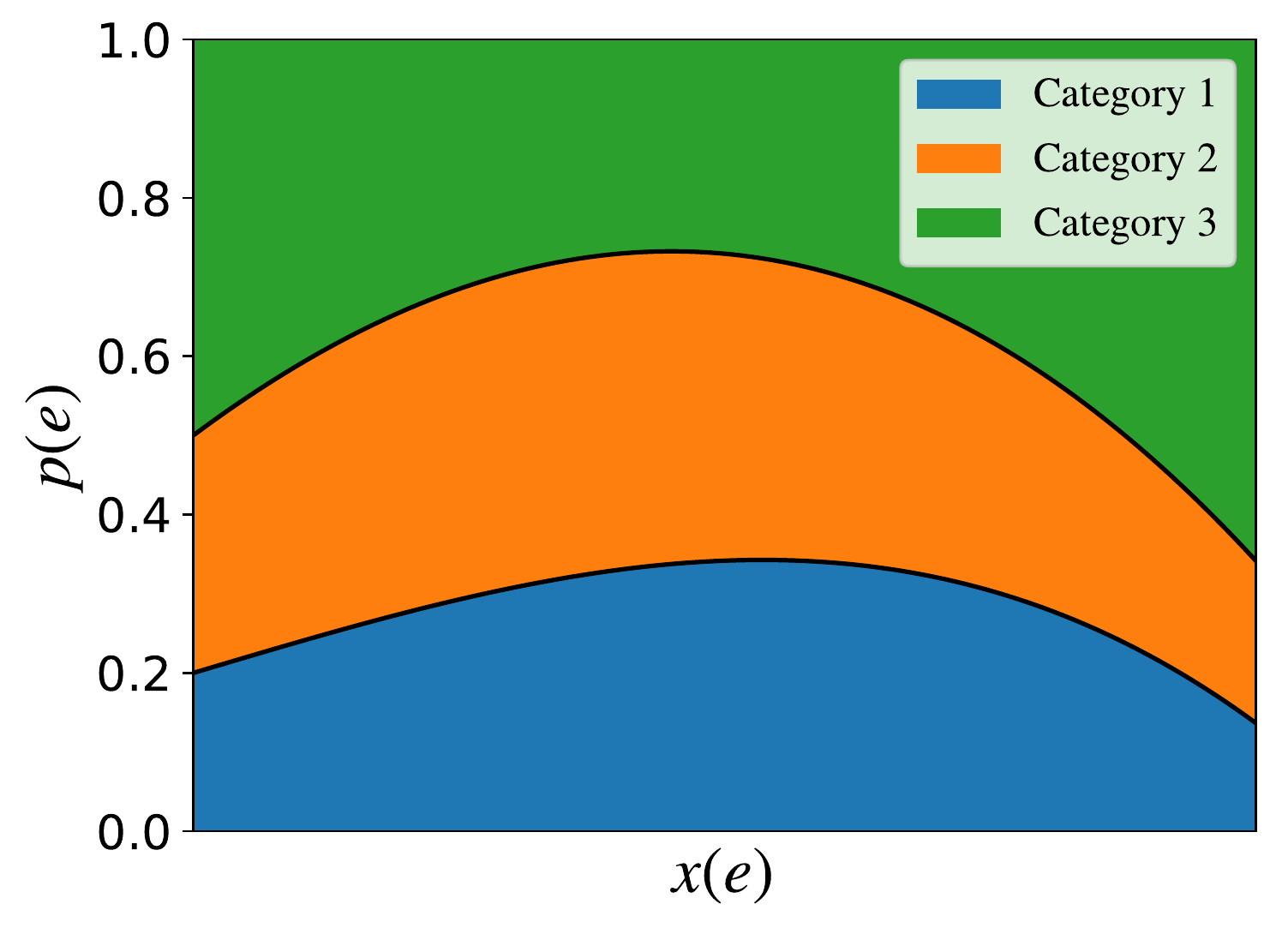}
 \caption{\label{fig:cartoon} A cartoon depiction of $\x$ dependent boundaries on $p(\e)$ between different categories.}
\end{figure}

But first, to gain some intuition, let us take a closer look at what the iterative prescription described above does operationally. From a given categorizer $\eta$ with corresponding signal-fraction functions $p_c(\x)$, we build a new categorizer $\bar{\eta}$ as\footnote{To keep the presentation simple, we are ignoring the case where event rejection is allowed. $\bar{\eta}(\e)$ will be \emph{rejected} or \code{None} if every category leads to a negative contribution.}
\begin{equation}
\bar{\eta}(\e) = \argmax_{c} \bigg[\edneym\big(p_c(\x), p(\e)\big)\bigg]~~, \label{eqn:nextneym}
\end{equation}
where $\edneym(U, V) = -U^2+2UV$. Note that $\edneym(U, V)$ has a global maximum in the first argument at $U=V$, and monotonically decreases as $U$ moves away from $V$. It is also symmetric\footnote{Maximum in the first argument at $U=V$ and monotonic decrease as $U$ moves away from this maximum will hold for the eventwise contributions for other statistical distances as well, but not the symmetricity.} in $U$ about this maximum. Combining this observation with \eref{eqn:nextneym}, we can see that $\bar{\eta}$ places each event in the category with $p_c(\x)$ closest to $p(\e)$. We can think of $\bar{\eta}$ as grouping events by their $p(\e)$ values with $\x$ dependent boundaries between different categories---\fref{fig:cartoon} shows a cartoon of what such a categorization could look like. This makes sense considering that the purpose of categorization is to reduce the information loss arising from mixing up regions of phase-space with different $p(\e)$ values when reducing the full event $\e$ to the event variable $\x$.

Recall that $p_c(\x)$ is the mean $p$ in category $c$ conditional on $\x$. This process of iteratively placing events in the category with the closest mean, and then recomputing the mean for each category could be reminiscent of Lloyd's k-means clustering algorithm \cite{Lloyd:1982kmeans, Forgy:1965kmeans}. This connection can be made clearer by noting from \eref{eqn:edneym} that
\begin{equation}
 \edneym\big(p_c(\x), p(\e)\big) = -\big[p(\e)-p_c(\x)\big]^2 + p(\e)^2~~.
\end{equation}
The first term is congruous to $[X-\langle X\rangle_c]^2$, the objective function k-means clustering seeks to minimize the average of, where $\langle X\rangle_c$ is the sample mean within cluster $c$ which contains $X$, and the last term is independent of the event categorization criteria. More explicitly, the objective function to be minimized in k-means clustering can be written as
\begin{equation}
\sum\limits_{X}~\sum\limits_{\substack{\text{cluster}\\c}} \delta\left(c,~\genfrac{}{}{0pt}{0}{\text{cluster containing}}{X}\right) \Big[X-\langle X\rangle_c\Big]^2~~,
\end{equation}
which bears close resemblance to the right-hand side of \eref{eqn:eventwiseneym}. In fact, for the case of $\dneym$, our iterative training procedure can be thought of as the k-means clustering algorithm with events within each $\x$-``bin'' being clustered based on their $p(\e)$ values. But instead of doing this bin-by-bin in $\x$, we can take an unbinned approach to the estimation of $p_c(\x)$. We will elaborate more on this in \sref{sec:prescription}.

Note that we have not yet proved that the categorizer will ``improve'' during each iterative step or that it will converge on an optimal categorizer in a finite number of steps---at this point the algorithm is held together by the heuristic argument presented in \eref{eqn:heuristic} and the connection to k-means clustering. In \aref{appendix:c}, we present a proof of the correctness of the iterative training algorithm (for $\dneym$, as well as the other statistical distances). The proof will be similar to that of the generic k-means clustering algorithm, and will rely on the fact that $\edneym\big(p_c(\x), p(\e)\big)$ and its counterparts for the other distances are linear in $p(\e)$ and have a global maximum in the first argument at $p_c(\x) = p(\e)$.

\subsection{Generalizing to other statistical distances}
We can repeat the process in \sref{subsec:dneym} for the other statistical distances in \eref{eqn:d} and generate eventwise contributions to the other statistical distances as well. We will skip the derivation and only present the relevant results here. First let us define $\tdstat$ as
\begin{subequations} \label{eqn:td}
\begin{align}
\tdneym &\equiv ~~\frac{\dneym}{N}~~,\\
\tdpear &\equiv ~~\frac{\dpear}{N}~~,\\
\tdkl &\equiv 2~\frac{\dkl}{N}~~,\\
\tdrevkl &\equiv 2~\frac{\drevkl}{N}~~,\\
\tdjeff &\equiv ~~\frac{\djeff}{N}~~,\\
\tdb &\equiv 8~\frac{\db}{N}~~.
\end{align}
\end{subequations}
The numerical factors of $2$ and $8$ are introduced to make all the $\tdstat$-s converge to each other in the limit $s_c(\x)\,/\,n_c(\x) \rightarrow 0$ for all $c$ and almost all $\x$. 
For all $\tdstat$-s, we have
\begin{equation}
 \tdstat = \mean_{\e\,\sim\,\Ha}~\left[\sum\limits_{c=1}^C~\delta(c, \eta(\e))~\edstat\big(p_c(\x), p(\e)\big)\right]~~, \label{eqn:eventwisestat}
\end{equation}
where the functions $\edstat: [0, 1)\times[0, 1] \rightarrow \mathbb{R}$ are given by

\begin{subequations} \label{eqn:ed}
\begin{alignat}{2}
\edneym(U, V)~&= ~~~~~~~~~~~~~-U^2~&&+ ~~~~~~~~~~~~~2UV~~, \label{eqn:ed1}\\
\edpear(U, V)~&= ~~~~~~~~-\frac{U^2}{(1-U)^2}~&&+ ~~~~~~~~\frac{(2-U)UV}{(1-U)^2}~~,\\
\edkl(U, V)~&= -2\left[\ln(1-U) + \frac{U}{1-U}\right]~&&+ ~~~~~~~~~~~\frac{2UV}{1-U}~~,\\
\edrevkl(U, V)~&= ~~~~~2\big[\ln(1-U) + U\big]~&&- ~~~~~~~~2\ln(1-U)V~~,\\
\edjeff(U, V)~&= ~~~~~~~~~-\frac{U^2}{1-U}~&&+ ~\left[\frac{U}{1-U} - \ln(1-U)\right]~V~~,\\
\edb(U, V)~&= ~~~~~~4\left[2-\frac{2-U}{\sqrt{1-U}}\right]~&&+ ~~~~~4\left[\frac{1}{\sqrt{1-U}}-1\right]~V~~. \label{eqn:ed6}
\end{alignat}
\end{subequations}
With these choices, the validity of \eref{eqn:eventwisestat} can be verified (after some manipulation) by noting that because $\edstat\big(p_c(\x), p(\e)\big)$ is linear in $p(\e)$, it can be replaced by $\edstat\big(p_c(\x), p_c(\x)\big)$ in \eref{eqn:eventwisestat}.

For each $\edstat$ it can be shown that for all $V\in[0, 1]$, $\edstat(U, V)$ is monotonically increasing in $U$ for $0\leq U\leq V$ and monotonically decreasing in $U$ for $V \leq U < 1$, with global maximum in $U$ attained at $U=V$. This means that in each iterative step, the ``next'' categorizer will place event $\e$ in one of the two categories with ``current'' $p_c(\x)$ closest to $p(\e)$ on either side, or reject the event if that option is allowed.

\subsection{The rectified ``misalignment''} \label{subsec:rectmisalign}

In the introductory \sref{intro:dimreduce}, the \emph{source} of misalignment between the physics goals and typical ML-based event selectors
was attributed to not ensuring that categorization is based only on information complementary to the event variable $\x$. Now we are in a position to point out \emph{what} that misalignment is. In typical ML-based selector training, all signal events are treated on the same footing and all background events are treated on the same footing, in the sense that the reward/penalty for including/excluding an event depends only on whether the event is from the signal or background sample. On the other hand, our eventwise contribution functions $\edstat(p_c(\x), p(\e))$ suggest that the reward/penalty should also depend on the $\x$ value of the event. For the rest of this subsection, let us restrict our attention to event selectors only, for simplicity.

\begin{figure}[htp]
 \includegraphics[width=.48\columnwidth]{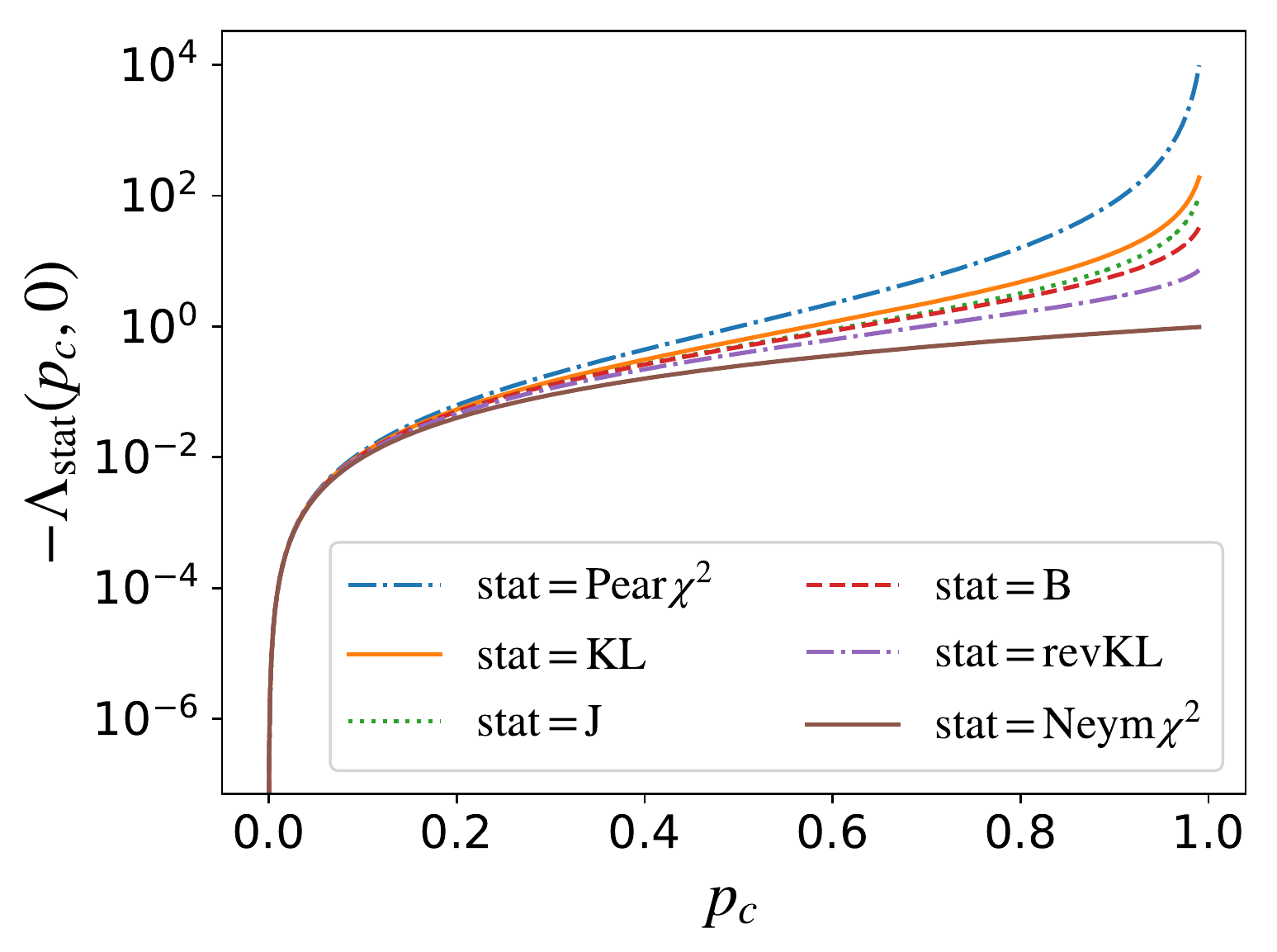}
 \hskip 3mm
 \includegraphics[width=.48\columnwidth]{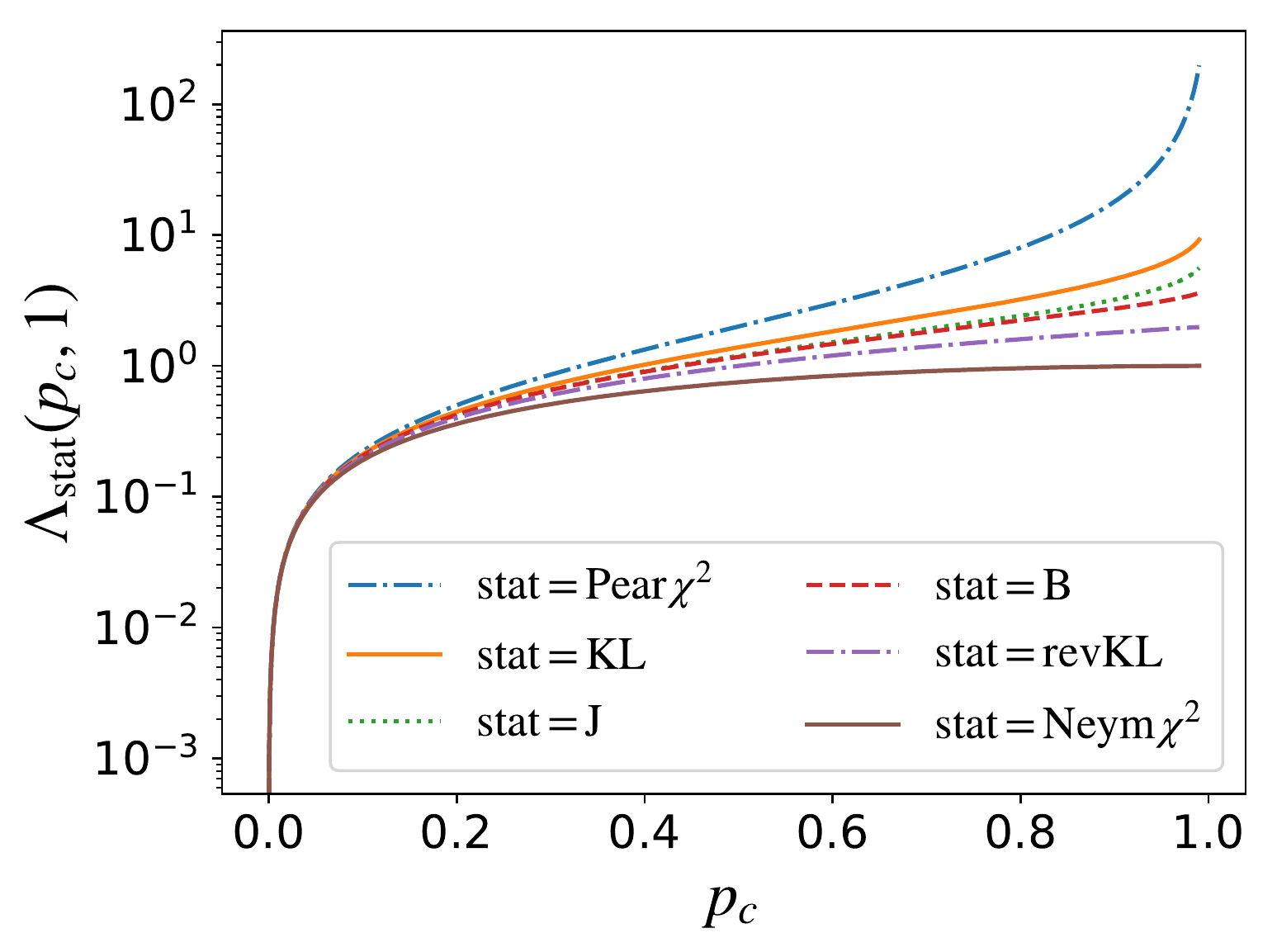}
 \caption{\label{fig:themisalignment} The penalty (left panel) and the reward (right panel) associated with selecting a background and a signal event, respectively, into an $\x$-bin of purity $p_c$. The $y$-axes of the plots are in log-scale.}
\end{figure}

Negative $\edstat(p_c, 0)$ represents the penalty for selecting a background event in an $\x$-``bin'' of signal purity $p_c$. As can be seen from the left panel of \fref{fig:themisalignment}, this penalty increases with $p_c$ for each of our distance measures. Similarly, $\edstat(p_c, 1)$ represents the reward for selecting a signal event in an $\x$-``bin'' of signal purity $p_c$. As can be seen from the right panel of \fref{fig:themisalignment}, this reward also increases with $p_c$.

\begin{figure}[htp]
\centering
 \includegraphics[width=.55\columnwidth]{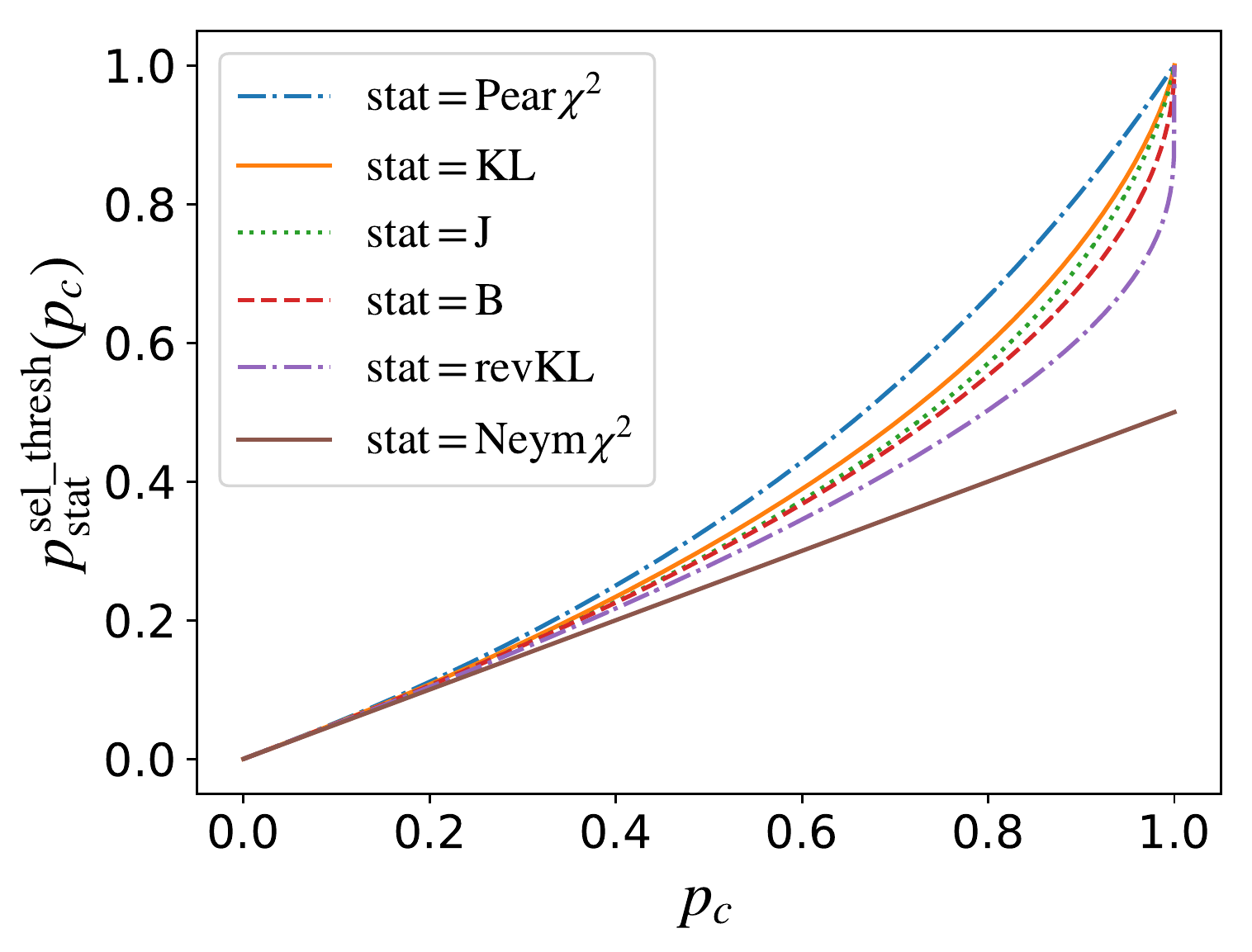}
 \caption{\label{fig:themisalignment2} The selection threshold $p^\text{sel\_thresh}_\text{stat}\left(p_c\right)$ for the different statistical distances.}
\end{figure}

An event $\e$ will warrant selection only if the following net-reward corresponding to $p(\e)$ is non-negative:
\begin{equation}
\text{net-reward} = \Big(1-p(\e)\Big) \edstat\big(p_c(\x), 0\big) + p(\e) \edstat\big(p_c(\x), 1\big)
\end{equation}
This net-reward is precisely $\edstat\big(p_c(\x), p(\e)\big)$. Let $p^\text{sel\_thresh}_\text{stat}(p_c)$ be the selection threshold on $p(\e)$ above which an event warrants being selected into a bin of purity $p_c$. This threshold is defined by the equation
\begin{equation}
\edstat\left[p_c,~p^\text{sel\_thresh}_\text{stat}\left(p_c\right)\right] = 0~~. \label{eqn:selthresh}
\end{equation}
\Fref{fig:themisalignment2} shows that this threshold is an increasing function of $p_c$ for each of our statistical distances, implying that the threshold is stronger in $\x$-``bins'' of higher purity. This is how our prescription naturally leads to decorrelation of the event selection criteria from the event variable $\x$.

Failure to incorporate these $\x$ dependences, or more precisely the $p_c(\x)$ dependences, is the misalignment present in typical ML-based selector training that our prescription rectifies. The fact that $p_c(\x)$ itself depends on the event selector is the reason our training prescription changes the per-event reward function being maximized at each iterative step.

\section{Prescription for training an optimal event categorizer}\label{sec:prescription}

In this section we provide a description of our training prescription.
Operationally, \erefs\eqref{eqn:eventwisestat} and \eqref{eqn:ed}, along with the definitions of $p(\e)$ and $p_c(\x)$ in \tref{tab:notation}, capture almost all the information we need to carry from the previous section (groundwork) into this one.

Let us begin with a sample of simulated events containing both signal and background, split into training and validation sets. For every simulated event $i$:
\begin{itemize}
 \item Let $\e_i$ represent all the observable information from the event.
 \item Let $w_i$ be the non-negative weight associated with the event.\footnote{We accommodate the possibility of weighted events in order to allow for easy scans of BSM parameter spaces by simple reweighting of training events using the  matrix element \cite{Albertsson:2018maf, Gainer:2014bta, Mattelaer:2016gcx}.}
 \item Let $y_i$ be the label specifying the subsample the event is from: 1 for signal, 0 for background.
 \item Let $\x_i = \x(\e_i)$ be the value of the event variable $\x$ whose distribution is to be analyzed to search for the presence of a signal.
\end{itemize}
Let $\tilde{S}/\tilde{B}$ be the overall signal to background ratio in the \textbf{training} sample. We will \textbf{not} assume that this matches the overall signal to background ratio $S/B$ expected in an experiment under the hypothesis $\Ha$. This is to allow for the fact that this ratio could be a priori unknown in the search (if the signal cross-section is an unspecified parameter in the theory). Even in cases when the ratio $S/B$ is known, working with a highly imbalanced sample with very different values for
$\tilde{S}$ and $\tilde{B}$ can make learning $p(\e)$ using ML techniques difficult. So it is often preferable to use training samples with $\tilde{S}/\tilde{B}$ close to 1.

\subsection{Preprocessing: Learning \texorpdfstring{$p(e)$}{p(e)}} \label{subsec:learn_p}

From the training data we need to estimate $p(\e)$. If the dimensionality of $\e$ is large, this can be done with machine learning techniques \cite{Cranmer:2015bka}. The accuracy of the estimate will depend on several factors including, but not limited to, the amount of computational resources and training data available.

We can address the limitation posed by inaccuracies in the estimation of $p(\e)$ by treating the output of the classifying algorithm (BDT or neural network for example) not as $p(\e)$ but simply as a property of the event to base our categorization on.
Let $O_\text{ML}(\e)$ be the machine learning classifier output. We can estimate $p(\e)$ as the probability that an event sampled from $\Ha$ will be a signal event conditional on \textbf{both} $O_\text{ML}(\e)$ and $\x(\e)$, as opposed to conditional on $\e$. Presumably the dimensionality of $\x(\e)$ is small enough to perform this estimation with sufficient accuracy. This can be done either using traditional regression methods (parameterized or unparameterized), or with another machine learning layer (with $O_\text{ML}$ and $\x$ as inputs). Henceforth, $p(\e_i)$ and $p_i$ will both refer to this \textbf{\emph{estimate}} of $p(\e_i)$.\footnote{For convenience, we are not introducing new notation, like $\hat{p}$, to refer to the \emph{estimate} of $p$.}

Out of the six statistical distance measures, only $\tdpear$ maximization leads to a categorizer that is independent of the value of the overall signal to background ratio $S/B$ under $\Ha$. Training using the other five measures is sensitive to $S/B$; if training is to be done using these measures, $w_i$ and $p_i$ both need to be appropriately scaled before proceeding further. To scale the signal cross-section by a factor of $\lambda$ (i.e., if $S/B = \lambda\tilde{S}/\tilde{B}$), we need to apply the following transformations\footnote{In our python implementation ThickBrick \cite{TB} there is an option to handle this scaling internally.}
\begin{subequations} \label{eqn:transform}
\begin{align}
w_i &\rightarrow \lambda w_i~~~\text{ if } y_i=1 \text{, i.e., for signal events}\label{eqn:transform1}\\
p_i &\rightarrow \frac{\lambda p_i}{1 + (\lambda-1) p_i}~~~~\text{ for all events } i\label{eqn:transform2}
\end{align}
\end{subequations}
In searches where there is no a priori expectation for the size of the signal cross-section\footnote{SUSY and UED searches would be examples where there exists an a priori expectation for the signal cross-section, which is governed by the known gauge couplings and BSM partner spins.}, one option is to use $\tdpear$ as the distance measure or choose a very small value of $\lambda$ (which is equivalent to using $\tdpear$). Another option is to choose a value of $\lambda$ that puts the signal cross-section either near the low end of discoverability or near the potential upper-limit on the signal cross-section (low end of rejectability) expected from the available experimental data.

At the end of this preprocessing, each training event should have $w$, $y$, $\x$, and $p$ associated with it (possibly scaled). In addition, we should have a way of computing $\x$ and $p$ for validation MC events as well as real events from the experiment (by first running the event $\e$ through the machine learning classifier to get $O_\text{ML}(\e)$ and using that to get $p$, performing the scaling operations in \eref{eqn:transform} if applicable).

It is important that the following equation approximately holds true for validation MC events sampled from $\Ha$:
\begin{equation}
p \approx \frac{\mean\limits_{\e\,\sim\,\Ha}\Big[~wy~\big|~\x, p~\Big]}{\mean\limits_{\e\,\sim\,\Ha}\Big[~w~\big|~\x, p~\Big]} \label{eqn:preprocessing}
\end{equation}
\Eref{eqn:preprocessing} just means that $p$ approximately equals the probability of an event sampled from $\Ha$ being a signal event, conditional on $(\x, p)$, with the inexactness resulting from training limitations.

\subsection{Choices to be made} \label{subsec:choices}
\begin{enumerate}
\item \textbf{\emph{Choose a distance measure to train with.}} As mentioned earlier, our prescription involves choosing a distance measure between the `background only' and `background+signal' hypotheses to be maximized by the event categorizer. The possible options are $\tdneym$, $\tdpear$, $\tdkl$, $\tdrevkl$, $\tdjeff$, and $\tdb$.

As mentioned above in \sref{subsec:learn_p}, maximizing $\tdpear$ leads to a categorizer whose optimality is independent of the overall signal to background ratio. This is because scaling the signal number density by \textbf{an overall factor} ($s(\e) \rightarrow \lambda s(\e)$) only changes $\dpear$ by a multiplicative factor ($\dpear \rightarrow \lambda^2 \dpear$). This makes $\tdpear$ a good option when there is a priori no expectation for the potential signal cross-section.

We show in \aref{appendix:b} that training using $\tdrevkl$ (with training signal cross-section near the low end of discoverability) is theoretically optimal for signal discovery. Similarly, training using $\tdkl$ (with training signal cross-section near the expected upper-limit on it) is theoretically optimal for upper-limit setting.

In practice, the main difference between the different distance measures lies in the ($\x$ dependent) boundaries on the values of $p$ between different categories---they have different thresholds to move events from a low purity category to a high purity category. Let us call a distance measure that leads to higher thresholds as `stricter'. We show in \aref{appendix:d} that the distance measures arranged in the increasing order of strictness are as follows (left to right, least strict to most strict):
\begin{equation}
\tdneym \myprec \tdrevkl \myprec \tdb \myprec \tdjeff \myprec \tdkl \myprec \tdpear \label{eqn:strictorder}
\end{equation}
Here, the notation $\tdneym \myprec \tdrevkl$ just means that $\tdneym$ precedes $\tdrevkl$ in order of strictness (as described above).
Note that the hierarchy~\eqref{eqn:strictorder} is reflected in the ordering of the selection thresholds in \fref{fig:themisalignment2}.

One might want to try out different distance measures, and pick one based on practical considerations like having sufficient number of events in each category, shape features of the distribution of $\x$ post-categorization, etc.---it is possible that background MC validation (post-categorization) is easier when using certain distance measures than others.

Henceforth, where applicable, $\tdstat$ will refer to the chosen distance measure.
\item \textbf{\emph{Choose the number of categories }}\code{(int C $\geq$ 1)}\textbf{\emph{ and whether or not to allow event rejection }}\code{(bool allow\_rejection)}\textbf{\emph{.}}

To use our prescription to train an event selector (as opposed to a categorizer), set the number of categories \code{C} to \code{1}, and set \code{allow\_rejection} to \code{True}.

Theoretically, in the asymptotic (large number of events) limit, having more categories will lead to higher sensitivity. In fact, in the infinite categories limit, the sensitivity of the search will approach the sensitivity of a likelihood-ratio test performed in the full phase-space $\e$, which is the theoretical upper-limit of the sensitivity of the search. Also, for a given number of categories, allowing event rejection will lead to higher sensitivity than disallowing it. In practice, however, increasing the number of categories will lead to data getting diluted out across categories, which can lead to problems.

Fortunately, though, continually increasing the number of categories typically leads to diminishing returns.\footnote{The diminishing returns can be attributed to the differentiable behavior of $\edstat(U, V)$ in parameter $U$ at/near the maximum $U=V$. As a result, the returns from creating more categories to reduce the distance between $p(\e)$ and the closest $p_c(\x)$ are diminishing.} In \sref{subsec:performance} we will show how to estimate the value of $\tdstat$ (post training) 
in the \code{C}$\,\,\,\rightarrow\infty$ limit. This can be used to assess the utility of increasing the number of categories and weigh it against the corresponding disadvantages, in order to settle on a good value of \code{C}.
\end{enumerate}

\subsection{Iterative training prescription} \label{subsec:iterative_algorithm}
\begin{algorithm}[htp]
\label{algorithm}
\DontPrintSemicolon
\SetKwProg{Fn}{define}{:}{end~definition}
\SetKwFunction{FNext}{Next\_categorizer}
\SetKw{Update}{update:}
\SetKw{Initialize}{initialize:}
\SetAlgoLined

\BlankLine
\KwResult{Trained categorizer $\eta$}
 \BlankLine
 Preprocess training events (get $w$, $y$, $\x$, and $p$ for all events)\;
 Choose $\tdstat$, \code{C}, \code{allow\_rejection}\;
 \Initialize categorizer $\eta$, a function of $\x$ and $p$\;
 \BlankLine\BlankLine
 \Repeat{{\normalfont appropriate termination condition based on training convergence}}{
  \tcc{Code block for one iterative step}
  Categorize events in the training set based on their $\x$ and $p$ values, as per $\eta$\;
  Estimate the functions $p_c(\x)$ for $c\in \{1,...,\code{C}\}$ (for current $\eta$)\;
  \BlankLine\BlankLine\BlankLine
  \Fn{\FNext{$\x$, $p$}}{
  \code{$p_c\text{\_list}$\,=\,\textbf{sorted}}\Big($\big[p_c(\x)$ for $c \in \{1,...\normalfont\code{C}\}\big]\Big)$\;
  {\normalfont\code{max\_$\Lambda$ = }}\code{$\max\limits_{\text{i}\in \{1,...\text{C}\}}\bigg[\edstat\Big[p_c\text{\_list[i]}, p\Big]\bigg]$}\;
  {\normalfont\code{best\_c = }}\code{$\argmax\limits_{\text{i}\in \{1,...\text{C}\}}\bigg[\edstat\Big[p_c\text{\_list[i]}, p\Big]\bigg]$}\;
  \eIf{\normalfont\code{(max\_$\Lambda \geq 0$)}}
  {\Return {\normalfont\code{best\_c}}}
  {
  \eIf{\normalfont\code{(allow\_rejection == True)}}
  {\Return {\normalfont\code{rejected}}}
  {\Return {\normalfont\code{best\_c}}}
  }
  }
  \BlankLine\BlankLine\BlankLine
  \Update $\eta$ = \FNext\;
  \BlankLine\BlankLine
 }
 \caption{Iterative prescription to train $\eta$}
\end{algorithm}
Algorithm~\ref{algorithm} shows the iterative event categorizer training prescription introduced in \sref{subsec:iterative}. We have already motivated the algorithm in \sref{sec:groundwork}, and the proof of correctness has been relegated to \aref{appendix:c}. Here, we will elaborate on different parts of the algorithm from an implementation point of view, as well as provide diagnostic notes from a practical usage point of view.

Lines 1 and 2 refer to the preprocessing covered in \sref{subsec:learn_p} and the choices covered in \sref{subsec:choices}.

\textbf{Initialization (line 3):} The categorizer $\eta$ will categorize events based on their $\x$ and $p$ values. It can be initialized to apply arbitrary cutoffs/thresholds on the $p$ values. The initial cutoffs/thresholds could possibly be $\x$ dependent, though this is not necessary. $\eta$ an also be initialized to a randomized categorizer or to a constant function.

Note that $\eta$ being a function of $\x$ and $p$ is consistent with our original notation of $\eta$ being a function of event $\e$, since $\x(\e)$ and $p(\e)$ are part of the event description after the preprocessing step. Also note that in our description here, $\eta$ is expected to return an integer between 1 and \code{C} or, alternatively, ``\code{rejected}'' if \code{allow\_rejection} is set to \code{True}.

Lines 5--21 contain the code block for one iterative training step.

\textbf{Estimation of $p_c(\x)$ (lines 5, 6):}\footnote{Note that the estimation of $p_c(\x)$ could be split between line 6 and line 8, with line 6 representing any required preprocessing and line 8 representing the actual estimation for particular values of $\x$ and $c$. This is typically the case in kernel regression.} To estimate the signal-fraction functions $p_c(\x)$ for a given categorizer $\eta$, training events are first categorized based on $\eta$. Then, from training events in a given category $c$, $p_c(\x)$ can be estimated as the expected value of either $p$ or $y$ conditional on $\x$. This regression problem can be solved using a plethora of techniques---binned and unbinned, parametric and non-parametric (like kernel regression).

The estimation of $p_c(\x)$ for a given categorizer comes with the typical challenges of machine learning, including the bias--variance trade-off. One might want to employ cross-validation techniques (for example, in picking the bandwidth for kernel regression) to get the best results. If the estimation of $p_c(\x)$ is affected in some regions of $\x$ by a low density of training events, it might be helpful to increase the sampling density in those regions (and appropriately decrease the event weights). Of course, with a reasonably large number of training events, and a reasonably small dimensionality of $\x$ and number of categories, these measures will be rendered inessential.

\textbf{Defining the next categorizer (lines 7--20):} The next categorizer is constructed so as to reject events if $\edstat(p_c(\x), p) < 0$ for all categories $c$ (and if event rejection is allowed). Otherwise, the next categorizer groups events at a given value of $\x$ into different categories, based on their $p$ values.

The sorting in line 8 ensures that at each value of $\x$, events in a \emph{higher} category will have higher value of $p$. In other words, line 8 ensures that category boundaries (on values of $p$) do not interweave or cross each other when $\x$ is varied. Note that, our distance measures $\tdstat$ are all invariant under an $\x$ dependent cyclic permutation of categories. But having the categories ordered by their $p_c$ at each $\x$ makes the categorization more meaningful. This also helps with the estimation of the signal-fraction functions $p_c(\x)$ by reducing their sensitivity to changes in $\x$.

Note that during each iteration, the next categorizer depends implicitly on the estimated functions, $p_c(\x)$, for the current categorizer. This means that the estimated functions $p_c(\x)$ will need to be packaged into the final product of the training, which is the trained categorizer. The details of this packaging will depend on the particular regression technique employed.

\textbf{Termination condition (line 21):} 
Lloyd's k-means clustering algorithm converges and terminates when the category assignments do not change in a training step. This convergence is guaranteed within a finite number of steps, and in practice the fraction of events whose category assignments change during a step drops significantly over only a few tens of iterations.

The guarantee of convergence within in a finite number of steps extends to our algorithm as well (as argued in \aref{appendix:c}) if the estimation of $p_c(\x)$ is performed by binning events into different (fixed) $\x$-bins and finding the expected value of $p$ or $y$ in each bin. But if one takes an unbinned approach to the estimation of $p_c(\x)$, it is possible to run into cycles near the end of the training process due to artifacts in the regression technique. This can be handled by terminating the iterative training when the fraction of events reassigned in some step drops below a certain threshold.

One can also terminate the algorithm after a predetermined number of steps---the convergence of the algorithm will typically be fast enough (for small values of \code{C}) to warrant this. Alternative to these, we can also employ a stopping condition based on the saturation of $\tdstat$, as explained below.

\subsection{Performance evaluation} \label{subsec:performance}
For any given categorizer $\eta$ (intermediate or final), $\tdstat$ can be estimated from \eref{eqn:eventwisestat}, after first estimating the signal-fraction functions $p_c(\x)$. It can also be estimated as
\begin{equation}
 \tdstat = \mean_{\e\,\sim\,\Ha}~\left[\sum\limits_{c=1}^C~\delta(c, \eta(\e))~\edstat\big(p_c(\x), p_c(\x)\big)\right]~~.
\end{equation}
This can serve multiple purposes. First, we can use it to devise a termination condition---terminate the training when the fractional or absolute increase in the (estimated) value of $\tdstat$ drops below a certain threshold.\footnote{Near the end of the training process, it is possible for the estimated $\tdstat$ value to decrease over training steps. This is caused purely by artifacts in the regression technique used in the estimation of $p_c(\x)$. Such a decrease is otherwise not possible, as proved in \aref{appendix:c}.}

Secondly, the values of $\tdstat$ for categorizers trained with different values of \code{C} (estimated from validation datasets) can be used to choose between different values of \code{C}. Furthermore, in the $\code{C}\rightarrow \infty$ limit, the value of $\tdstat$ (for the optimal categorizer) can be estimated as
\begin{equation}
 \lim_{\code{C}\rightarrow\infty} \Big(\text{optimal }\tdstat\Big) = \mean_{\e\,\sim\,\Ha}~\left[\edstat\big(p(\e), p(\e)\big)\right]~~. \label{eqn:infC}
\end{equation}
This is also theoretical upper-limit on $\tdstat$ achievable by analyzing the distribution of events in the full-phase space $\e$. As mentioned in \sref{subsec:choices} this can be used to assess the improvement one stands to gain by continually increasing \code{C}. \Eref{eqn:infC} follows from \eref{eqn:eventwisestat} by noting that the optimal categorizer in the infinite categories limit will keep events with different values of $p$ from mixing.

Note that comparisons between categorizers based on $\tdstat$ will only be meaningful if a common performance measure is employed, regardless of the performance measure the individual categorizers were trained to maximize. In fact, it can be shown that for any given $\Hn$, $\Ha$, and categorizer $\eta$, the different $\tdstat$ values will satisfy (same as order of strictness in \eref{eqn:strictorder})
\begin{equation}
\tdneym \leq \tdrevkl \leq \tdb \leq \tdjeff \leq \tdkl \leq \tdpear~~. \label{eqn:strictorder3}
\end{equation}

\subsection{Miscellaneous notes}
\begin{itemize}
\item Note that the iterative training procedure will converge on a \textbf{locally optimal} categorizer. To ensure that the training does not lead to a categorizer much worse than the globally optimal one, one may have to repeat the training procedure with different initial categorizers.
\item It is possible that one is not interested in $\x$ dependent cutoffs and thresholds on $p$, and simply wants to maximize the overall $\sum\limits_c S^2_c/N_c$ (or the equivalent for a different statistical distance), using flat boundaries between categories. In such situations, a dummy event variable $\x$ can be used with each event having the same value of the event variable. Alternatively, instead of estimating the signal-fraction functions $p_c(\x)$ in line 6 of algorithm~\ref{algorithm}, one can estimate the expected value of $p$ conditional only on $c$. Either of these simplifications leads to an elegant solution to the problem of finding the optimal cutoffs or ``working points'' on ML classifier outputs.
\item It is possible that the training process converges poorly in terms of fraction of events reassigned in each step, despite the estimated value of $\tdstat$ saturating. It is also possible for different repetitions of the training process (with different training datasets or different initializations of $\eta$) to converge on categorizers that are substantially different from one another, despite having similar values of estimated $\tdstat$.

This leeway in categorizer specification can be interpreted as certain decisions regarding category assignments of events \emph{not mattering much}, and could be a sign that the number of categories \code{C} is too high (to be warranted by the amount of training data used). In such situations, it may be possible to reduce the variability in the trained categorizer by reducing the number of categories \code{C} without a significant drop in performance. Alternatively, one can reduce the variability in the trained classifier by sufficiently increasing the number of training events, in order to reliably probe the small variation in performance.
\item Note that one has no control over the fraction of events that get placed in different categories by the trained categorizer. This could lead to very lopsided categorizations, leading to situations where increasing the number of categories \code{C} makes the expected number of events in some categories inconveniently small while other categories still have sufficient number of events to warrant further splitting. In such situations, one can freeze certain categories and force others to be split further by passing them through another categorizer. This is similar to the sequential splitting of events into categories used in \cite{Bourilkov:2019als}.

One could also use different values of \code{C} in different regions of $\x$, with each region using a different categorizer.
\item It is possible to train a categorizer stochastically instead of iteratively, by maintaining a moving estimate of $p_c(\x)$, updated based on a few randomly sampled training events (mini-batch) in each step. Similar to stochastic k-means clustering algorithms \cite{Bottou:1994onlinekmeans, Sculley:2010batchkmeans}, such a stochastic implementation will scale up better with large training datasets and large number of categories \code{C}.
\end{itemize}

As one can see, our training prescription is not hard-and-fast. It is more of a paradigm with several choices and variations, not all of which have been covered here, and a comprehensive exploration of these variations is beyond the scope of this work.

\section{ThickBrick: A public implementation in Python} \label{sec:thickbrick}
\href{\thickbrickurl}{ThickBrick} \cite{TB} is an open source, Python 3 package which provides an implementation of the prescription introduced in this paper. In this section we will demonstrate our prescription with a toy example analyzed using ThickBrick, and the code used in the analysis can be found among the usage examples on the \href{\thickbrickurl}{package website}.

\subsection*{Toy example} \label{subsec:toy}

\begin{figure}[tp]
\centering
 \includegraphics[width=.5\columnwidth]{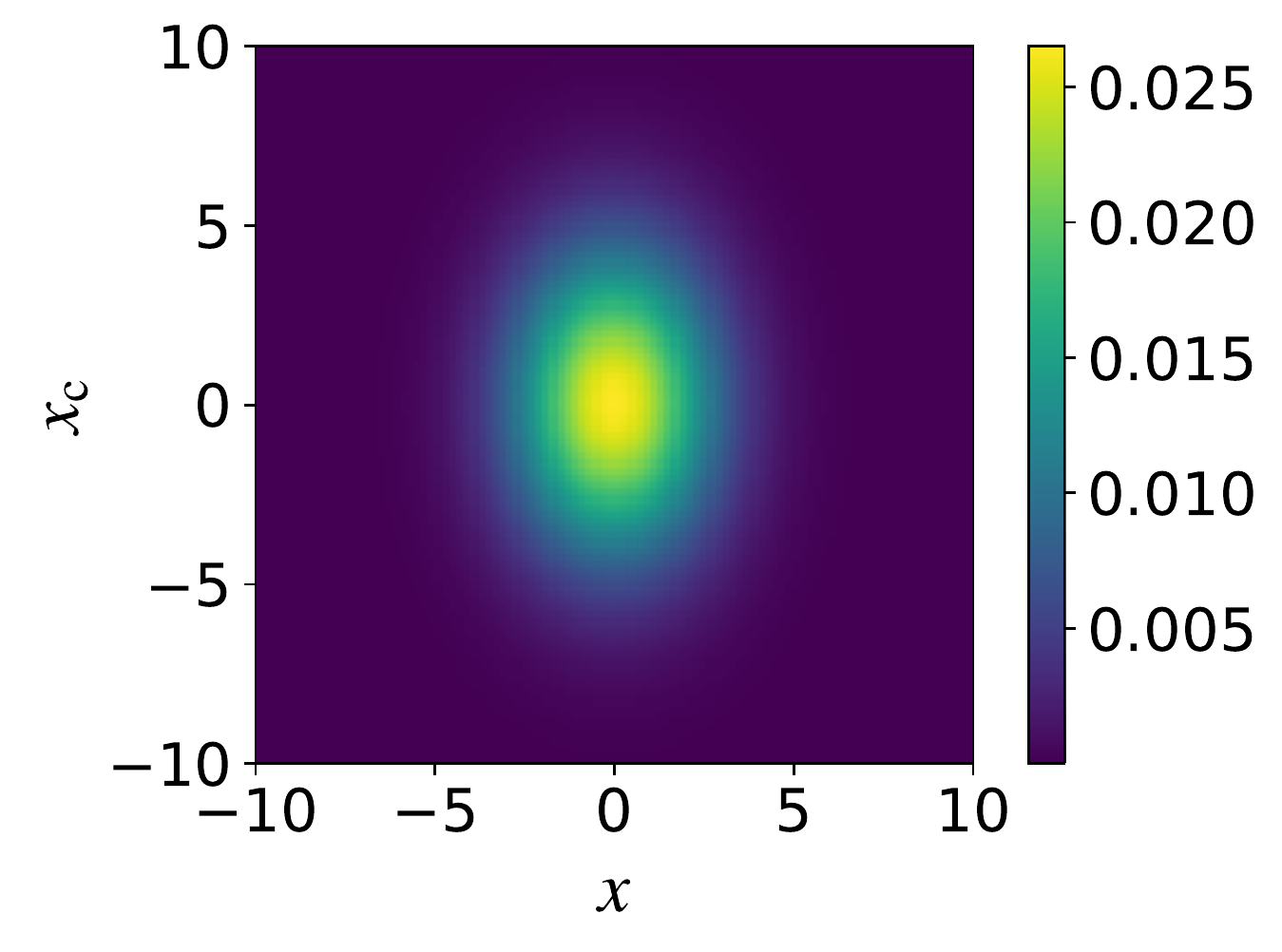}
 \caption{\label{fig:signal_full} A 2d histogram of the signal distribution in \eref{eqn:signal_dist} normalized to 1 (setting $S/B=1$).}
\end{figure}

\begin{figure}[tp]
\centering
 \includegraphics[width=.5\columnwidth]{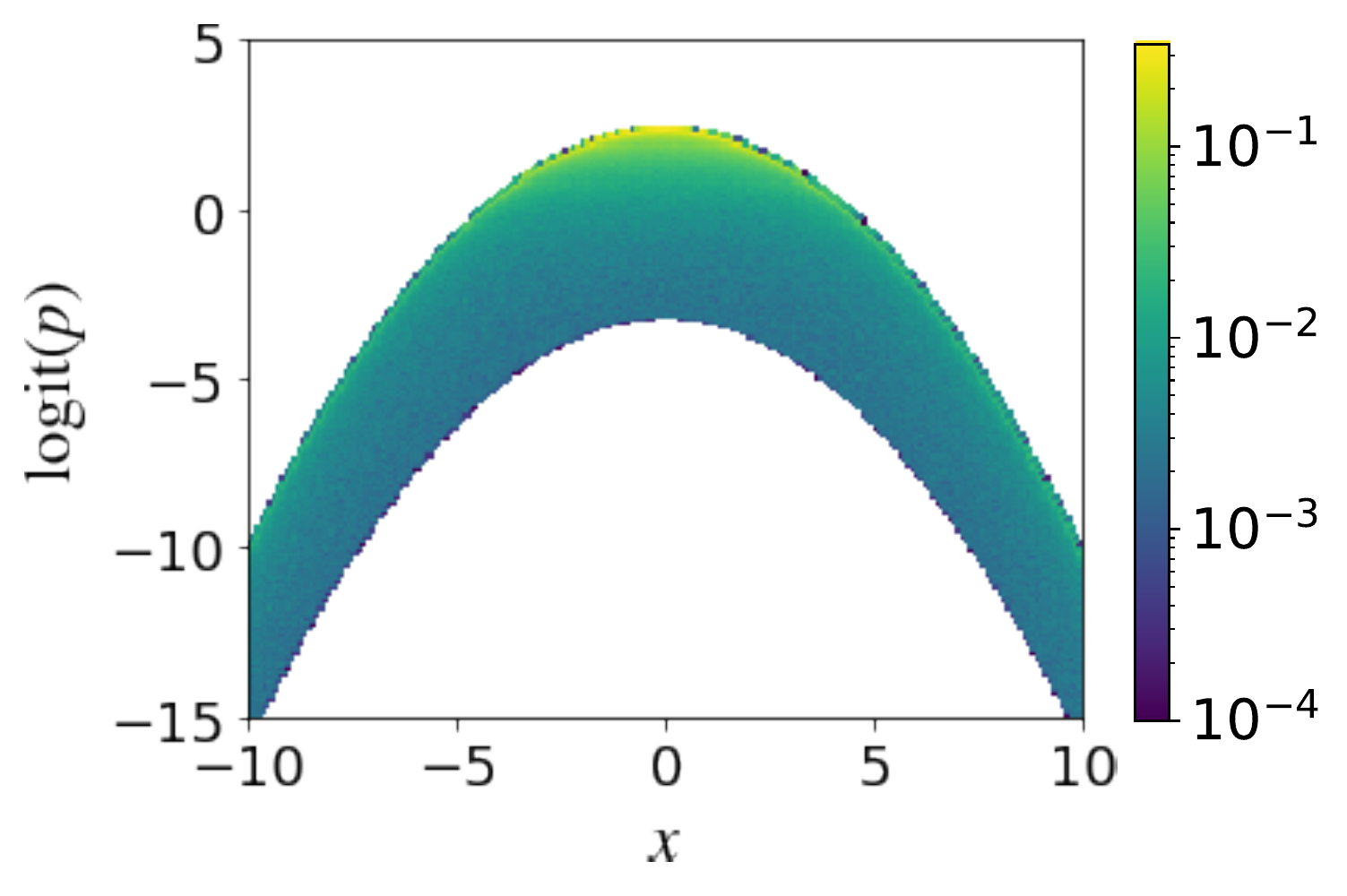}
 \hskip 8mm
 \includegraphics[width=.43\columnwidth]{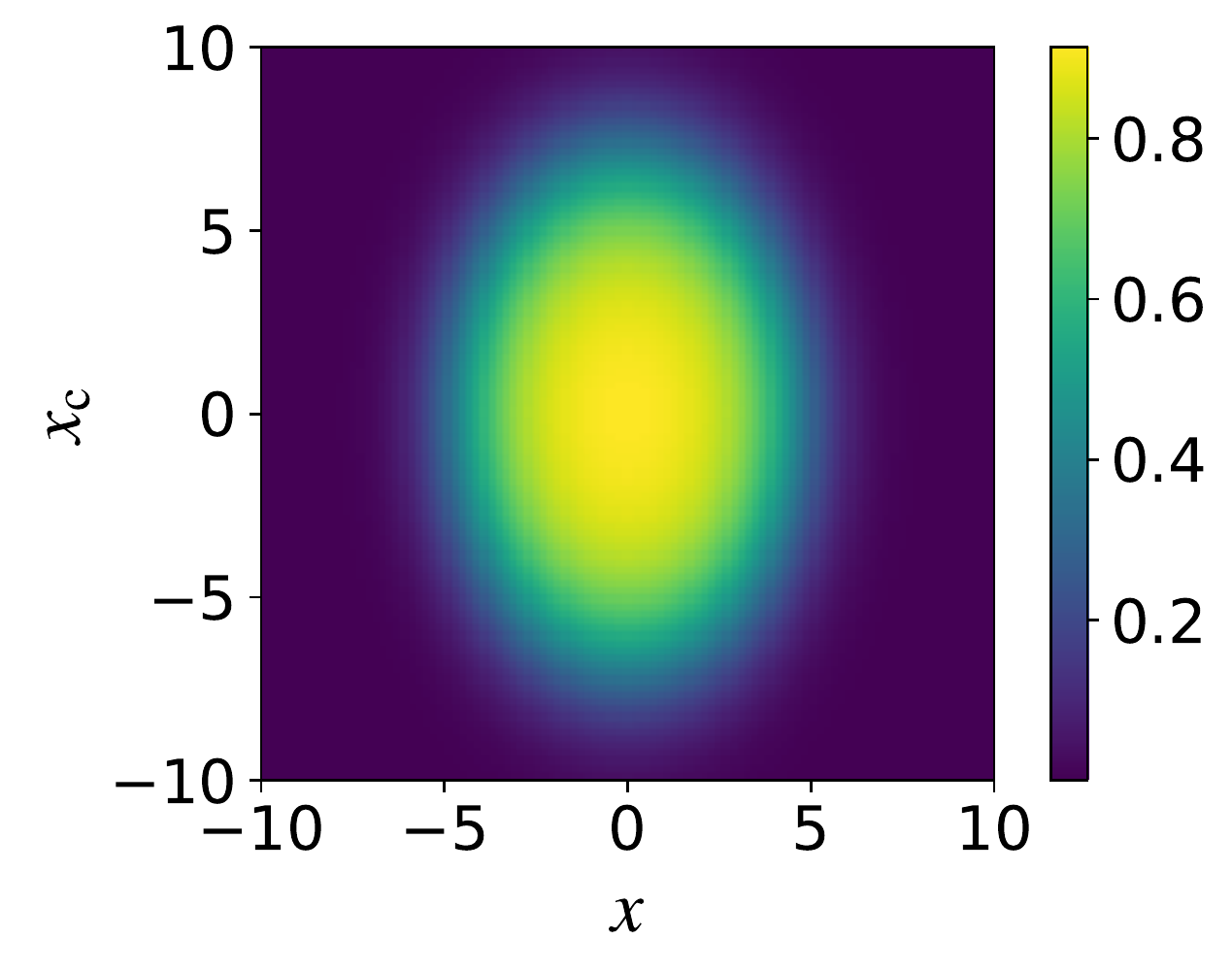}
 \includegraphics[width=.45\columnwidth]{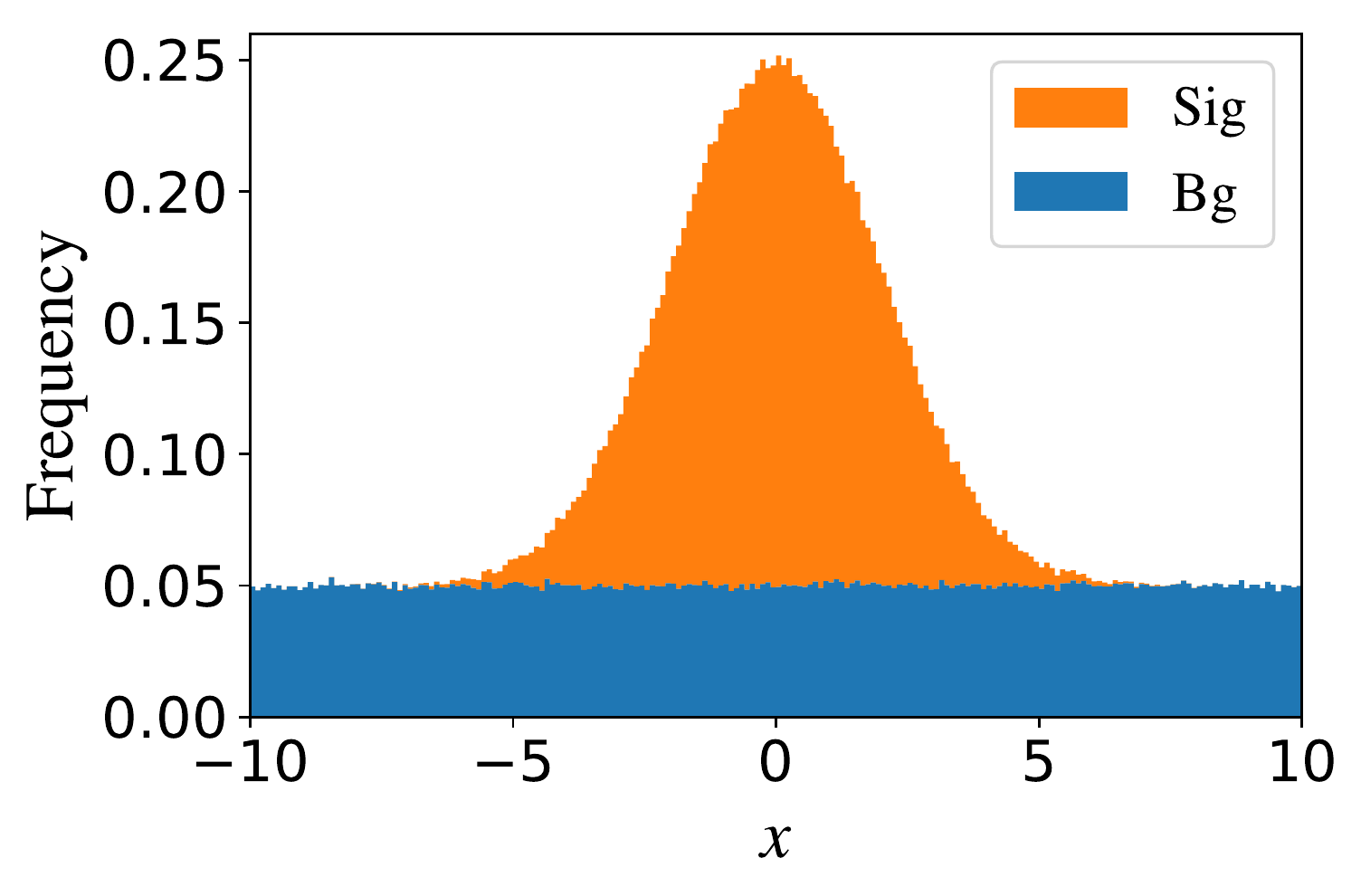}
 \hskip 10mm
 \includegraphics[width=.45\columnwidth]{before_selection_10percent.pdf}
 \caption{\label{fig:signal} Plots illustrating the toy data under consideration. The top-left panel shows the unit-normalized 2d distribution of $(x,~\logit{(p)})$. The top-right panel shows a plot of $p(x, x_\text{c})$ given in \eref{eqn:learned_p}. The bottom two panels show the distribution of signal events (orange) and background events (blue) in the event variable $x$ before any event selection. The background distribution is normalized to 1 in both, and the signal-to-background ratio is taken to be 1 (bottom-left panel, the signal is 50\% of the total) and $1/9$ (bottom-right panel, the signal is 10\% of the total).}
\end{figure}

For our toy example, we will use two dimensional data $\e \equiv (x, x_\text{c})$ generated as follows. The background is taken to be uniformly distributed over the square-shaped region defined by $-10 \leq x, x_\text{c} \leq 10$.
\begin{equation}
 b(x_1, x_2) = \begin{cases}
 B/400, &\text{if }|x|, |x_\text{c}| \leq 10\\
 0, &\text{otherwise}
 \end{cases}
\end{equation}
The signal is taken to be constrained within the same region with a truncated bivariate normal distribution given by
\begin{equation} \label{eqn:signal_dist}
 s(x, x_\text{c}) = \begin{cases}
 S A e^{-x^2/(2\sigma^2)}~e^{-x_\text{c}^2/(2\sigma_\text{c}^2)}, &\text{if }|x|, |x_\text{c}| \leq 10\\
 0, &\text{otherwise}
 \end{cases}~~,
\end{equation}
with $\sigma = 2$ and $\sigma_\text{c} = 3$, and $A$ is the normalization factor given by
\begin{equation}
A = \frac{2}{\pi\sigma\sigma_\text{c}}~\left[\erf\left(\frac{10}{\sigma\sqrt{2}}\right) - \erf\left(\frac{-10}{\sigma\sqrt{2}}\right)\right]^{-1}~\left[\erf\left(\frac{10}{\sigma_\text{c}\sqrt{2}}\right) - \erf\left(\frac{-10}{\sigma_\text{c}\sqrt{2}}\right)\right]^{-1}~~.
\end{equation}
\Fref{fig:signal_full} shows the unit normalized signal distribution.
In this example, we will use $x$ as the event variable $\x$ which will be used to search for the signal. $x_\text{c}$ represents the information in $\e$ complementary to $\x$. In the preprocessing step, we can take training events from the background and signal distribution in a 1:1 ratio and estimate the probability $p$ that an event came from the signal distribution condition on $(x, x_\text{c})$ using a ML classifier training algorithm as per \sref{subsec:learn_p}. But since the focus of our work here is not this preprocessing step, for simplicity we will use our knowledge of the underlying distributions to directly calculate $p(x, x_\text{c})$\footnote{Note that the quality of the event selection/categorization is dependent on the quality of estimation of $p(\e)$, and here we are bypassing this step.} as
\begin{equation}
 p(x, x_c) = \frac{s(x, x_\text{c})}{s(x, x_\text{c}) + b(x, x_\text{c})} \label{eqn:learned_p}~~,
\end{equation}
with $S/B$ set to 1. This $p(x, x_\text{c})$ is plotted in the top-right panel of \fref{fig:signal}. In this section, we will consider other values of the overall signal-to-background ratio $S/B$ as well, but we will let ThickBrick handle the scaling internally. Therefore, in all plots in this section $p$ will refer to the value corresponding to $S/B = 1$. The top-left panel of \fref{fig:signal} shows the unit normalized 2d distribution of $\Big(x, \logit{\big((p(x, x_\text{c})\big)}\Big)$. For better visualization of the spread of $p$ at low values of $|x|$, here we use the $\logit$ function given by
\begin{equation}
\logit(p) \equiv \ln\left(\frac{p}{1-p}\right)~~,
\end{equation}
which also has the convenient property that scaling the signal cross-section or, equivalently, changing $S/B$, only changes $\logit{(p)}$ by an additive constant. In the top-left panel of \fref{fig:signal}, the function $\logit{(p)}$ appears to be constrained from below only because of the truncation of the data at the boundaries $x_\text{c} = \pm 10$. Finally, 
the bottom two panels of \fref{fig:signal} show the distribution of signal events (orange) and background events (blue) in the event variable $x$ before any event selection. The bottom-left panel has $S/B = 1$ (the signal is 50\% of the total) and the bottom-right panel has $S/B = 1/9$ (the signal is 10\% of the total), with the background distributions in both panels normalized to 1.

\subsubsection*{Event selection}
\begin{figure}[tp]
\centering
 \includegraphics[width=.5\columnwidth]{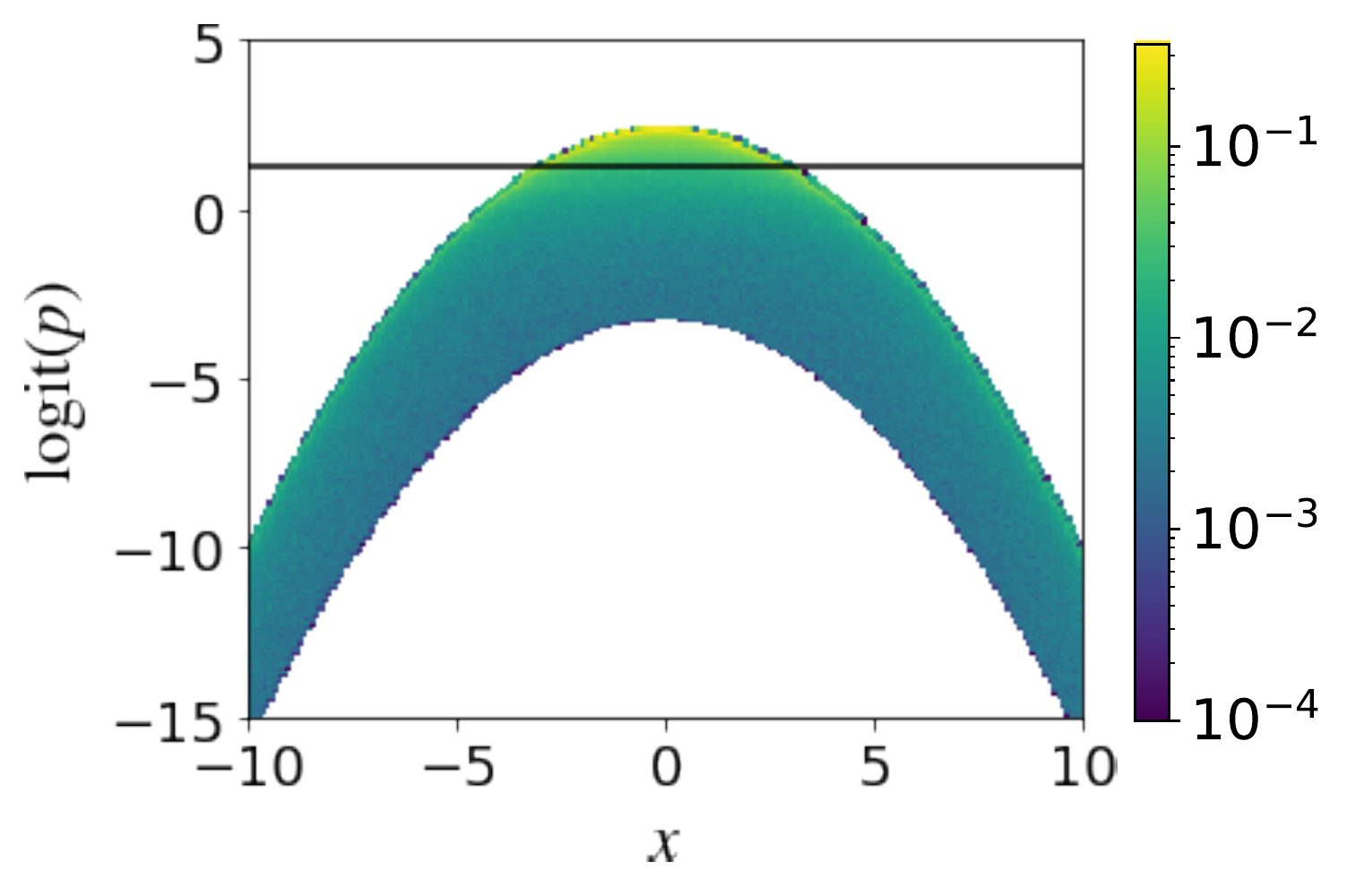}
 \hskip 8mm
 \includegraphics[width=.43\columnwidth]{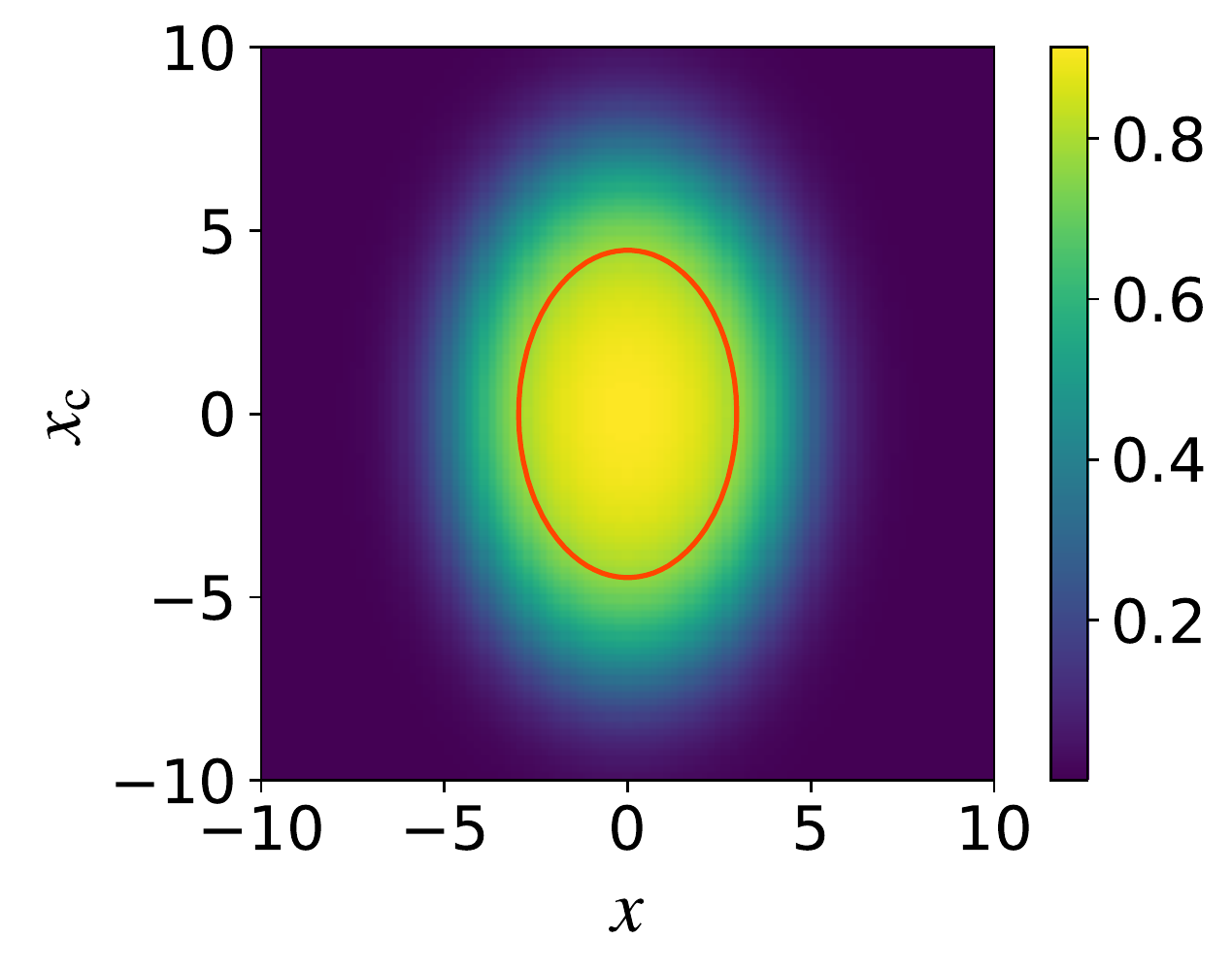}
 \vskip 3mm
 \includegraphics[width=.45\columnwidth]{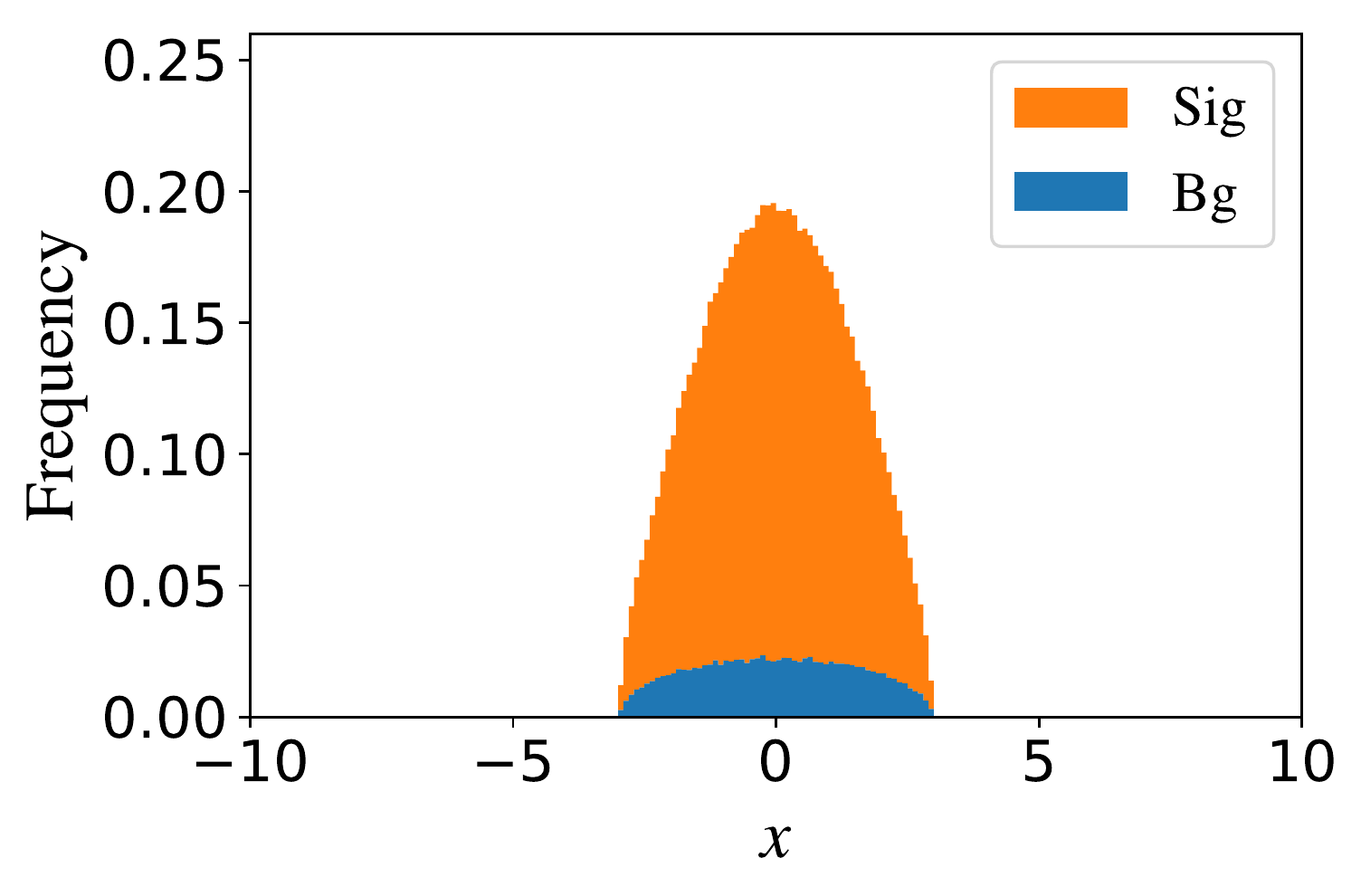}
 \hskip 10mm
 \includegraphics[width=.45\columnwidth]{flat_selection_10percent.pdf}
 \caption{\label{fig:flatcut} \textbf{Best flat-threshold selector}. The black horizontal line in the top-left panel shows the constant selection threshold that maximizes $\tdpear$. The region bounded by the orange curve in the top-right panel is the region in $(x, x_\text{c})$ selected by this threshold. The colors in the plot have the same meaning as in the top-right plot of \fref{fig:signal}. The bottom two panels show the signal and background distributions in $x$ after event selection. The background distribution \textbf{before selection} is normalized to 1 in both, and the signal-to-background ratio \textbf{before selection} is taken to be 1 (bottom-left panel) and $1/9$ (bottom-right panel). This allows for direct comparisons with the plots in \fref{fig:signal}.}
\end{figure}

Let us start off with event selection (\code{C = 1}, \code{allow\_rejection = True}) with $\tdpear$ as the distance measure (so that we do not have to consider the effect of changing $S/B$). \Fref{fig:flatcut} illustrates the best event selector with an $x$-independent threshold on $p$. The black horizontal line in the top-left panel shows the constant threshold that maximizes $\tdpear$---events above this threshold will be selected. The top-right panel shows the region in the full phase-space that is selected by this threshold (the region bounded by the orange curve). The bottom two panels show the signal and background distributions in $x$ after event selection. The background distribution \textbf{before selection} is normalized to 1 in both, and the signal-to-background ratio \textbf{before selection} is taken to be 1 (bottom-left panel) and $1/9$ (bottom-right panel). These plots can therefore be directly compared to the bottom panels of \fref{fig:signal}. Notice the strong correlation between the selection criterion and the event variable $x$ in the top-right plots, and the resulting background shaping in the bottom panels.

\begin{figure}[tp]
\centering
 \includegraphics[width=.5\columnwidth]{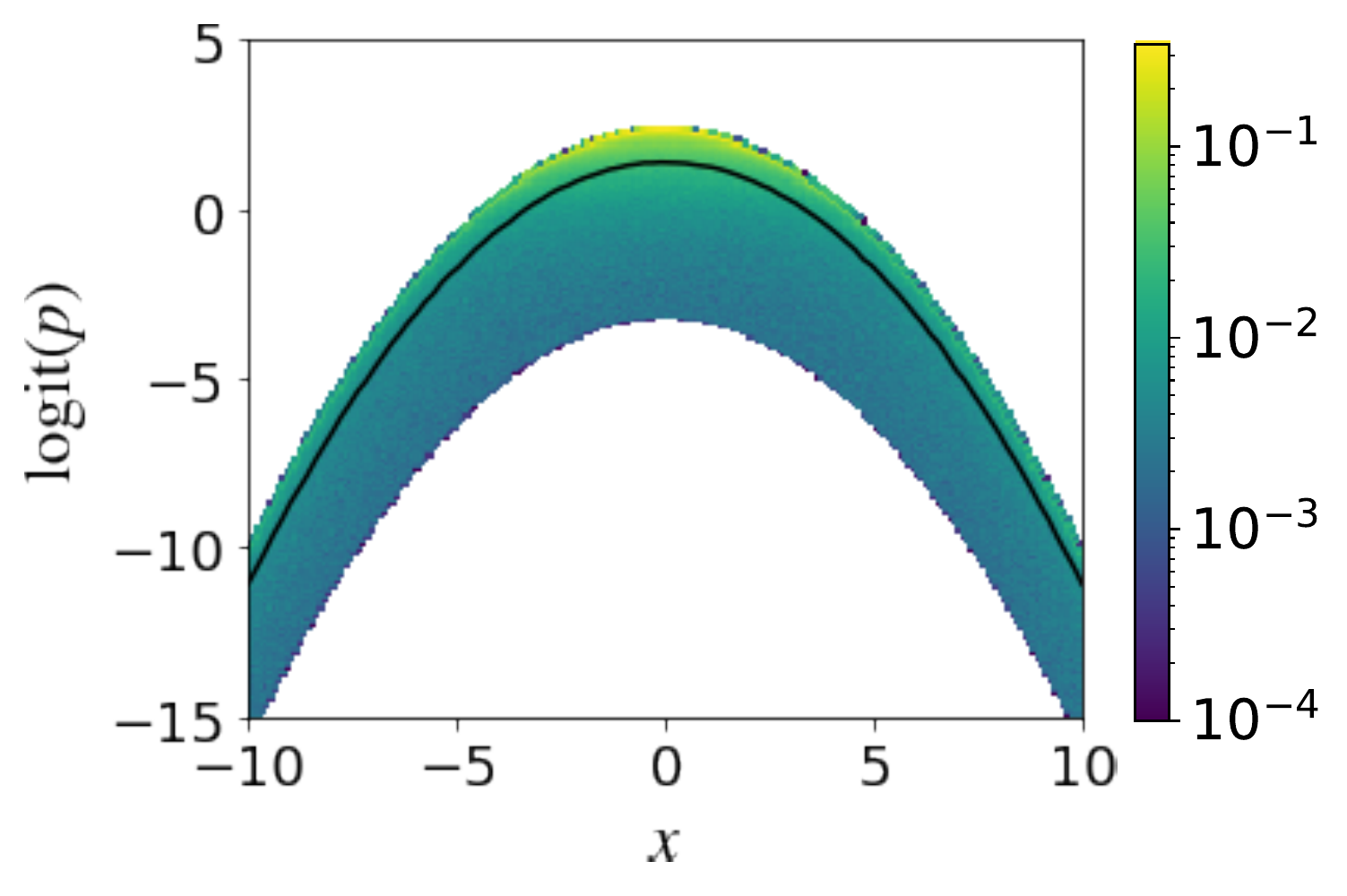}
 \hskip 8mm
 \includegraphics[width=.43\columnwidth]{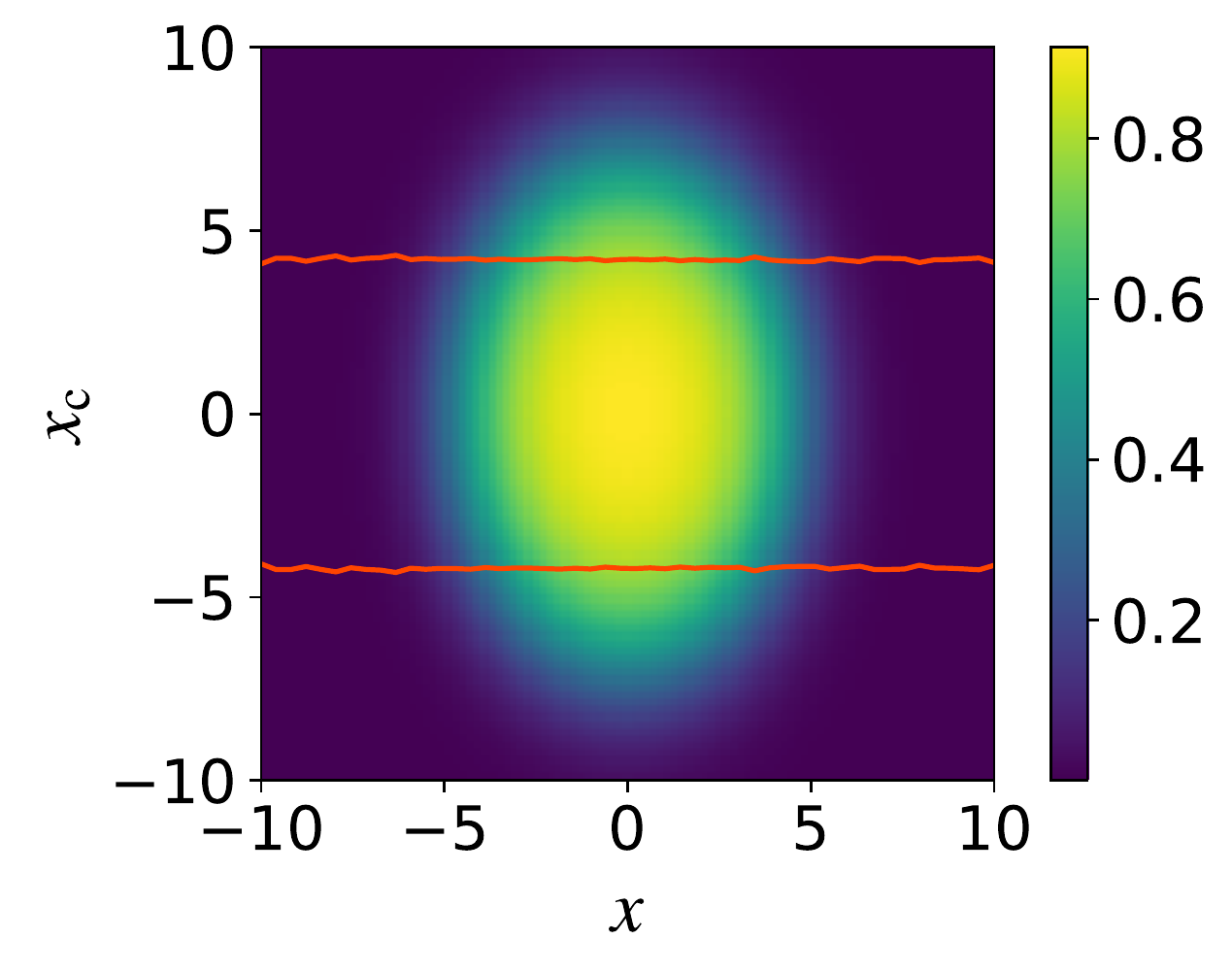}
 \vskip 3mm
 \includegraphics[width=.45\columnwidth]{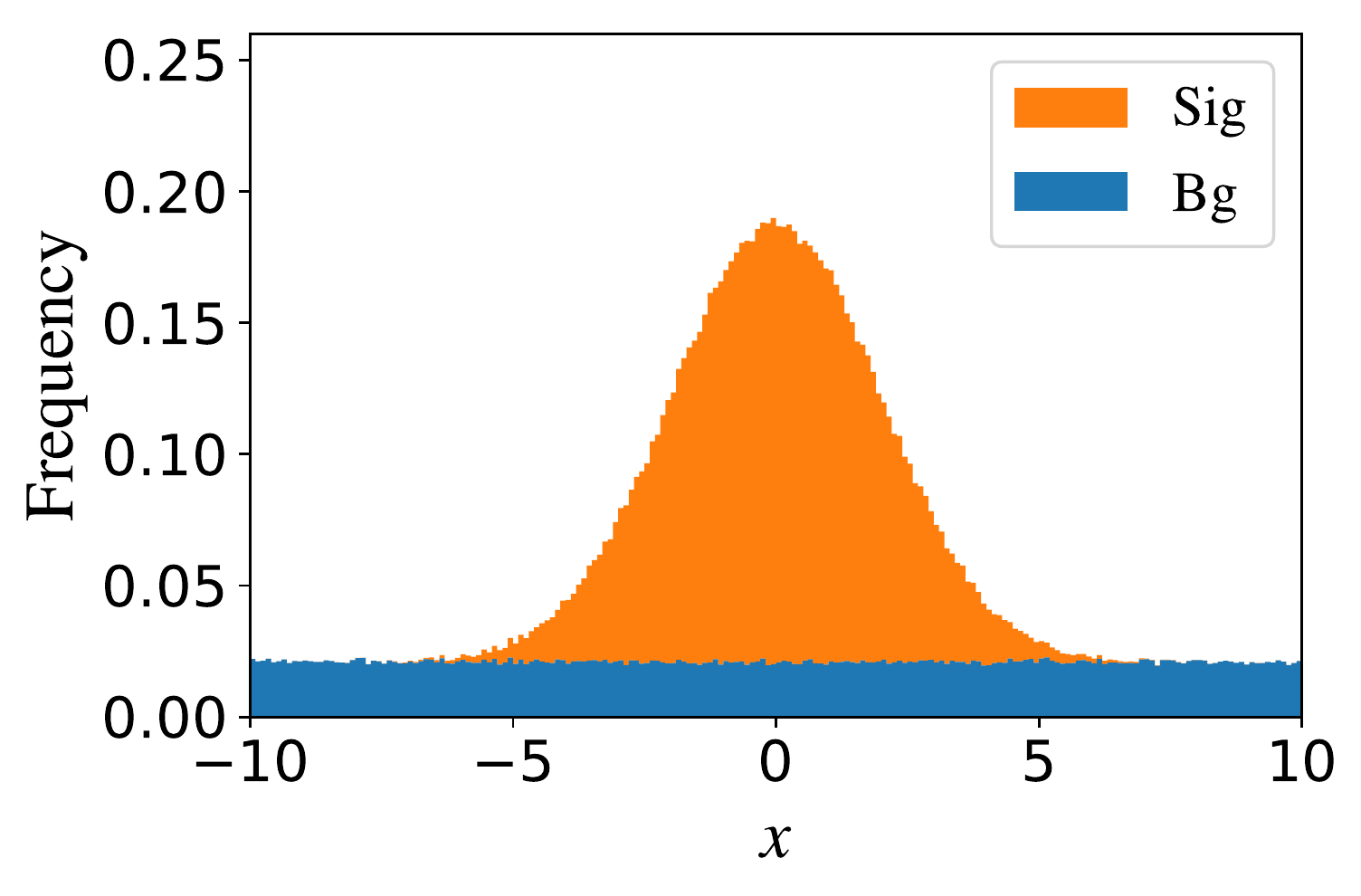}
 \hskip 10mm
 \includegraphics[width=.45\columnwidth]{dep_selection_10percent.pdf}
 \caption{\label{fig:depcut} \textbf{Event selector trained by ThickBrick} to maximize $\tdpear$ with $S/B$ set to 1. All panels are analogous to their counterparts in \fref{fig:flatcut}. The region bounded by the orange horizontal lines in the top-right panel is selected.}
\end{figure}

Next, let us consider a selector trained as per our prescription to maximize the same distance measure $\tdpear$. We used the 1d Nadaraya--Watson kernel regression technique \cite{Nadaraya:1964kerreg, Watson:1964kerreg} to estimate $p_c$ in each iterative step (implemented within ThickBrick). A training sample of 50000 background and signal events each was used to train the selector, and the training converged in 10 training steps. \Fref{fig:depcut} illustrates the trained event selector, and all the panels are analogous to their counterparts in \fref{fig:flatcut}---in the top-right panel, the region within the two horizontal orange lines is selected. Note how the selection is decorrelated from the event variable, confirming that event selection is occurring only based on the complementary information contained in $x_c$.

\subsubsection*{A word on decorrelation}

In this toy example $x_\text{c}$ and $x$ were constructed to be independent of each other under both the signal and background distributions for illustrative purposes only---in our training process, we never use this fact. The optimal selection thresholds were learned only from the distribution of events in the $(x,p)$ space. One can see that our event selector uses only complementary information from the fact that the selection cuts in the top-right panel of \fref{fig:depcut} are horizontal. 

\begin{figure}[tp]
\centering
 \includegraphics[height=.32\columnwidth]{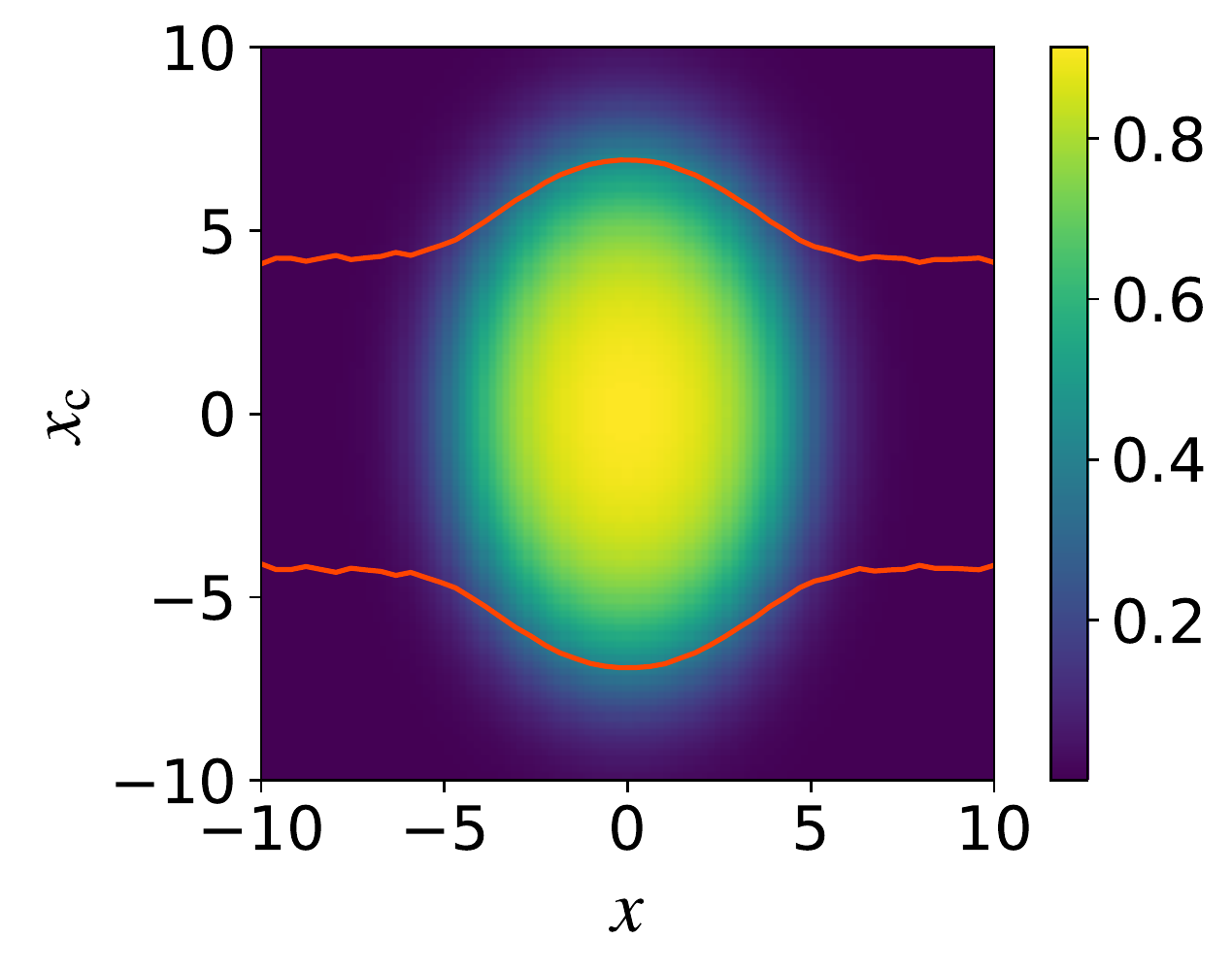}
 \hskip 5mm
 \includegraphics[height=.32\columnwidth]{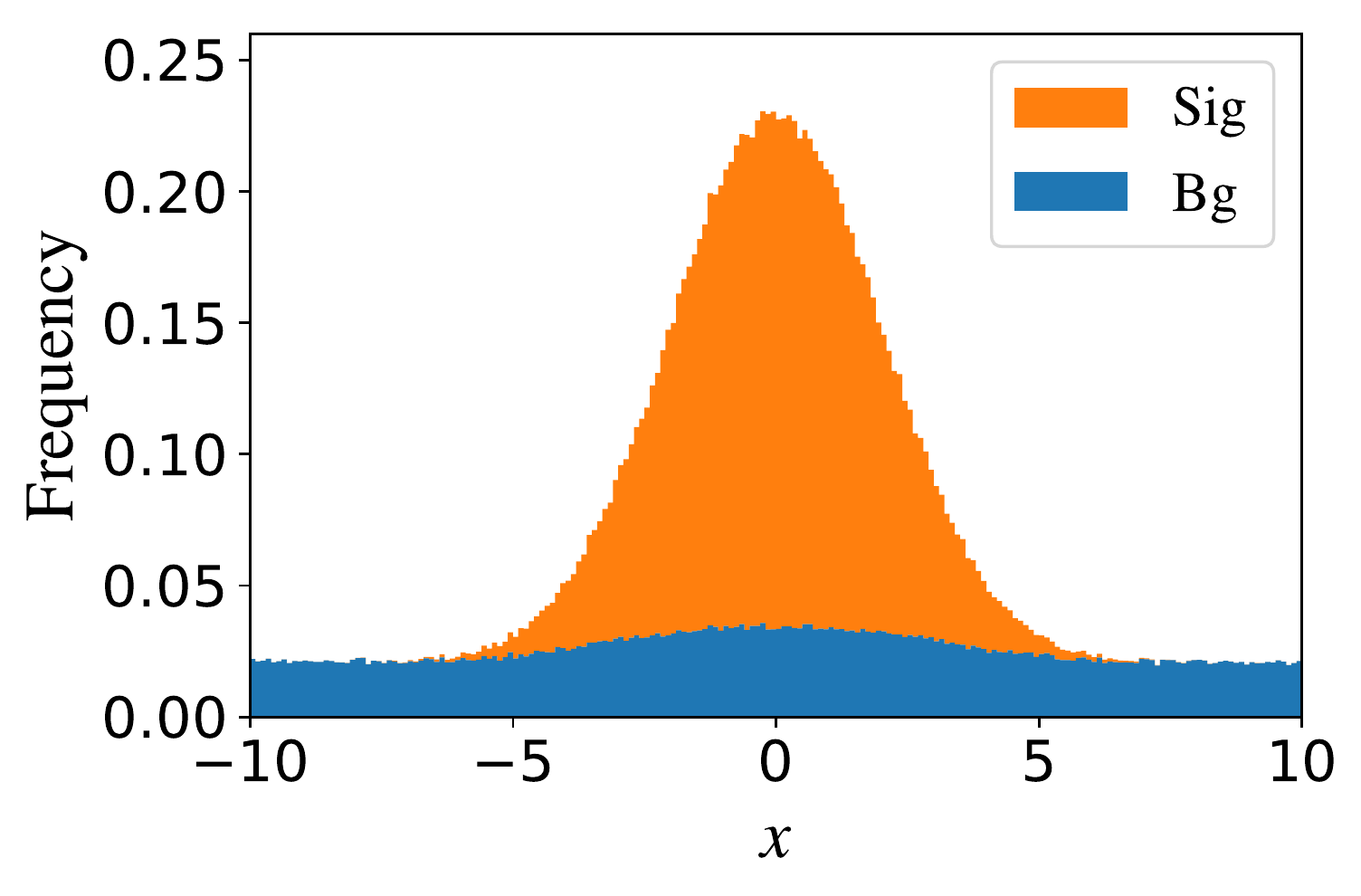}
 \caption{\label{fig:comparechi} \textbf{Event selector trained by ThickBrick} to maximize $\tdneym$. The left and right panels are analogous to the top-right and bottom-left panels of \fref{fig:depcut}, respectively.}
\end{figure}

Note that different distance measures will decorrelate the selection criteria from the event variable $\x$ to different degrees, and $\tdpear$ is a special distance measure that leads to categorizers that are \textbf{completely} decorrelated from the event variable $\x$. Since the decorrelation properties of our distance measures are not the main focus of this paper, we only briefly discuss this point in section~\ref{subsec:epi1}. In \fref{fig:comparechi}, we illustrate a selector trained to maximize $\tdneym$, which is at the other end of the spectrum in terms of degree of decorrelation. Since training with distances other than $\tdpear$ is sensitive to the value of $S/B$, and the contrast between different distance measures increases with increasing $S/B$, we will use a high signal-to-background ratio of 1 for illustrative purposes.

Note that the implication of \fref{fig:comparechi} is \textbf{not} that the corresponding event selector bases its decision partially on information non-complementary to $\x$. In our toy example, the threshold on $p$ at a given value of $x$ will not change if we scale up the number of signal events at other values of $x$. The correct interpretation is that the selector bases its decision on the \textbf{number} density distributions of $x_\text{c}$ under $\Hn$ and $\Ha$ conditional on $x$, instead of the \textbf{probability} distributions of $x_\text{c}$ conditional on $x$---the background and signal distributions in $x$ affect the selection criteria through this subtle difference.

\subsubsection*{Event categorization}
\begin{figure}[tp]
\centering
 \includegraphics[width=.5\columnwidth]{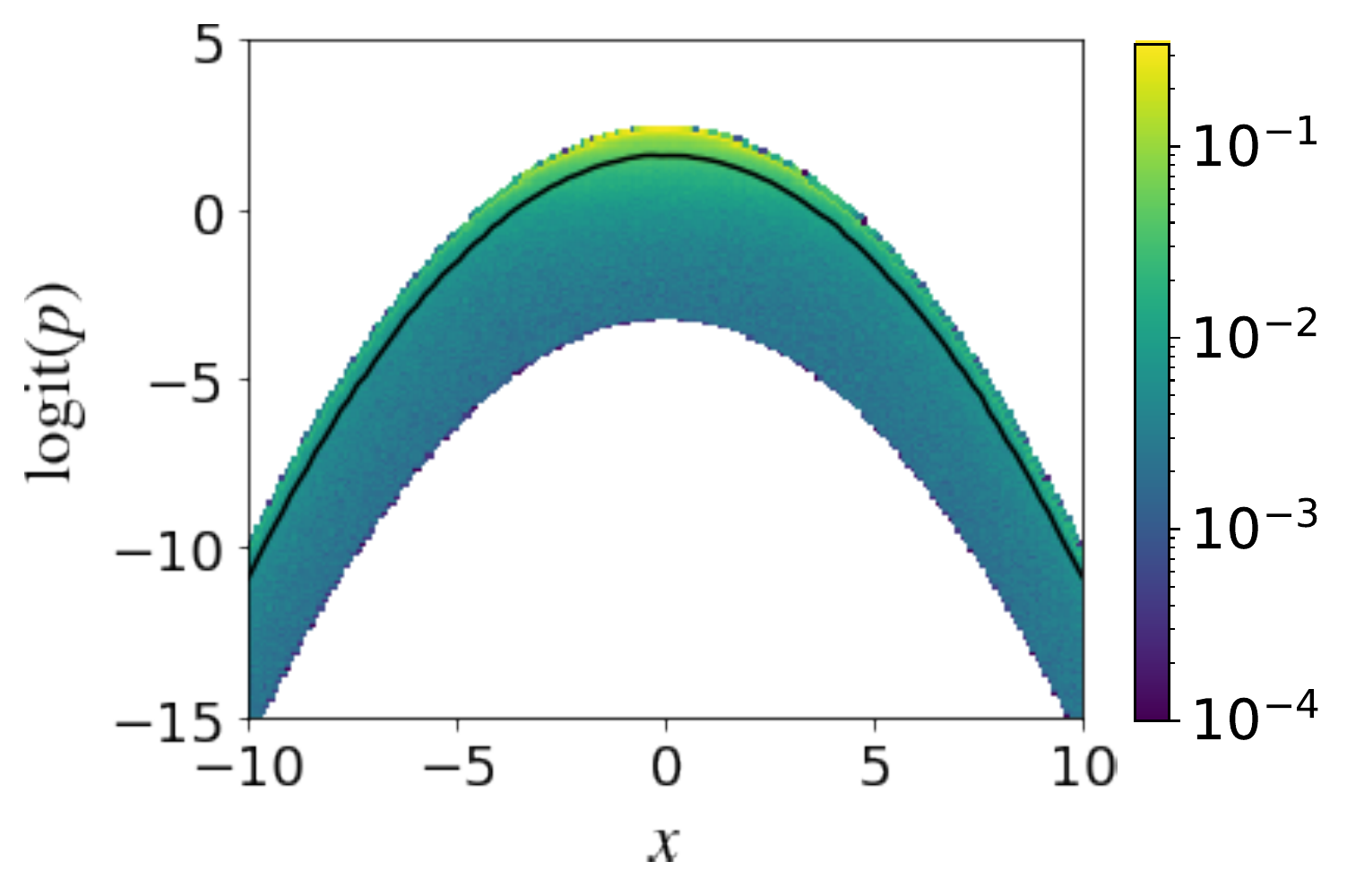}
 \hskip 8mm
 \includegraphics[width=.43\columnwidth]{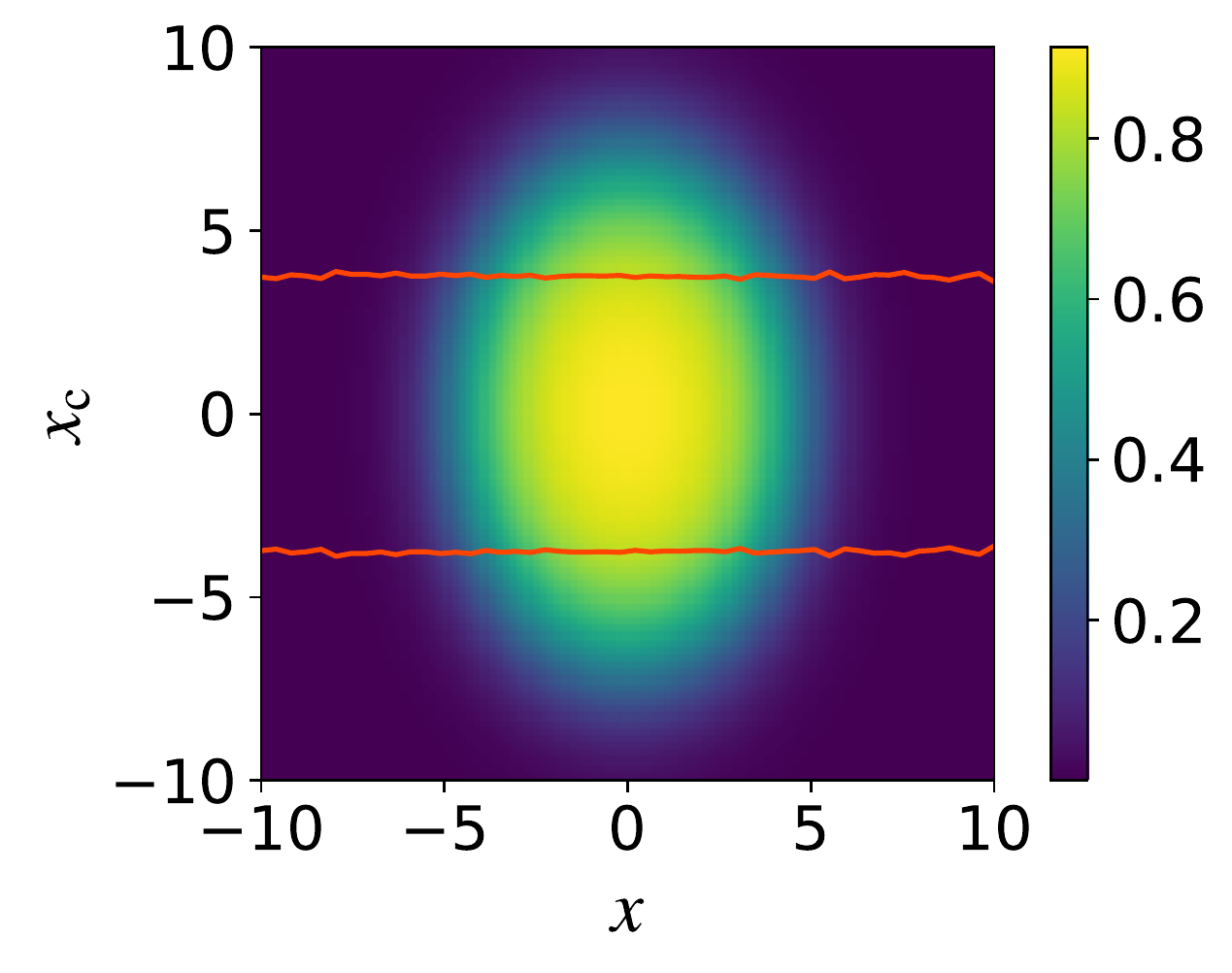}
 \includegraphics[width=.5\columnwidth]{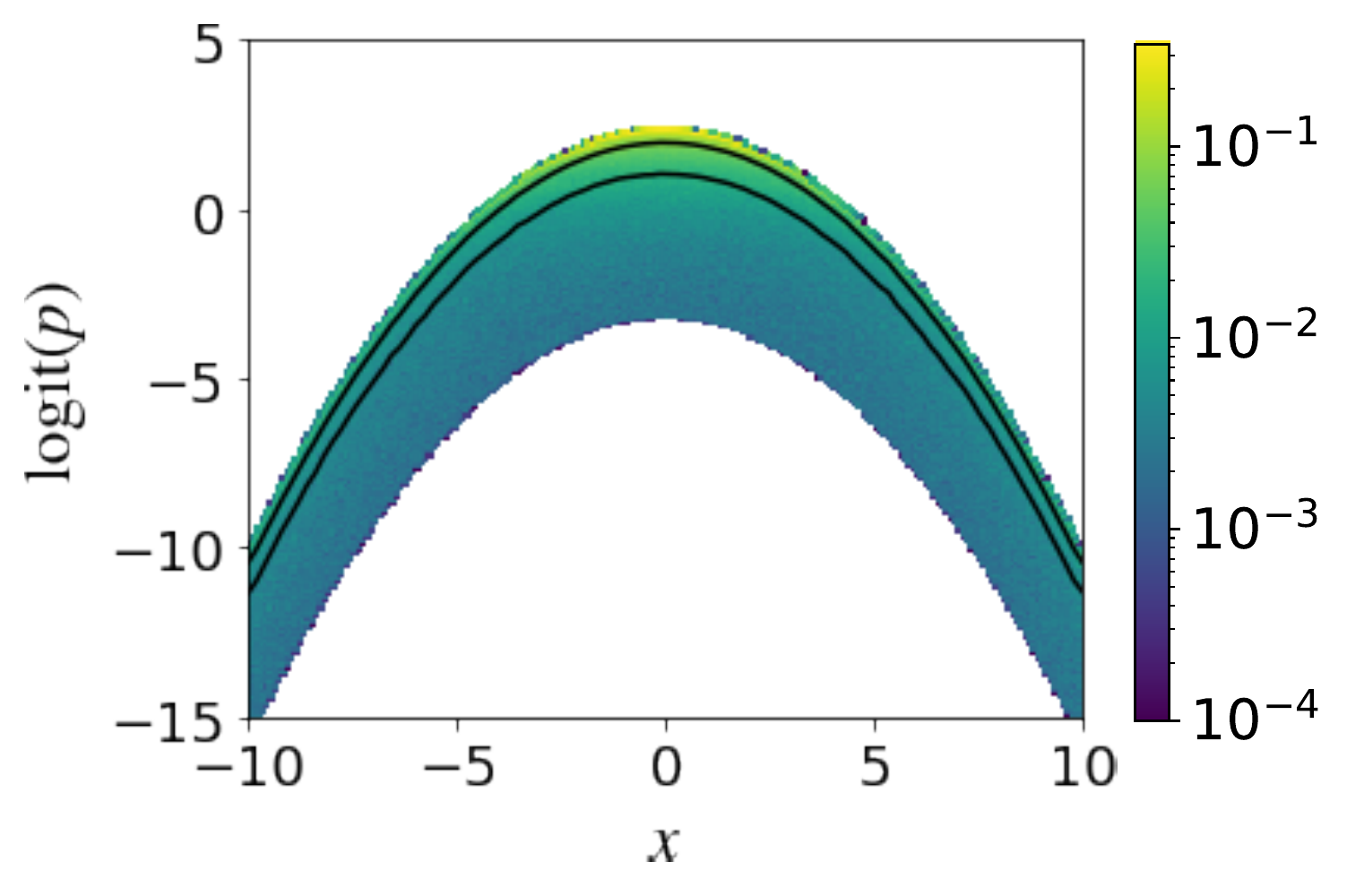}
 \hskip 8mm
 \includegraphics[width=.43\columnwidth]{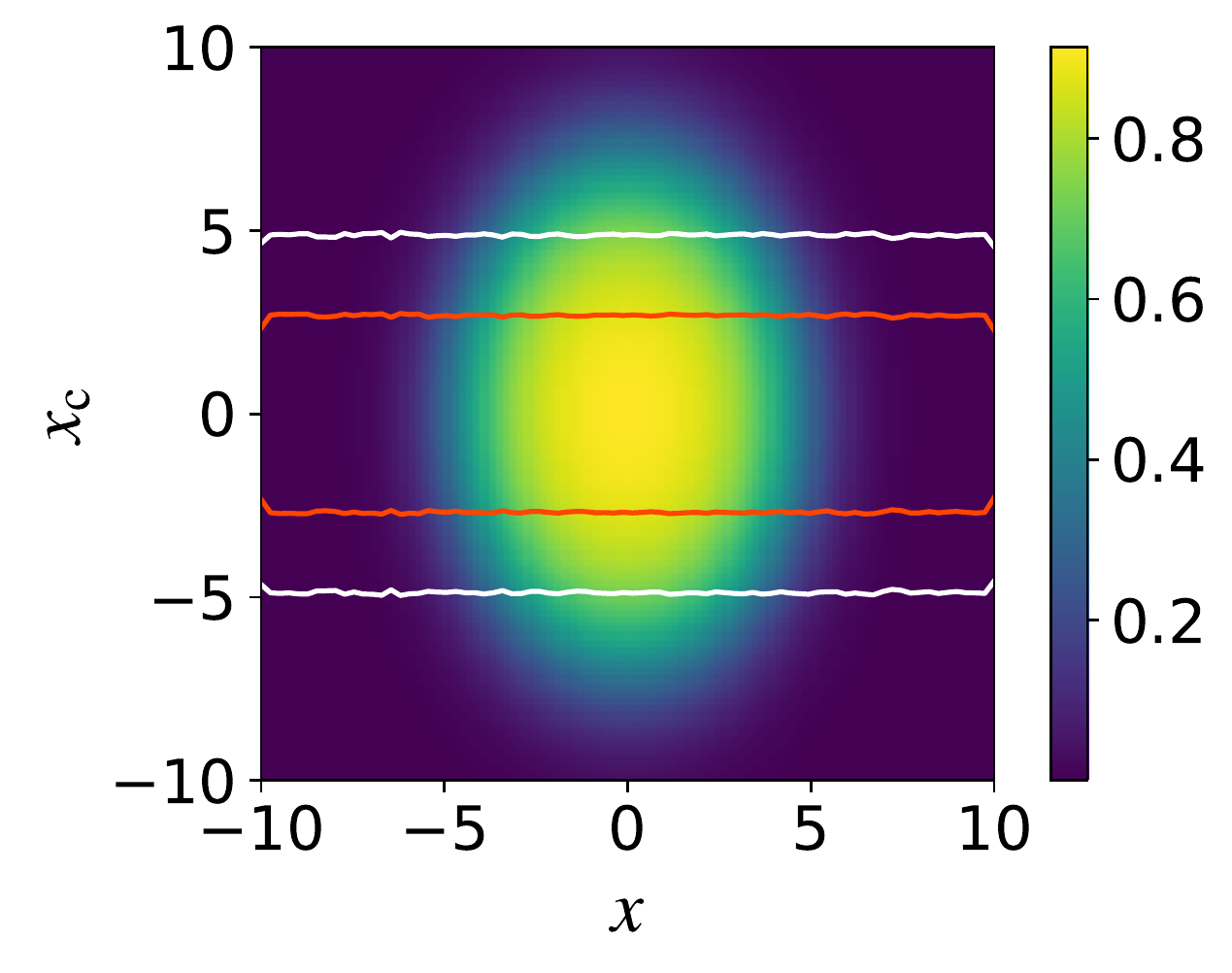}
 \caption{\label{fig:eventcategorization} The same as the top two panels in \fref{fig:depcut}, but for the case of event categorizers trained by ThickBrick. Event rejection was turned off and the number of categories was two (three) in the top (bottom) row. In the top-right panel, the region within the orange horizontal lines comprises category 2,  while the two regions outside the orange lines comprise category 1.
 In the bottom-right panel, the region within the orange horizontal lines comprises category 3, the two regions bounded by an orange and a white line comprise category 2, and the two regions outside the white lines comprise category 1.
 }
\end{figure}

Next we will consider event categorization with event rejection turned off (\code{allow\_rejection = False}). As before, we will only consider $\tdpear$ and use a training sample of 50000 background and signal events each. The top and bottom rows of \fref{fig:eventcategorization} illustrate categorizers trained with number of categories set to 2 and 3, respectively. In both cases, the black curves in the plots on left show the threshold between categories. The right panels show the corresponding boundaries in the $(x, x_\text{c})$ phase-space. In the top-right panel, the region within the orange horizontal lines comprises category 2,  while the two regions outside the orange lines comprise category 1. In the bottom-right panel, the region within the orange horizontal lines comprises category 3, the two regions bounded by an orange and a white line comprise category 2, and the two regions outside the white lines comprise category 1.

\subsubsection*{Performance}
\begin{figure}[tp]
\centering
 \includegraphics[width=.6\columnwidth]{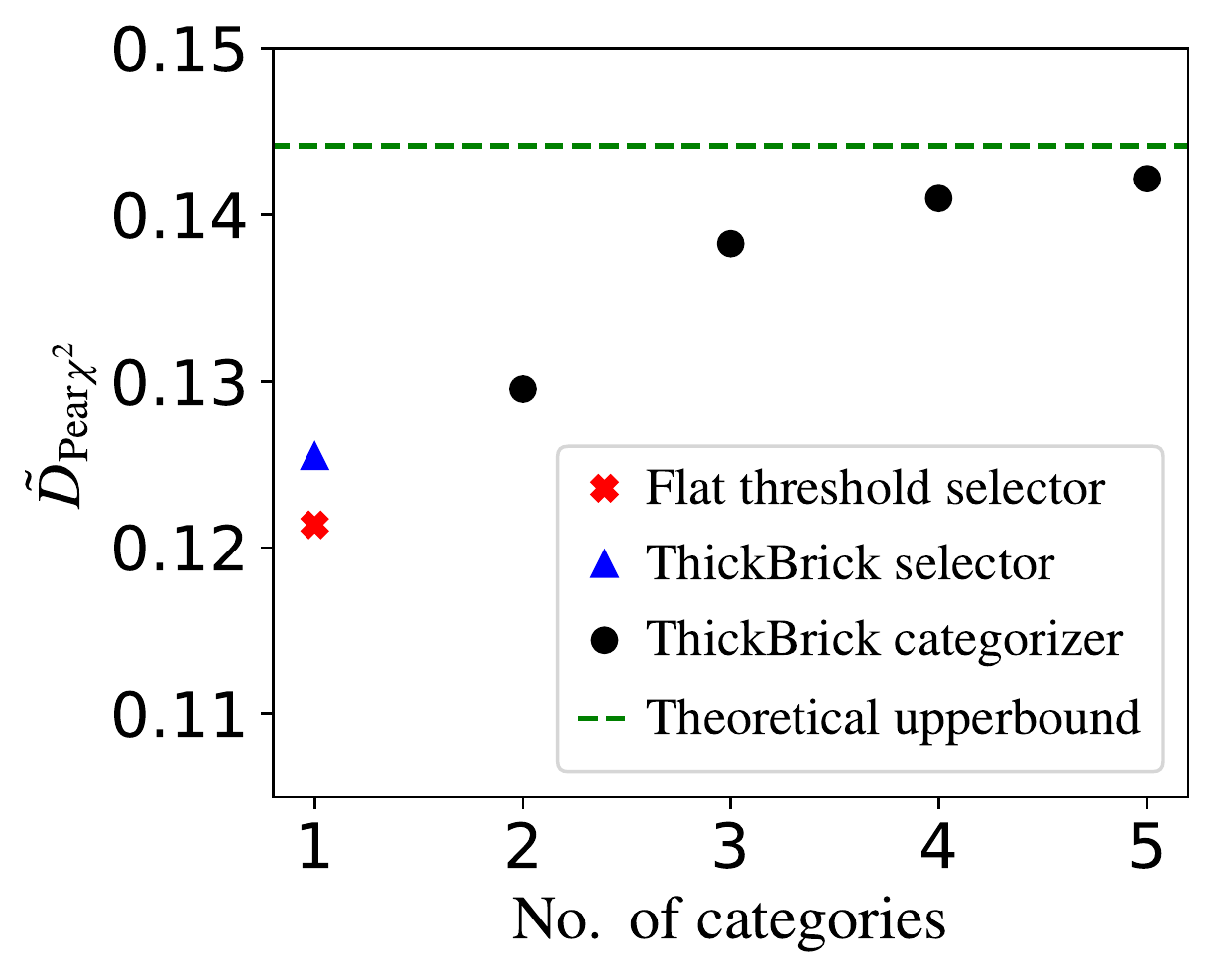}
 \caption{\label{fig:moneyplot} \textbf{Comparison of different event selectors and categorizers} in terms of the estimated $\tdpear$ with $S/B=1/9$. 
 The red cross corresponds to the best flat-threshold selector, while the blue triangle corresponds to an event selector trained using our prescription. The three black dots all correspond to event categorizers with different values of $\code{C}$ (all with \code{allow\_rejection = True}). The horizontal green dashed line shows
 the theoretical upper-limit on $\dpear$ for $\code{C}\rightarrow\infty$.}
\end{figure}

\Fref{fig:moneyplot} illustrates the performance of the event selectors and categorizers trained using our prescription, as captured by the estimated $\tdpear$ with $S/B$ set to $1/9$. For comparison, the best event selector that uses a flat threshold on $p(\e)$ is also shown, along with the theoretical upper-limit on $\dpear$ achievable by analyzing the distribution of events in the full phase-space $(x, x_\text{c})$. The $x$-axis of the plot shows the number of selected categories $\code{C}$. The red cross corresponds to the best flat-threshold selector, while the blue triangle corresponds to an event selector trained using our prescription. The three black dots all correspond to event categorizers with different values of $\code{C}$ (all with \code{allow\_rejection = True}). There are no appreciable error bars to report in \fref{fig:moneyplot}.

Note that, as expected, the event selector trained with ThickBrick performs better than a flat-threshold selector. Also note the diminishing returns with increasing values of \code{C}, as anticipated in \sref{subsec:choices}.

\section{Conclusions} \label{sec:conclusions}
In this paper we have provided an optimal event selection and categorization prescription to maximize the statistical significance of a potential signal excess (in the distribution of a low dimensional event variable). This statistical significance is captured by one of six different statistical distance measures given in \eref{eqn:d}.

Our prescription calls for first learning the probability that an event (sampled from the ``sig+bg'' hypothesis, $\Ha$) is a signal event, possibly using machine learning methods. Let us refer to the probability thus learned as ``the machine learning output''. Our prescription then finds the optimal selection thresholds (for event selectors) or boundary between categories (for event categorizers) in the machine learning output in an iterative manner. These thresholds and boundaries are dependent on the event variable. The thresholds can be thought of as chosen to extract the best sensitivity out of each $\x$-``bin''.

Event selectors (categorizers) trained using our prescription are expected to outperform typical ML-classifier-based selectors (categorizers) in terms of statistical sensitivity, as demonstrated using a toy example in section~\ref{subsec:toy}. In addition, our event selectors and categorizers will also be decorrelated from the event variable used in the analysis, which can help reduce the impact of systematic uncertainties on signal search analyses.

\section{Epilogue} \label{sec:epilogue}
In this section we provide some outlook on potential extensions and applications of our paradigm of \textbf{iteratively training \emph{on} ML output to use it optimally}, as well as provide some perspective on the origin of this work.

\subsection{Decorrelated classifiers and taggers} \label{subsec:epi1}
The distance measure $\dpear$ from \eref{eqn:dpear} is given by
\begin{equation}
\dpear = \sum\limits_{c=1}^C~\int\limits_{\Omega_\x} d\x~\frac{s_c^2(\x)}{b_c(\x)}~~.
\end{equation}
As noted in section~\ref{subsec:choices}, scaling the signal number density by \textbf{an overall factor} ($s(\e) \rightarrow \lambda s(\e)$) only changes $\dpear$ by a multiplicative factor ($\dpear \rightarrow \lambda^2 \dpear$). As a result, the ordering of categorizers by their $\tdstat$ values is independent of $S/B$, and hence, so is the optimal categorizer.

What is interesting is that the optimality of the categorizer that maximizes $\tdpear$ is also unaffected by an $\x$ dependent scaling given by
\begin{equation}
s(\e) \rightarrow \lambda\big(\x(\e)\big)~s(\e)~~. \label{eqn:xdep_transform}
\end{equation}
This is because at a given value of $\x$, the transformation in \eref{eqn:xdep_transform} only changes $\sum\limits_c s_c^2(\x)/b_c(\x)$ by the multiplicative factor $\lambda^2\big(\x\big)$, and the optimal categorizer maximizes $\sum\limits_c s_c^2/b_c$ \textbf{at each value of} $\x$.

This means that the categorizer which maximizes $\tdpear$ will be \textbf{completely} decorrelated from $\x$---it will not ``learn'' from the distributions of signal and background over $\x$ in the training data, and will only learn from their respective probability distributions conditional on $\x$ (see figures~\ref{fig:depcut} and \ref{fig:eventcategorization}). This property is often desirable, especially when the distribution over $\x$ (of signal and/or background) is uncertain \cite{Shimmin:2017mfk, Dolen:2016kst, Louppe:2016ylz}. For example, this decorrelation can be useful in jet substructure based boosted-decay taggers (decorrelation with jet mass is desirable). Another potential application is in decorrelating quark and gluon jet classifiers trained on ``naturally mixed samples'' (like quark enriched Z+jet samples and gluon enriched dijet samples) \cite{Dery:2017fap,Metodiev:2017vrx,Komiske:2018oaa} from observables like jet energy and pseudorapidity, which are controlled by the different underlying kinematics of the corresponding mixed samples.

While $\dpear$ is a good performance measure from the standpoint of signal discovery, there may be other performance measures better suited to other applications, which share the same decorrelation property. For example, for a classification problem which places signal and background on the same footing, it may be better to use $D_\mathrm{classify}$, defined below, as the performance measure to maximize (iteratively).
\begin{equation}
D_\mathrm{classify} = -\sum\limits_{c=1}^{C}~\int\limits_{\Omega_\x} d\x~\sqrt{s_c(\x) b_c(\x)}.
\end{equation}
This form of $D_\mathrm{classify}$ has connections to the Bhattacharyya coefficient \cite{Kailath:1967bhatt}.

In fact, we can generalize $\dpear$ and $D_\mathrm{classify}$ to a family of performance measures parameterized by $\nu$, all of which share the decorrelation property,\footnote{Although the utility of such a generalization is presently unclear.} as follows:
\begin{equation}
D_{\mathrm{classify},\nu} = \mathrm{sgn}\big[\nu(\nu-1)\big]~\sum\limits_{c=1}^{C}~\int\limits_{\Omega_\x} d\x~s_c^\nu(\x) b_c^{1-\nu}(\x) \text{~~,~~~~~~~~~for } \nu(\nu-1) \neq 0~~.
\end{equation}
Similarly, we can also generalize $D_\mathrm{classify}$ to multi-class classification problems. We will explore these ideas in a future publication.

\subsection{Nixing the preprocessing step} \label{subsec:nix}
In the way our prescription is presented here, we extract all the information contained in the event $\e$ (for the purpose of signal detection) into the quantity $p(\e)$ in the preprocessing step, and then use that information for categorization purposes. But this preprocessing is not really necessary since we already have access to $\e$, which tautologically contains all the information contained in $\e$.

In fact, our method originally started life
as an iterative training algorithm to train a categorizer that directly bases its decision/output on $\e$. In each iterative step, cost functions can be chosen to mimic the iterative step presented in this paper. The cost functions will depend on categorizer output, bg/sig label of training event, and $p_c(\x)$, but not $p(\e)$. The existence of such cost functions can be seen by noting that
\begin{enumerate}
\item For a given categorizer, $p_c(\x)$ can be computed directly from the bg/sig label without using $p(\e)$.
\item Any function linear in $p(\e)$ (like our eventwise contribution functions $\edstat$) can be estimated by replacing $p(\e)$ with the bg/sig label (0/1) and averaging over the event sample.
\end{enumerate}

While theoretically both approaches are equivalent (in the no training-error limit), one potential advantage of training without the preprocessing step is that we can train our categorizer until the validated improvement in our chosen distance measure saturates, instead of stopping the training process (for the preprocessing ML classifier) when the validated improvement in the prediction of $p(\e)$ saturates. If $p(\e)$ can be learned without errors, both approaches are equivalent. But if there will be errors in the estimation of $p(\e)$, then nixing the preprocessing step could prove to be a better use of computational training resources in terms of performance of the final product, which is the trained categorizer.

Since its conception, the presentation of our method has morphed significantly as a result of our interactions with the experimental and data analysis community. The current form provides for easier description of our idea as well as (hopefully) easy adoption. Instead of implementing an iterative training algorithm with brand new cost functions, users can simply use our one-size-fits-almost-all implementation on top of a preprocessing step which is already commonplace in high energy data analysis. The performance difference between the two approaches is likely to be small considering that in several applications, the dimensionality of HEP data is low enough for a fairly accurate estimation of $p(\e)$ using ML techniques, given the typical computational power at our disposal these days. Nevertheless, for the sake of completeness we might elaborate on the ``categorization training prescription sans preprocessing'' in a future publication.

\acknowledgments
The work of KM and PS is supported in part by the United States Department of Energy under Grant No. DE-SC0010296. PS is grateful to the LHC Physics Center at Fermilab for hospitality and financial support as part of the Guests and Visitors Program in the summer of 2019.
PS would like to thank Prof. Arunava Banerjee whose course on machine learning at the University of Florida jump-started this project. The authors would like to thank the following colleagues for useful discussions and inputs: Darin Acosta, Felix Kling, Jacobo Konigsberg, Danny Noonan, Andrzej Novak, Brendan O’Brien, Nathaniel Joseph Pastika, Jake Rosenzweig, Reinhard Schwienhorst, and Sean-Jiun Wang.

\appendix
\section{Statistical distances} \label{appendix:a}

In this section, we will derive the expressions for the different statistical distances given in \eref{eqn:d}. Let $\myS$ be an ordered\footnote{\label{foot:ordering}Ordering of events can be derived from some form of event id. Ordering is demanded just so we don't have to worry about combinatorial factors when writing down the probability (density) of a particular instance of $\myS$, $\myS_c$, or $\Upsilon_c$, since the events are distinguishable.} collection of collider events. A given instance of $\myS$ will be treated as one of the possible outcomes of the experiment performed to produce it. The events in $\myS$ are independent and identically distributed. The total number of events in $\myS$ is a Poisson distributed random variable whose mean could possibly depend on the underlying theory.

After categorizing events in $\myS$ into $C$ categories (and possibly rejecting some), let $\myS_c$ be the ordered\footref{foot:ordering} collection of events in category $c$. Let $\Upsilon_c\equiv[k_c, (\x_{c,1}, \x_{c,2},\dots,\x_{c,k_c})]$ represent the list of $\x$ values for events in category $c$, where $k_c$ is the number of events in category $c$.

Since hypothesis testing is to be performed by analyzing the $\Upsilon$-s, we will next write down their probability densities under the two hypotheses. Let $\mathcal{P}_\text{b}(\,\bm{\cdot}\,)\equiv\mathcal{P}(\,\bm{\cdot}\,\cond\Hn)$ and $\mathcal{P}_\text{b+s}(\,\bm{\cdot}\,)\equiv\mathcal{P}(\,\bm{\cdot}\,\cond\Ha)$ represent the probability density of $\,\bm{\cdot}\,$ under the null and alternative hypotheses respectively. We can write the probability densities of $\Upsilon_c$ as
\begin{subequations}
\begin{align}
\mathcal{P}_\text{b}(\Upsilon_c) &= \frac{e^{-B_c} B_c^{k_c}}{k_c!}\prod\limits_{i=1}^{k_c} \frac{b_c(\x_{c,i})}{B_c} \label{eqn:PbUp}~~,\\
\mathcal{P}_\text{b+s}(\Upsilon_c) &= \frac{e^{-N_c} N_c^{k_c}}{k_c!}\prod\limits_{i=1}^{k_c} \frac{n_c(\x_{c,i})}{N_c} \label{eqn:PbsUp}~~.
\end{align}
\end{subequations}
These probability densities are normalized as\footnote{\Eref{eqn:normalization} implicitly specifies the reference measure with respect to which the probability densities of $\Upsilon_c$ and $\{\Upsilon\}$ are defined.}
\begin{equation}
\int d\Upsilon_c~\mathcal{P}(\Upsilon_c) \equiv \sum\limits_{k_c=0}^{\infty}~\int\limits_{\Omega_\x}d\x_{c,1}\dots~\int\limits_{\Omega_\x}d\x_{c,k_c}~\mathcal{P}(\Upsilon_c) = 1~~. \label{eqn:normalization}
\end{equation}
Since the different $\Upsilon$-s are independent of each other, their joint probability density is given by
\begin{subequations}
\begin{align}
\mathcal{P}_\text{b}(\{\Upsilon\}) &\equiv \mathcal{P}_\text{b}(\Upsilon_1,\dots,\Upsilon_C) = \prod\limits_{c=1}^C \mathcal{P}_\text{b}(\Upsilon_c) \label{eqn:PbUpset}~~,\\
\mathcal{P}_\text{b+s}(\{\Upsilon\}) &\equiv \mathcal{P}_\text{b+s}(\Upsilon_1,\dots,\Upsilon_C) = \prod\limits_{c=1}^C \mathcal{P}_\text{b+s}(\Upsilon_c) \label{eqn:PbsUpset}~~.
\end{align}
\end{subequations}
From \erefs\eqref{eqn:PbUp}, \eqref{eqn:PbsUp}, \eqref{eqn:PbUpset}, and \eqref{eqn:PbsUpset}, the ratio of the probability densities can be written as
\begin{equation}
\frac{\mathcal{P}_\text{b+s}(\{\Upsilon\})}{\mathcal{P}_\text{b}(\{\Upsilon\})} = \prod\limits_{c=1}^C \left[e^{-N_c+B_c} \prod\limits_{i=1}^{k_c} \frac{n_c(\x_{c, i})}{b_c(\x_{c, i})}\right]~~.\label{eqn:likeratio}
\end{equation}
This expression will be used in the derivation of some of our statistical distances.

\subsection{\texorpdfstring{$\dneym$}{DNeym} and \texorpdfstring{$\dpear$}{DPear}: Asymptotic expected values of Neyman-\texorpdfstring{$\chi^2$}{chisq} and Pearson-\texorpdfstring{$\chi^2$}{chisq} statistics}
In binned analysis of the distribution of $\x$ within each category, the Neyman-$\chi^2$ and Pearson-$\chi^2$ test statistics are given by
\begin{subequations}
\begin{align}
\text{Neym}\chi^2 = \sum\limits_{c=1}^C~\sum\limits_{i\in \x\text{\,bins}} \frac{\left(\text{obs}_{c,i} - \text{exp}_{c,i}\right)^2}{\text{obs}_{c,i}}~~,\\
\text{Pear}\chi^2 = \sum\limits_{c=1}^C~\sum\limits_{i\in \x\text{\,bins}} \frac{\left(\text{obs}_{c,i} - \text{exp}_{c,i}\right)^2}{\text{exp}_{c,i}}~~.
\end{align}
\end{subequations}
where $\text{exp}_{c,i}$ and $\text{obs}_{c,i}$ are respectively the expected and observed number of events in bin $i$ of category $c$. The summation is only performed over bins with non-zero $\text{obs}_{c,i}$ for Neyman-$\chi^2$, and non-zero $\text{exp}_{c,i}$ for Pearson-$\chi^2$.

Let $N_{c,i}$, $B_{c,i}$, and $S_{c,i}$ be the number densities $n_c(\x)$, $b_c(\x)$, and $s_c(\x)$ integrated over the bin $i$ in category $c$. For both variants of the $\chi^2$ statistic, let $\chi^2_{b}$ and $\chi^2_{b+s}$ be the value of the statistic calculated with $\text{exp}_{c,i}$ set to $B_{c,i}$ and $N_{c,i}$ respectively.

In the asymptotic and fine binning limit, we can show for both variants of the $\chi^2$ statistic that
\begin{subequations}
\begin{align}
\E\left[~\chi^2_{b}~;~\Ha~\right] &\rightarrow D_{\chi^2} \label{eqn:dchisqinterpret1}~~,\\
\bigg[\E\left[~\chi^2_{b}~;~\Ha~\right] - \E\left[~\chi^2_{b}~;~\Hn~\right]\bigg] &\rightarrow D_{\chi^2} \label{eqn:dchisqinterpret2}~~,\\
\E\left[~\chi^2_{b} - \chi^2_{b+s}~;~\Ha~\right] &\rightarrow D_{\chi^2} \label{eqn:dchisqinterpret3}~~,
\end{align}
\end{subequations}
where $\E[\;\cdot\;;\;\Hn]$ and $\E[\;\cdot\;;\;\Ha]$ represent the expected value of $\cdot$ under $\Hn$ and $\Ha$ respectively, and the two variants of $D_{\chi^2}$ are the $\dneym$ and $\dpear$ defined in equations ~\eqref{eqn:dneym} and \eqref{eqn:dpear}.

The left-hand side of \eref{eqn:dchisqinterpret1} is a measure of how poorly $\Hn$ would fit data produced under $\Ha$. The left-hand side of \eref{eqn:dchisqinterpret2} is a measure of how much worse $\Hn$ would fit data produced under $\Ha$ than it would fit data produced under $\Hn$. And the left-hand side of \eref{eqn:dchisqinterpret3} is a measure of how much worse $\Hn$ would be than $\Ha$ at fitting data produced under $\Ha$.

We will omit an explicit proof of statements (\ref{eqn:dchisqinterpret1}-\ref{eqn:dchisqinterpret3}) here, and instead only provide an outline. We begin with the following observations: If $X$ is a Poisson distributed random variable with mean $\mu$, then the expected value of $X^2$ is $\mu^2+\mu$. Similarly, the expected value of $1/X$ for positive $X$ tends to $1/\mu + 1/\mu^2$ in the large $\mu$ limit \cite{Jones:2004negmom}, and the probability of $X$ being $0$ vanishes in the large $\mu$ limit. Using these ideas, we can show that asymptotically (up to the fastest growing term), the left-hand sides of eqns~(\ref{eqn:dchisqinterpret1}-\ref{eqn:dchisqinterpret3}) tend to
\begin{equation}
\sum\limits_{c=1}^C~\sum\limits_{i\in \x\text{\,bins}} \frac{\left(N_{c,i} - B_{c,i}\right)^2}{N_{c,i}}
\end{equation}
for Neyman-$\chi^2$ and
\begin{equation}
\sum\limits_{c=1}^C~\sum\limits_{i\in \x\text{\,bins}} \frac{\left(N_{c,i} - B_{c,i}\right)^2}{B_{c,i}}
\end{equation}
for Pearson-$\chi^2$. These are approximations to the expressions for $\dneym$ and $\dneym$ in \erefs\eref{eqn:dneym} and \eref{eqn:dpear}, and they can be made arbitrarily accurate in the fine binning limit.

\subsection{\texorpdfstring{$\dkl$}{DKL}: Kullback--Leibler divergence from null to alternative}
The Kullback--Leibler divergence from the null hypothesis to the alternative hypothesis (for the $\Upsilon$-s) is given by
\begin{equation}
\dkl \equiv \dkl\left(\Ha\parallel\Hn\right) = \int d\{\Upsilon\}~\mathcal{P}_\text{b+s}(\{\Upsilon\}) \ln\left[\frac{\mathcal{P}_\text{b+s}(\{\Upsilon\})}{\mathcal{P}_\text{b}(\{\Upsilon\})}\right]~~, \label{eqn:dkldef}
\end{equation}
where $\int d\{\Upsilon\}$ is a shorthand for
\begin{equation}
\int d\{\Upsilon\} \equiv \prod\limits_{c=1}^C\left[\sum\limits_{k_c=0}^{\infty}~\int\limits_{\Omega_\x}d\x_{c,1}\dots~\int\limits_{\Omega_\x}d\x_{c,k_c}\right]~~. \label{eqn:shorthand}
\end{equation}
Plugging \eref{eqn:likeratio} into \eref{eqn:dkldef}, and using $s_c = n_c-b_c$ and $S_c = N_C-B_c$, we get
\begin{equation}
\dkl = \int d\{\Upsilon\}~\mathcal{P}_\text{b+s}(\{\Upsilon\})~\sum\limits_{c=1}^C \left[-S_c -\sum\limits_{i=1}^{k_c}\ln\left[1-\frac{s_c(\x_{c,i})}{n_c(\x_{c,i})}\right]\right]~~.
\end{equation}
Using the fact that the different $\Upsilon$-s are independent of each other, and the fact that $\x_{c,i}$ for different $i$ values are independent of each other, we can simplify this to
\begin{align}
\dkl &=\sum\limits_{c=1}^C \vast[-S_c - \left[\sum\limits_{k_c=0}^{\infty} \frac{e^{-N_c}N_c^{k_c}}{k_c!}\,k_c\right] \frac{1}{N_c}\int\limits_{\Omega_{\x}}d\x~n_c(\x)\ln\left[1-\frac{s_c(\x)}{s_c(\x)}\right]\vast]\\
&=\sum\limits_{c=1}^C \vast[-S_c - \int\limits_{\Omega_{\x}}d\x~n_c(\x)\ln\left[1-\frac{s_c(\x)}{n_c(\x)}\right]\vast]\\
&=\sum\limits_{c=1}^C~\int\limits_{\Omega_{\x}}d\x~\left[-s_c(\x) - n_c(\x) \ln\left[1-\frac{s_c(\x)}{n_c(\x)}\right]\right]~~. \label{eqn:dklderive}
\end{align}
The last equation matches the expression for $\dkl$ given in \eref{eqn:dkl}.

\subsection{\texorpdfstring{$\drevkl$}{DrevKL}: Kullback--Leibler divergence from alternative to null}
The Kullback--Leibler divergence from the alternative hypothesis to the null hypothesis (for the $\Upsilon$-s) is given by
\begin{equation}
\drevkl \equiv \dkl\left(\Hn\parallel\Ha\right) = \int d\{\Upsilon\}~~\mathcal{P}_\text{b}(\{\Upsilon\}) \ln\left[\frac{\mathcal{P}_\text{b}(\{\Upsilon\})}{\mathcal{P}_\text{b+s}(\{\Upsilon\})}\right]~~. \label{eqn:drevkldef}
\end{equation}
Using manipulations similar to the ones we did for $\dkl$, we can show that
\begin{equation}
\drevkl = \sum\limits_{c=1}^C~\int\limits_{\Omega_{\x}}d\x~\left[s_c(\x) + b_c(\x) \ln\left[1-\frac{s_c(\x)}{n_c(\x)}\right]\right]~~, \label{eqn:drevklderive}
\end{equation}
which matches the expression for $\drevkl$ given in \eref{eqn:drevkl}. This expression can also be derived from \eref{eqn:dklderive} by making the substitutions $n_c \longleftrightarrow b_c$ and $s_c \rightarrow -s_c$.

\subsection{\texorpdfstring{$\djeff$}{DJ}: Jeffreys divergence}
Jeffreys divergence is a symmetrized version of Kullback--Leibler divergence given by the sum of the two asymmetric KL divergences. From \erefs\eqref{eqn:dklderive} and \eqref{eqn:drevklderive} we can see that
\begin{align}
\djeff &= \dkl + \drevkl\\
&=\sum\limits_{c=1}^C~\int\limits_{\Omega_{\x}}d\x~\left[-s_c(\x) \ln\left[1-\frac{s_c(\x)}{n_c(\x)}\right]\right]~~.
\end{align}
This matches the expression for $\djeff$ given in \eref{eqn:djeff}.

\subsection{\texorpdfstring{$\db$}{DB}: Bhattacharyya distance}
The Bhattacharyya distance between null and alternative hypotheses (for the $\Upsilon$-s) is given by
\begin{equation}
\db = -\ln \left[\int d\{\Upsilon\}~\sqrt{\mathcal{P}_\text{b}(\{\Upsilon\})~\mathcal{P}_\text{b+s}(\{\Upsilon\})}\right]~~.
\end{equation}
We can rewrite this as
\begin{equation}
e^{-\db} = \int d\{\Upsilon\}~\mathcal{P}_\text{b+s}(\{\Upsilon\})~\sqrt{\frac{\mathcal{P}_\text{b}(\{\Upsilon\})}{\mathcal{P}_\text{b+s}(\{\Upsilon\})}}~~. \label{eqn:exp_db}
\end{equation}
Plugging \eref{eqn:likeratio} into \eref{eqn:exp_db} we get
\begin{equation}
e^{-\db} = \Bigg[\prod\limits_{c=1}^C e^{S_c/2}\Bigg] \left[\int d\{\Upsilon\}~\mathcal{P}_\text{b+s}(\{\Upsilon\})\prod\limits_{c=1}^C\prod\limits_{i=1}^{k_c}\sqrt{\frac{b_c(\x_{c,i})}{n_c(\x_{c,i})}}~\right]~~.
\end{equation}
Plugging in \erefs\eqref{eqn:PbsUp}, \eqref{eqn:PbsUpset}, and \eqref{eqn:shorthand} here and using the fact that the different $\Upsilon$-s are independent of each other, and the fact that $\x_{c,i}$ for different $i$ values are independent of each other, we can simplify this to
\begin{equation}
e^{-\db} = \prod\limits_{c=1}^C \left[e^{S_c/2}\,e^{-N_c}\sum\limits_{k_c=0}^\infty \frac{1}{k_c!} \left[\;\int\limits_{\Omega_\x} d\x~n_c(\x) \sqrt{\frac{b_c(\x)}{n_c(\x)}}\right]^{k_c}\right]~~. \label{eqn:dbintermediate}
\end{equation}
Using the power series representation of the exponential function given by
\begin{equation}
e^A = \sum\limits_{k=0}^\infty \frac{A^k}{k!}~~,
\end{equation}
we can rewrite \eref{eqn:dbintermediate} as
\begin{align}
e^{-\db} &= \prod\limits_{c=1}^C \left[e^{S_c/2}\,e^{-N_c}\exp{\left[\;\int\limits_{\Omega_\x} d\x~n_c(\x) \sqrt{\frac{b_c(\x)}{n_c(\x)}}\right]}\right]\\
&= \exp{\left[\sum\limits_{c=1}^C~\int\limits_{\Omega_\x}d\x\left[\frac{s_c(\x)}{2} - n_c(\x) + n_c(\x) \sqrt{\frac{b_c(\x)}{n_c(\x)}}~\right]\right]}~~.
\end{align}
This matches the expression for $\db$ given in \eref{eqn:db} (after taking natural-log and multiplying by $-1$).

\section{An interpretation of Kullback--Leibler divergence} \label{appendix:b}

Let us consider a generic statistical test that decides between $\Hn$ or $\Ha$, based on data (post-categorization, i.e., after reduction of events to their $c$ and $\x$ values).

Power of the test, $1-\beta$, is the probability of the test choosing $\Ha$, given that $\Ha$ is true. It captures the ability of the test to detect the presence of a signal.

Size (or significance level) of the test, $\alpha$, is the probability of the test choosing $\Ha$, given that $\Hn$ is true. It captures the false alarm rate, and is typically set to \mbox{$\sim 3\times 10^{-7}$} (5$\sigma$) in high energy physics.

A statistical test that has the greatest power $1-\beta^*$ among all tests of a given size $\alpha^*$ is called a uniformly most powerful test or a UMP test (with $\alpha^*$ being a tunable parameter in the test). The likelihood-ratio test is a UMP for testing point hypotheses (like our $\Hn$ and $\Ha$). It can be shown for point hypotheses that
\begin{enumerate}
\item For a UMP test of \textbf{given power} $0 \leq 1-\beta^* < 1$,
\begin{equation}
\alpha^* \longrightarrow \left(1-\beta^*\right) e^{-\dkl} \text{~~~~asymptotically} \label{eqn:interpret1}
\end{equation}
\item For a UMP test of \textbf{given size} $0 < \alpha^* \leq 1$,
\begin{equation}
{1-\beta^*} \longrightarrow 1 - \left(1-\alpha^*\right) e^{-\drevkl} \text{~~~~asymptotically} \label{eqn:interpret2}
\end{equation}
\end{enumerate}
where ``asymptotically'' refers to the large number of events limit.

When the goal of event categorization is to maximize the chance of signal discovery (assuming its existence), we want to maximize the power for a given size (say, p-value corresponding to $5\sigma$). This can be done by maximizing $\drevkl$, according to statement~\eqref{eqn:interpret2}. When there is no a priori expectation for the size of the signal cross-section, it makes sense to work with a signal cross-section near the low end of discoverability.

On the other hand, when the goal of event categorization is to maximize the chance of ruling out the existence of a signal (assuming its absence), we want to minimize the size for a given power (say, $95\%$: this is the confidence that the signal does not exist---otherwise it would have been seen). This can be done by maximizing $\dkl$, according to statement~\eqref{eqn:interpret1}. When there is no a priori expectation for the size of the signal cross-section and the goal is to set the best upper-limit on it, it makes sense to work with a signal cross-section value near the expected upper-limit.

This discussion is purely theoretical and it is likely that the difference between training with $\tdrevkl$ and $\tdkl$ will not be practically significant in most situations. Nevertheless, we find this (slight) misalignment between the goals of event categorization for signal discovery and upper-limit setting a curiosity worth noting.

We will omit an explicit proof of statements~\eqref{eqn:interpret1} and \eqref{eqn:interpret2} here, and instead only provide an outline. Statement~\eqref{eqn:interpret1} can be proved by noting from eqn~\eqref{eqn:dkldef} that $\dkl$ is the expected value under $\Ha$ of the log-likelihood-ratio $\ln(\mathcal{P}_\text{b+s}\,/\,\mathcal{P}_\text{b})$ of $\{\Upsilon\}$. The other main ingredient in the proof is the fact that $\ln(\mathcal{P}_\text{b+s}\,/\,\mathcal{P}_\text{b})$ can be written as the sum of log-likelihood-ratios for the individual (independent and identically distributed) events. So, as per central limit theorem, $\ln(\mathcal{P}_\text{b+s}\,/\,\mathcal{P}_\text{b})$ will tend towards its expectation value, $\dkl$, in the asymptotic limit.\footnote{A nuance here is that the number of events in this summation is itself a random variable. But in the asymptotic limit, the number of events will tend towards its expectation value sufficiently fast that its variation can be ignored.} This means that for a UMP test of a given power\footnote{That the power is fixed forces us to have a critical region with fixed (non-vanishing) probability under $\Ha$ in the asymptotic limit. If instead we keep the size fixed, the critical region will have a fixed probability under $\Hn$, and the expectation value of log-likelihood-ratio under $\Hn$ ($\drevkl$) will become relevant in the asymptotic limit.}, asymptotically, the likelihood-ratio $\mathcal{P}_\text{b+s}\,/\,\mathcal{P}_\text{b}$ for all experimental outcomes $\{\Upsilon\}$ in the critical region of the test will tend to $e^{\dkl}$, and hence so will the ratio between the power and size of the test.

Statement~\eqref{eqn:interpret2} can be proved in a similar fashion. Alternatively, it can also be derived from \eqref{eqn:interpret1} by making the substitutions $\alpha^* \longleftrightarrow \beta^*$ (Type I error $\longleftrightarrow$ Type II error) and $\dkl \rightarrow \drevkl$.

\section{Proof of correctness of the iterative training algorithm} \label{appendix:c}
In this section we will prove the correctness as well as convergence of our event categorizer training prescription. The key ingredients to the proof are the following two properties
\begin{itemize}
 \item[(a)] $\edstat(U, V)$ is linear in $V$.
 \item[(b)] $\edstat(U, V)$ has a global maximum in $U$ at $U = V$.
\end{itemize}
The linearity property (a) is by construction of $\edstat$-s from $\dstat$-s. The global maximum property (b) is from the nature of $\dstat$-s, which penalize mixing of regions of phase-space with different $p(\e)$ values.

In each iterative step, from the current categorizer, $\eta$, a new categorizer, $\bar{\eta}$, is constructed. Let $p_c(\x)$ and $\bar{p}_c(\x)$ be the signal-fraction functions of $\eta$ and $\bar{\eta}$ respectively. Recall that $\bar{\eta}$ rejects an event (if allowed) if $\edstat\big(p_c(\x), p(\e)\big) < 0$ for all categories $c$. Otherwise, $\bar{\eta}$ places each event in the category that maximizes $\edstat\big(p_c(\x), p(\e)\big)$. From the construction of $\bar{\eta}$ it is clear that

\begin{equation}
 \mean_{\e\,\sim\,\Ha}~\left[\sum\limits_{c=1}^C~\delta\big(c, \eta(\e)\big)~\edstat\big(p_c(\x), p(\e)\big)\right] \leq \mean_{\e\,\sim\,\Ha}~\left[\sum\limits_{c=1}^C~\delta\big(c, \bar{\eta}(\e)\big)~\edstat\big(p_c(\x), p(\e)\big)\right] \label{eqn:c1}
\end{equation}
By treating ``averaging over all events'' as averaging over events with a given value of $(c, \x)$ followed by an (appropriately weighted) average over $(c, \x)$ values, we can write
\begin{equation}
 \mean_{\e\,\sim\,\Ha}~\left[\sum\limits_{c=1}^C~\delta\big(c, \bar{\eta}(\e)\big)~\edstat\big(p_c(\x), p(\e)\big)\right] = \mean_{\e\,\sim\,\Ha}~\left[\sum\limits_{c=1}^C~\delta\big(c, \bar{\eta}(\e)\big)~\edstat\big(p_c(\x), \bar{p}_c(\x)\big)\right] \label{eqn:c2}
\end{equation}
Here we have used the fact that all $\edstat$-s are linear in the second variable. Though we don't explicitly derive \eref{eqn:c2} here, it can be shown to follow from \eref{eqn:sampleavg1}. Next, using the fact that $\edstat(A, B) \leq \edstat(B, B)$ for all $A, B\in [0,1)$, we can write
\begin{equation}
 \mean_{\e\,\sim\,\Ha}~\left[\sum\limits_{c=1}^C~\delta\big(c, \bar{\eta}(\e)\big)~\edstat\big(p_c(\x), \bar{p}_c(\x)\big)\right] \leq \mean_{\e\,\sim\,\Ha}~\left[\sum\limits_{c=1}^C~\delta\big(c, \bar{\eta}(\e)\big)~\edstat\big(\bar{p}_c(\x), \bar{p}_c(\x)\big)\right] \label{eqn:c3}
\end{equation}
Using \eref{eqn:sampleavg1} and the linearity of $\edstat$ in the second variable again, we can bring $p(\e)$ back into the expression as
\begin{equation}
 \mean_{\e\,\sim\,\Ha}~\left[\sum\limits_{c=1}^C~\delta\big(c, \bar{\eta}(\e)\big)~\edstat\big(\bar{p}_c(\x), \bar{p}_c(\x)\big)\right] = \mean_{\e\,\sim\,\Ha}~\left[\sum\limits_{c=1}^C~\delta\big(c, \bar{\eta}(\e)\big)~\edstat\big(\bar{p}_c(\x), p(\e)\big)\right] \label{eqn:c4}
\end{equation}
Combining eqns~(\ref{eqn:c1}-\ref{eqn:c4}), we can see that
\begin{equation}
 \mean_{\e\,\sim\,\Ha}~\left[\sum\limits_{c=1}^C~\delta\big(c, \eta(\e)\big)~\edstat\big(p_c(\x), p(\e)\big)\right] \leq \mean_{\e\,\sim\,\Ha}~\left[\sum\limits_{c=1}^C~\delta\big(c, \bar{\eta}(\e)\big)~\edstat\big(\bar{p}_c(\x), p(\e)\big)\right] \label{eqn:c5}
\end{equation}
This can we rewritten using \eref{eqn:eventwisestat} as
\begin{equation}
 \left\{~\tdneym \text{ corresponding to } \eta~\right\}\leq \left\{~\tdneym \text{ corresponding to } \bar{\eta}~\right\}~~. \label{eqn:c6}
\end{equation}

\Eref{eqn:c6} shows that the categorizer ``improves'' in each iterative step. The only caveat here is the assumption that $p_c(\x)$ can be estimated accurately enough at each step.

To see that the algorithm will converge in a finite number of steps, note that during the training process a categorizer in the current step is defined completely by the assignment of training events to categories in the previous step. If the categorizer has to improve in each step, it has to be different in each step (no cycles), and there is only a finite number of possible category assignments for training events. So, in a finite number of iterations, the algorithm should converge on a categorizer that maximizes (the estimate from training data of) $\tdstat$ \textbf{locally}. 

A caveat to this argument is that due to artifacts in the unbinned estimation of the signal-fraction functions $p_c(\x)$, it is possible to run into cycles near the end of the training process. So it may make sense to terminate the iteration process when the fraction of events reassigned in some step drops below a certain threshold or if the (absolute or relative) gain in the estimated $\tdstat$ drops below a certain threshold. In binned analysis the convergence argument presented above holds without caveats, and it is guaranteed that $\bar{\eta}$ will be the same as $\eta$ in some training step---although it may take an unfeasible number of iterations to reach this stage.

The time complexity of the training process is $\mathcal{O}(C\mathcal{N}i + f(C, \mathcal{N}) i)$ where $\mathcal{N}$ is the number of training events, $C$ is the number of categories, $i$ is the number of iterative steps. $f(C, \mathcal{N})$ is the cost associated with estimating $p_c(\x)$, for each training event and $c$, from the training data (line 6 and line 8 (called from line 5) of algorithm~\ref{algorithm}).  
For binned analyses with $M$ $\x$-bins, $f(\mathcal{N}, C)$ is $\mathcal{O}(CM + C\mathcal{N})$, so the overall time complexity for training is $\mathcal{O}(C\mathcal{N} i)$ (assuming $\mathcal{N} > M$, i.e., there will be more training events than bins).

For unbinned analyses, it is possible to estimate $p_c(\x)$ using kernel regression (with appropriate choice of kernels) in $\mathcal{O}(2^d C\mathcal{N} \log \mathcal{N})$ 
time where $d$ is the dimensionality of $\x$ (using a k-d tree based implementation). This leads to an overall training time complexity of $\mathcal{O}(2^d C\mathcal{N} \log \mathcal{N} i)$.

While the number of iterations $i$ needed for convergence in the worst case is superpolynomial (unbounded by any polynomial), in practice, for reasonable values of $C$ and $d$, the value of $\tdstat$ flattens out after a few tens of iterations---just like in Lloyd's k-means clustering algorithm. So, with appropriate termination criteria, the iterative training process can be implemented to practically have a time complexity of $\mathcal{O}(C\mathcal{N})$ when using binned and $\mathcal{O}(C\mathcal{N}\log \mathcal{N})$ when using unbinned estimation of $p_c$ (ignoring $i$ and $2^d$).

\section{A comparison of the different distance measures} \label{appendix:d}

In a given iterative step of training, each event $\e$ will be categorized into one of the categories based on their $p_c(\x(\e))$ values from the previous step---event $\e$ will be placed into the one of the categories with previous $p_c(\x(\e))$ closest to $p(\e)$ on either side (if event rejection is allowed, rejected events can be thought of as forming a category of their own with $p_c(\x)=0$). The only difference between using different measures is in the threshold value of $p(\e)$, for a pair of categories, above which events will be placed in the ``higher purity'' category. 

Let $c_1$ and $c_2$ be two adjacent categories when ordered by $p_c(\x)$ for a given $\x$, with $p_{c_1}(\x) < p_{c_2}(\x)$. At the threshold value of $p(\e)$, say $\thstat$, between $p_{c_1}(\x)$ and $p_{c_2}(\x)$, $\edstat\Big[p_{c_1}(\x), \thstat\Big] = \edstat\Big[p_{c_2}(\x), \thstat\Big]$. For values of $p(\e)$ greater (lesser) than $\thstat$, $\edstat\big(p_{c_2}(\x), p(\e)\big)$ is greater (lesser) than $\edstat\big(p_{c_1}(\x), p(\e)\big)$.

Formally, let us define the threshold value $\thstat(p_1, p_2)$ of $p(\e)$, at a given value of $\x$, between two categories with $p_{c_1}(\x) = p_1$ and $p_{c_2}(\x) = p_2$, as satisfying
\begin{subequations}
\begin{align}
\min(p_1, p_2) \leq \thstat&(p_1, p_2) \leq \max(p_1, p_2)~~,\\
\edstat\big(p_1, \thstat(p_1, p_2)\big) &= \edstat\big(p_2, \thstat(p_1, p_2)\big)~~.
\end{align}
\end{subequations}
$\thstat$ is related to $p^\text{sel\_thresh}_\text{stat}$ defined \eref{eqn:selthresh} as $p^\text{sel\_thresh}_\text{stat}(p_c) = \thstat(0, p_c)$.
\vskip 5mm
It can be proved that $\forall~(p_1, p_2) \in [0, 1)\times[0, 1)$,
\begin{equation}
\thneym \leq \threvkl \leq \thb \leq \thjeff \leq \thkl \leq \thpear~~,\label{eqn:strictorder2}
\end{equation}
with equality iff $p_1=p_2$. 

\begin{figure}[htp]
\begin{tabular}{m{.27\textwidth} m{.3\textwidth} m{.3\textwidth}}
{$\thpear(p_1, p_2)$ (middle)\vskip 3mm
$\displaystyle\frac{\thpear(p_1, p_2)}{\thkl(p_1, p_2)}$ (right)\vskip 1mm} & \includegraphics[height=.14\textheight]{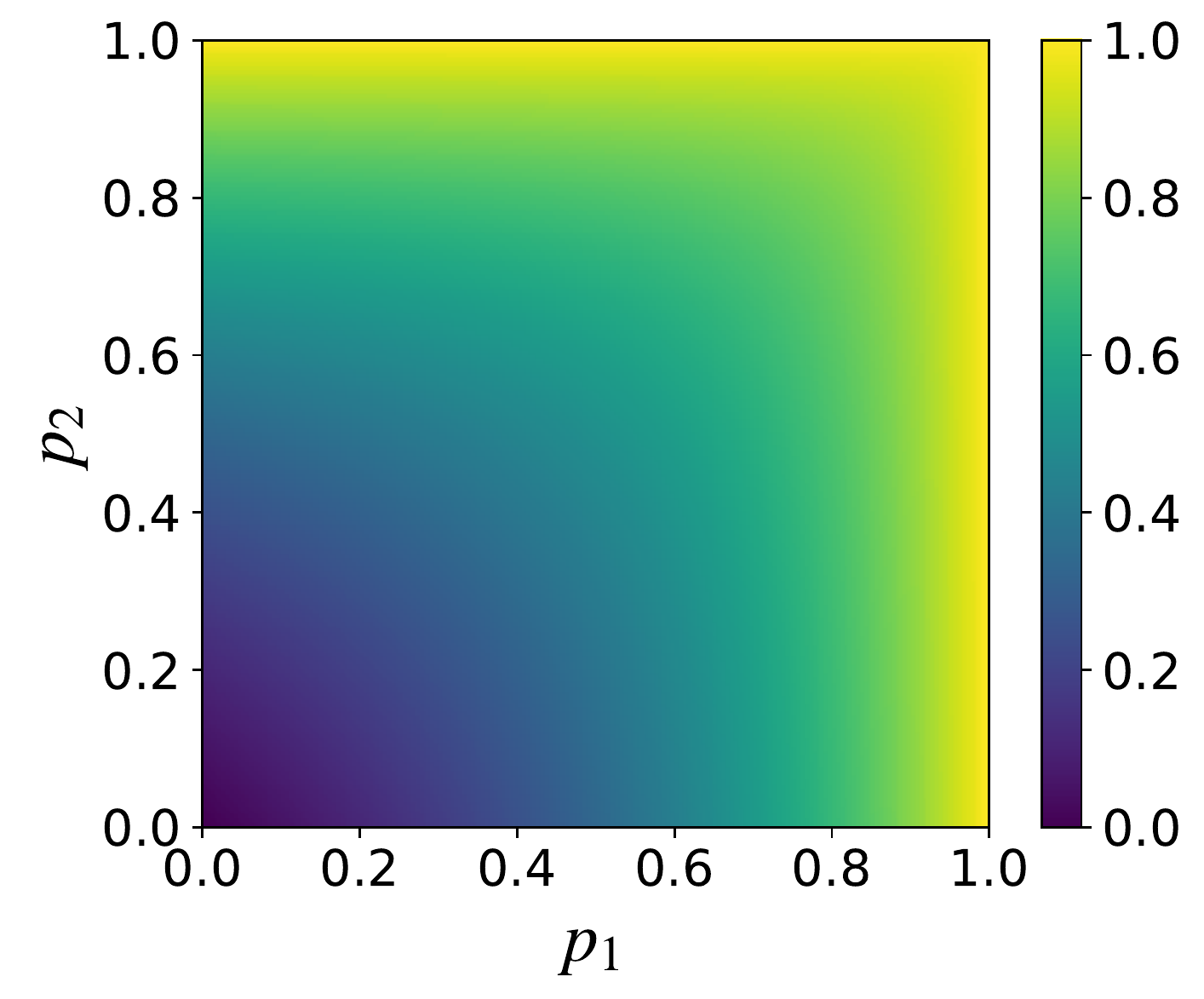} & \includegraphics[height=.14\textheight]{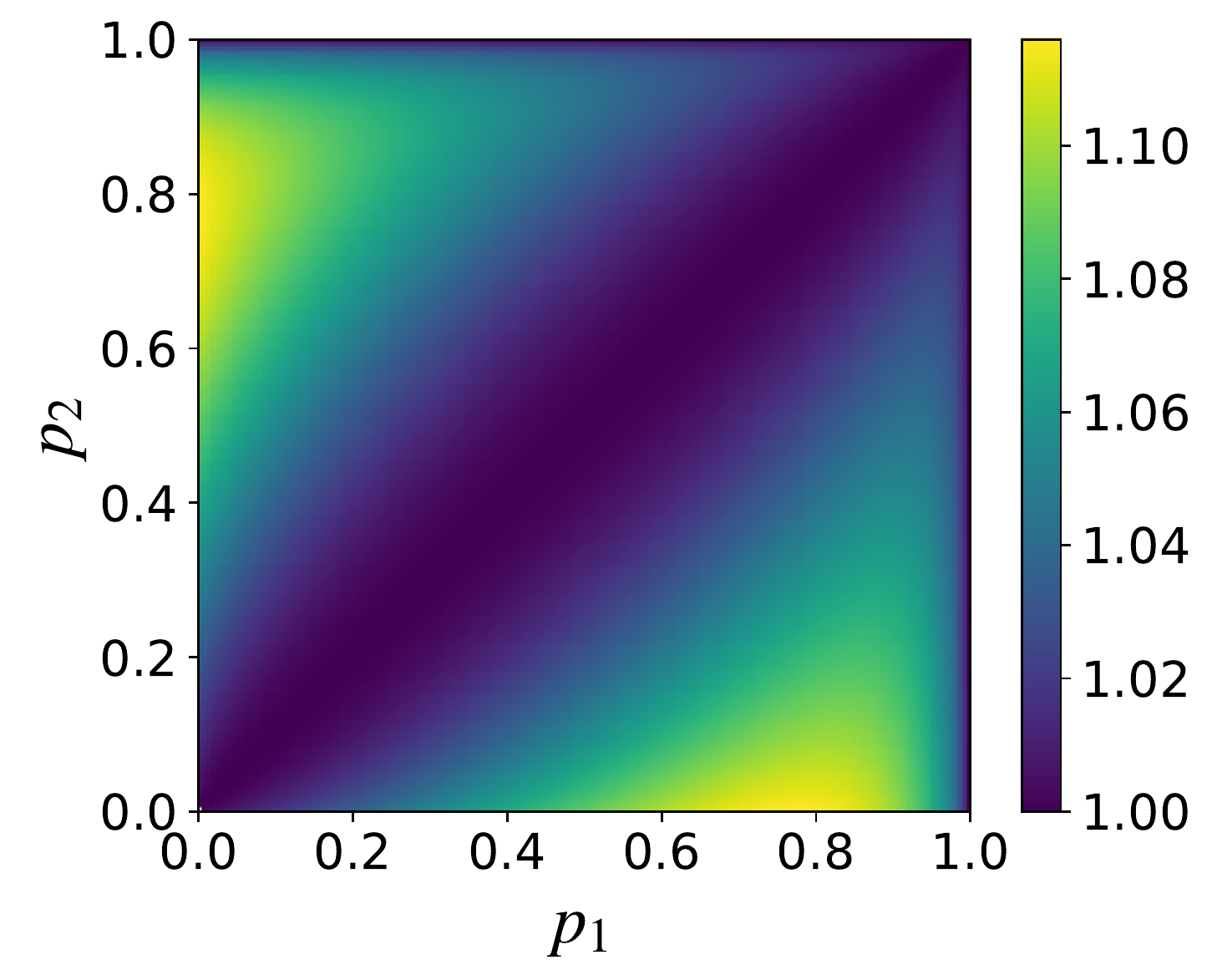}\\
{$\thkl(p_1, p_2)$ (middle)\vskip 3mm
$\displaystyle\frac{\thkl(p_1, p_2)}{\thjeff(p_1, p_2)}$ (right)\vskip 1mm} & \includegraphics[height=.14\textheight]{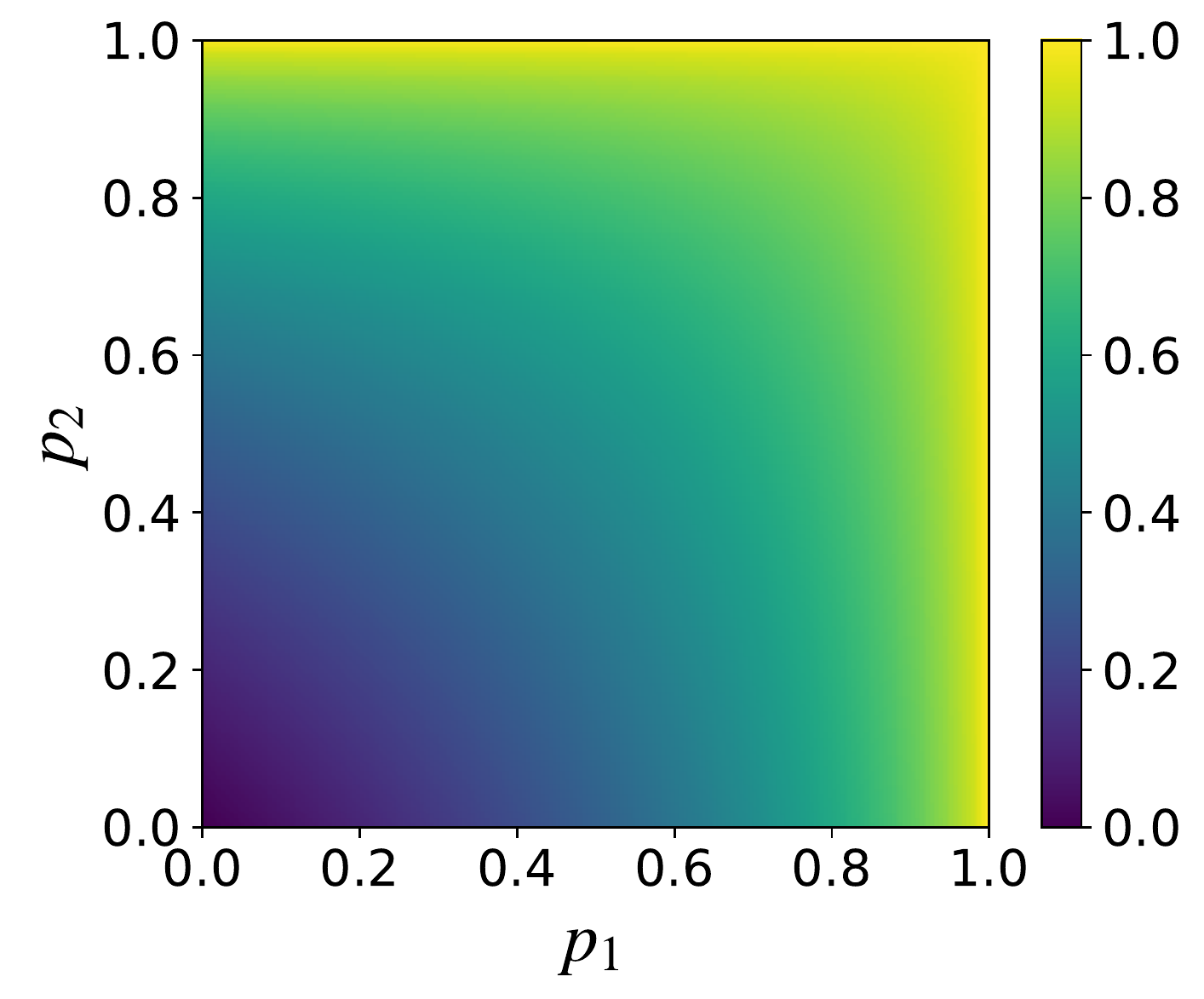} & \includegraphics[height=.14\textheight]{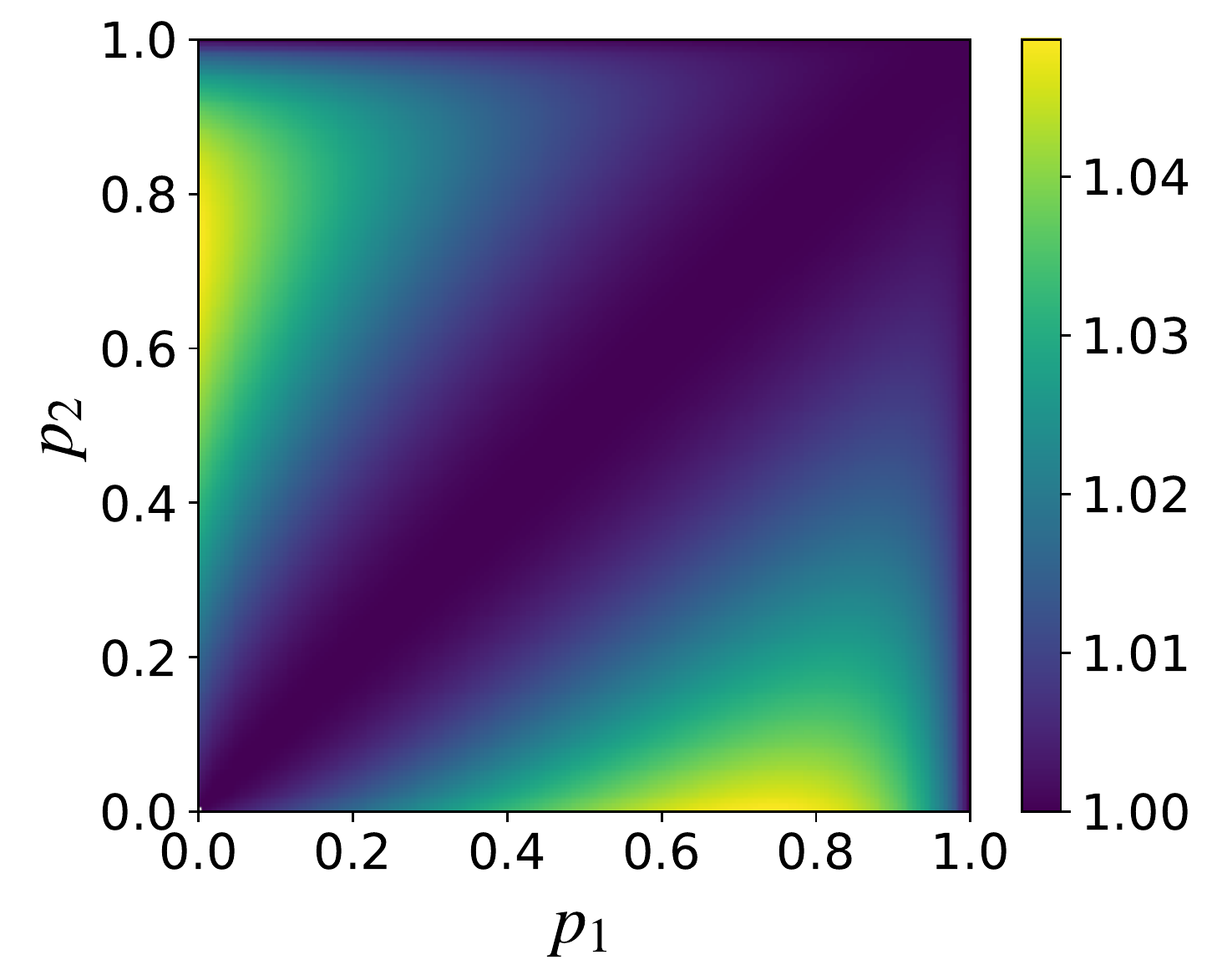}\\
{$\thjeff(p_1, p_2)$ (middle)\vskip 3mm
$\displaystyle\frac{\thjeff(p_1, p_2)}{\thb(p_1, p_2)}$ (right)\vskip 1mm} & \includegraphics[height=.14\textheight]{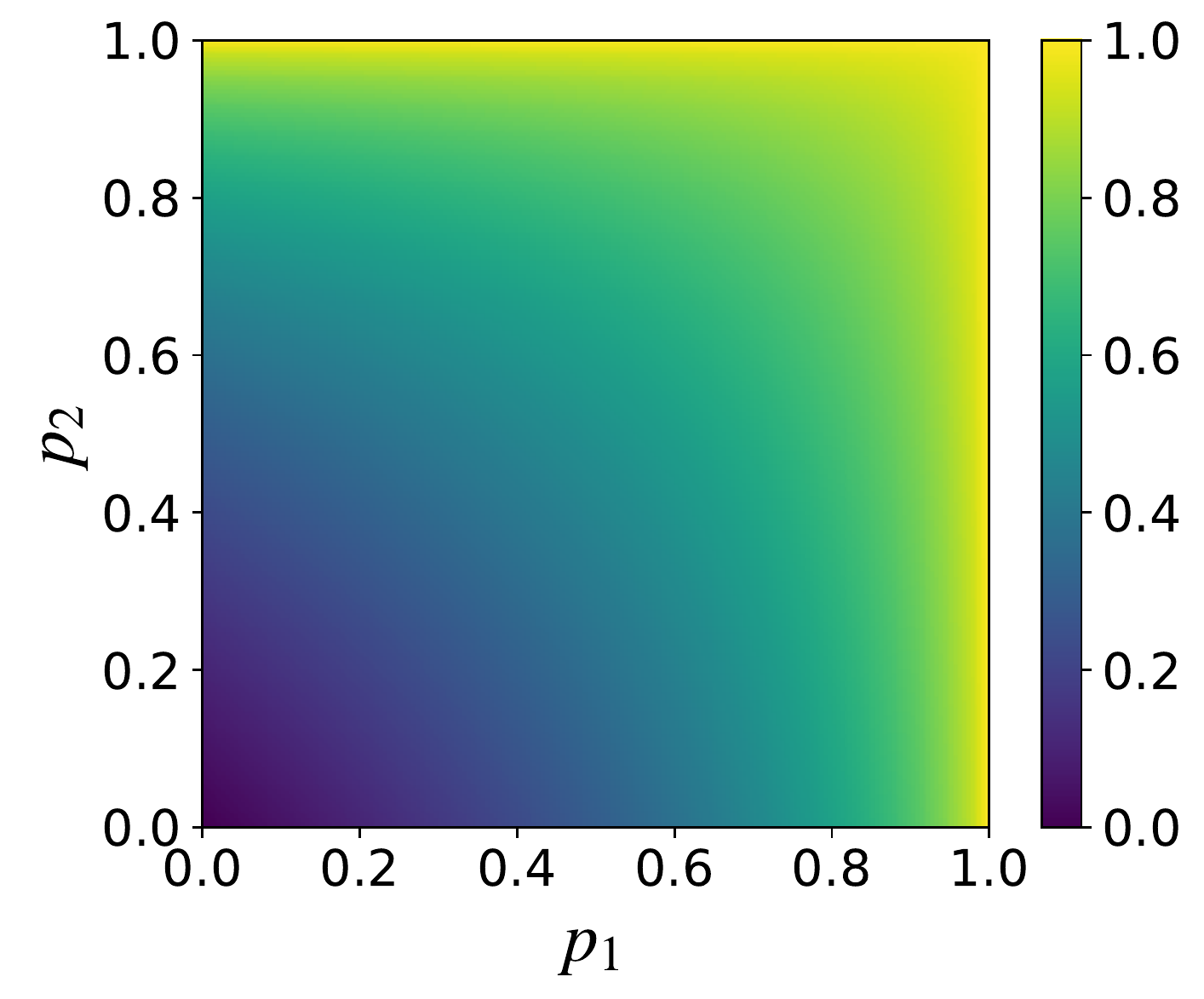} & \includegraphics[height=.14\textheight]{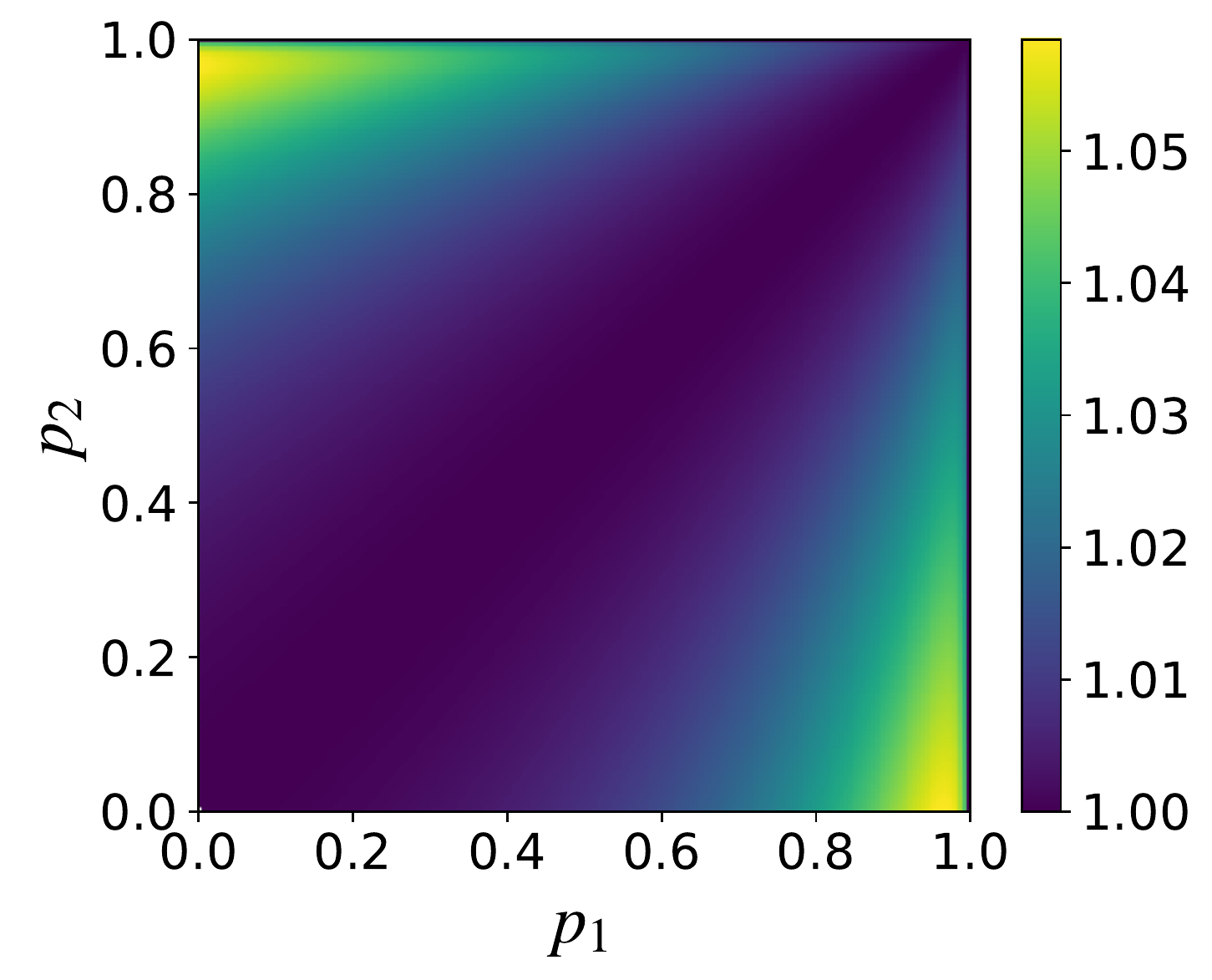}\\
{$\thb(p_1, p_2)$ (middle)\vskip 3mm
$\displaystyle\frac{\thb(p_1, p_2)}{\threvkl(p_1, p_2)}$ (right)\vskip 1mm} & \includegraphics[height=.14\textheight]{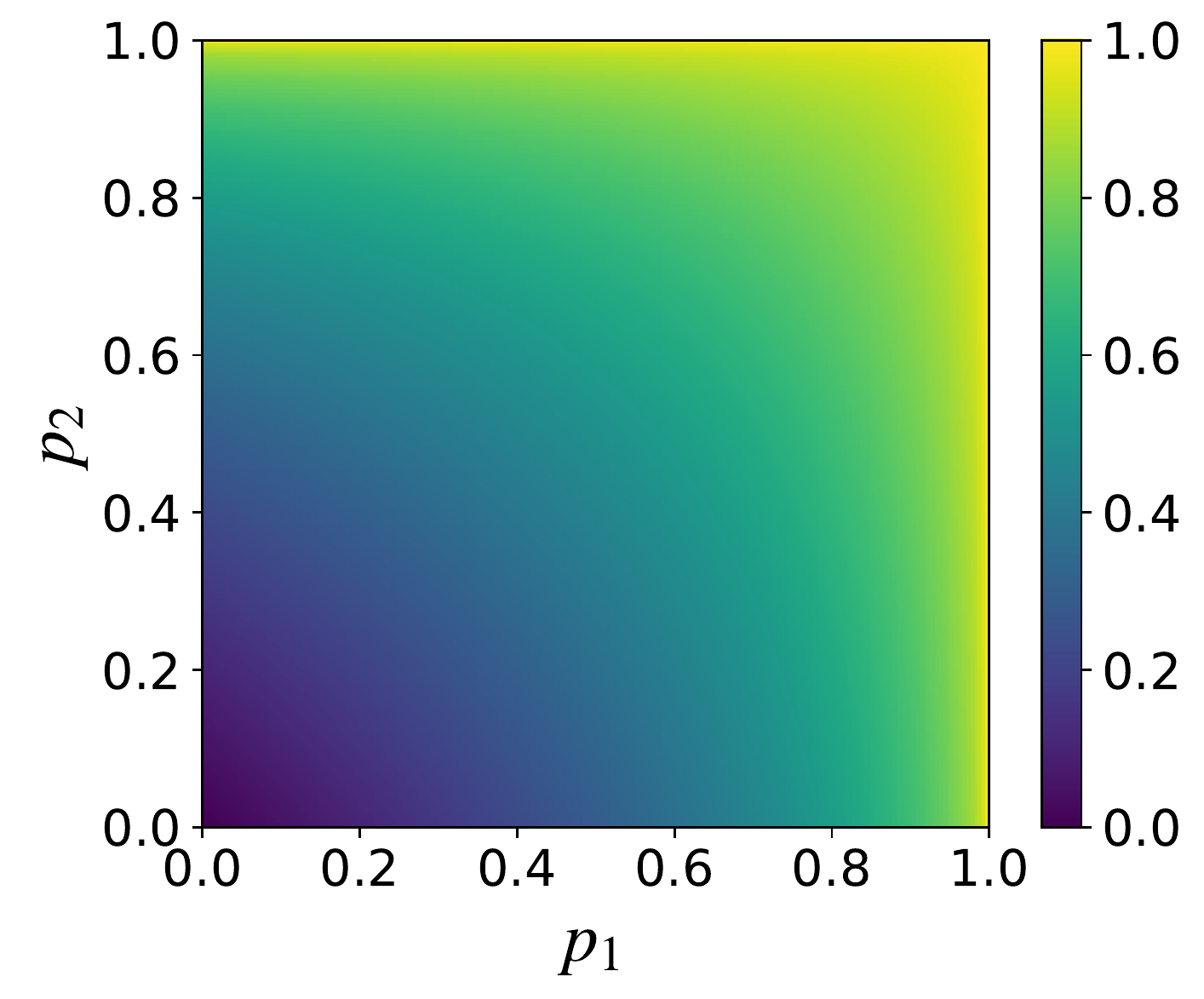} & \includegraphics[height=.14\textheight]{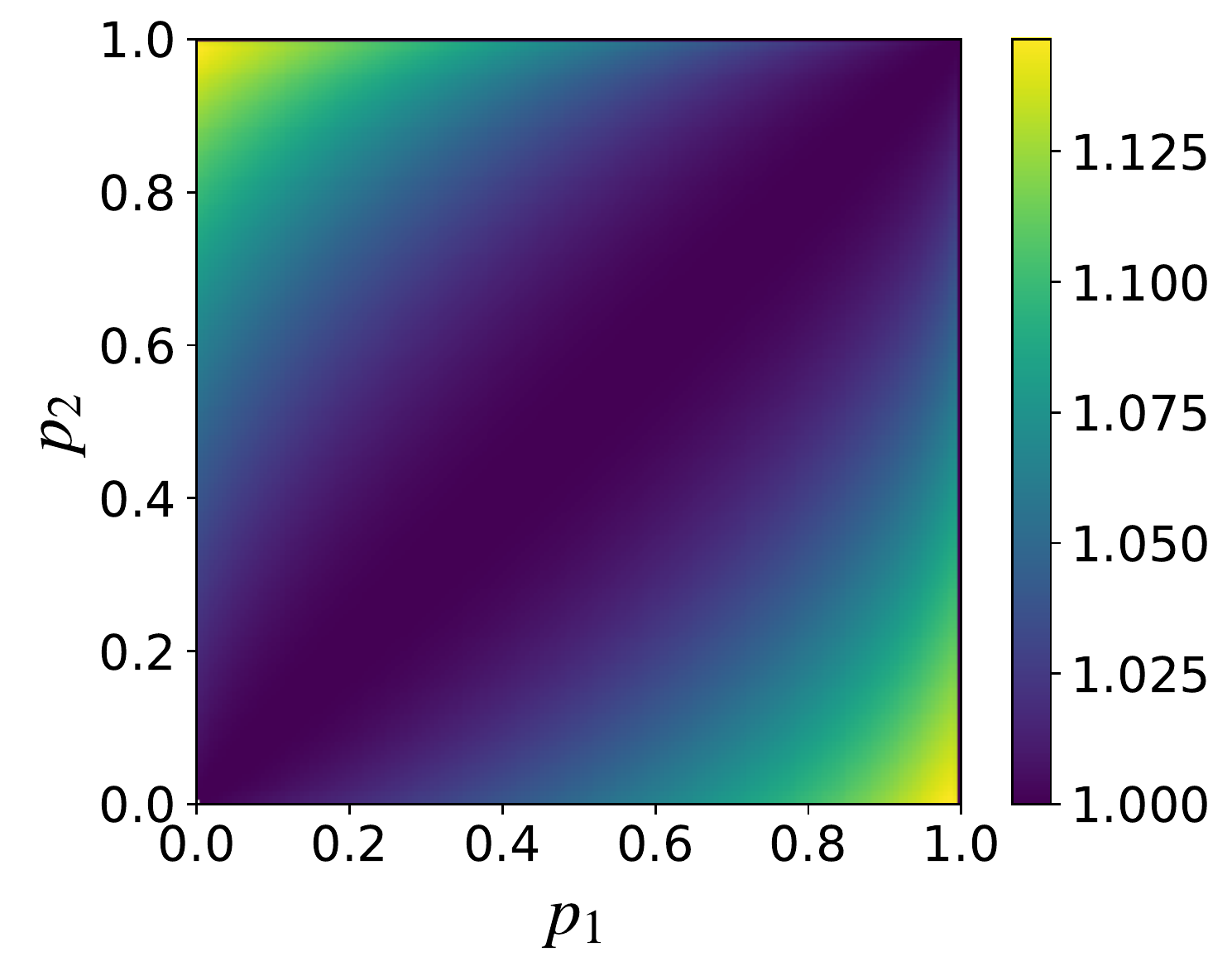}\\
{$\threvkl(p_1, p_2)$ (middle)\vskip 3mm
$\displaystyle\frac{\threvkl(p_1, p_2)}{\thneym(p_1, p_2)}$ (right)\vskip 1mm} & \includegraphics[height=.14\textheight]{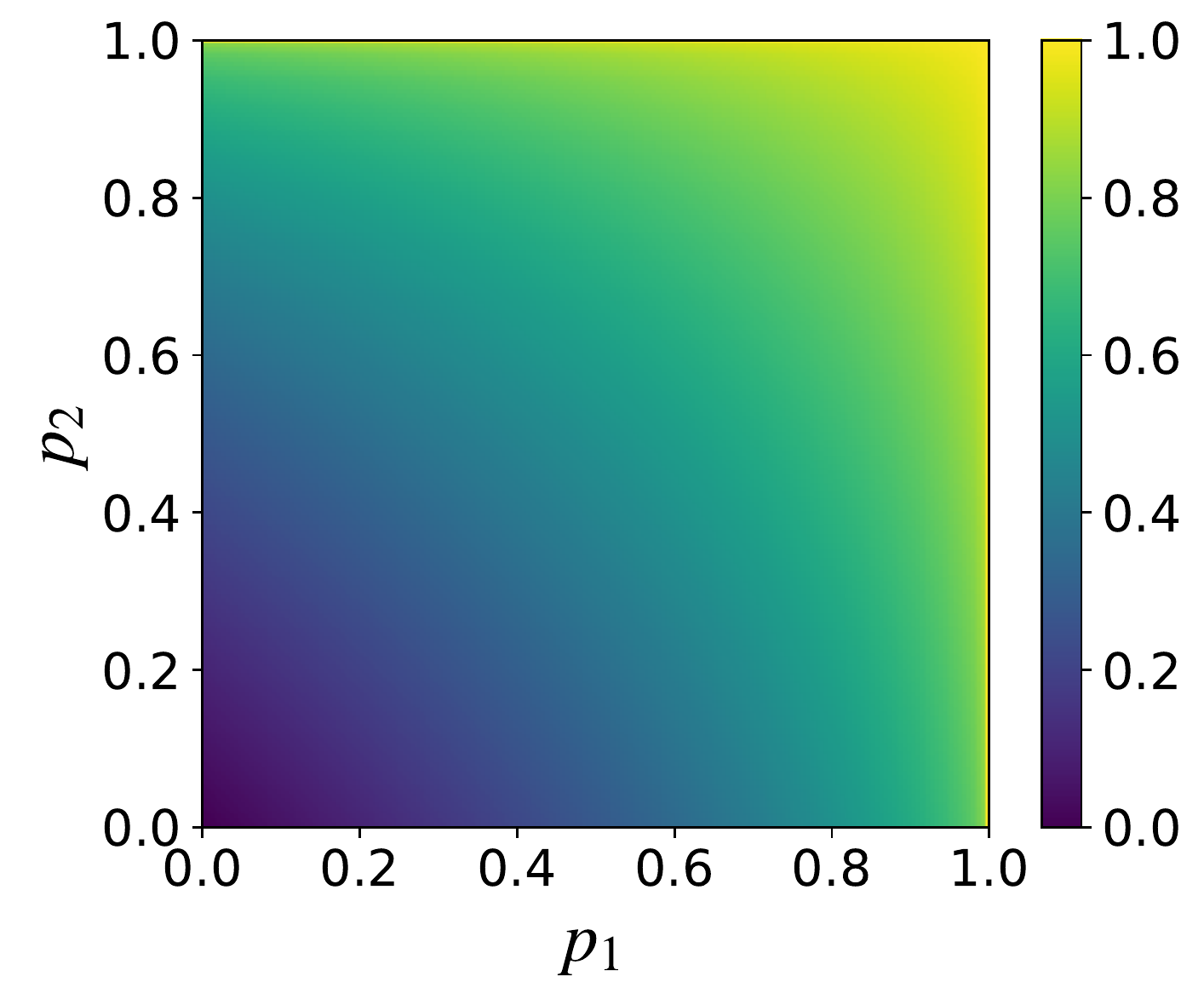} & \includegraphics[height=.14\textheight]{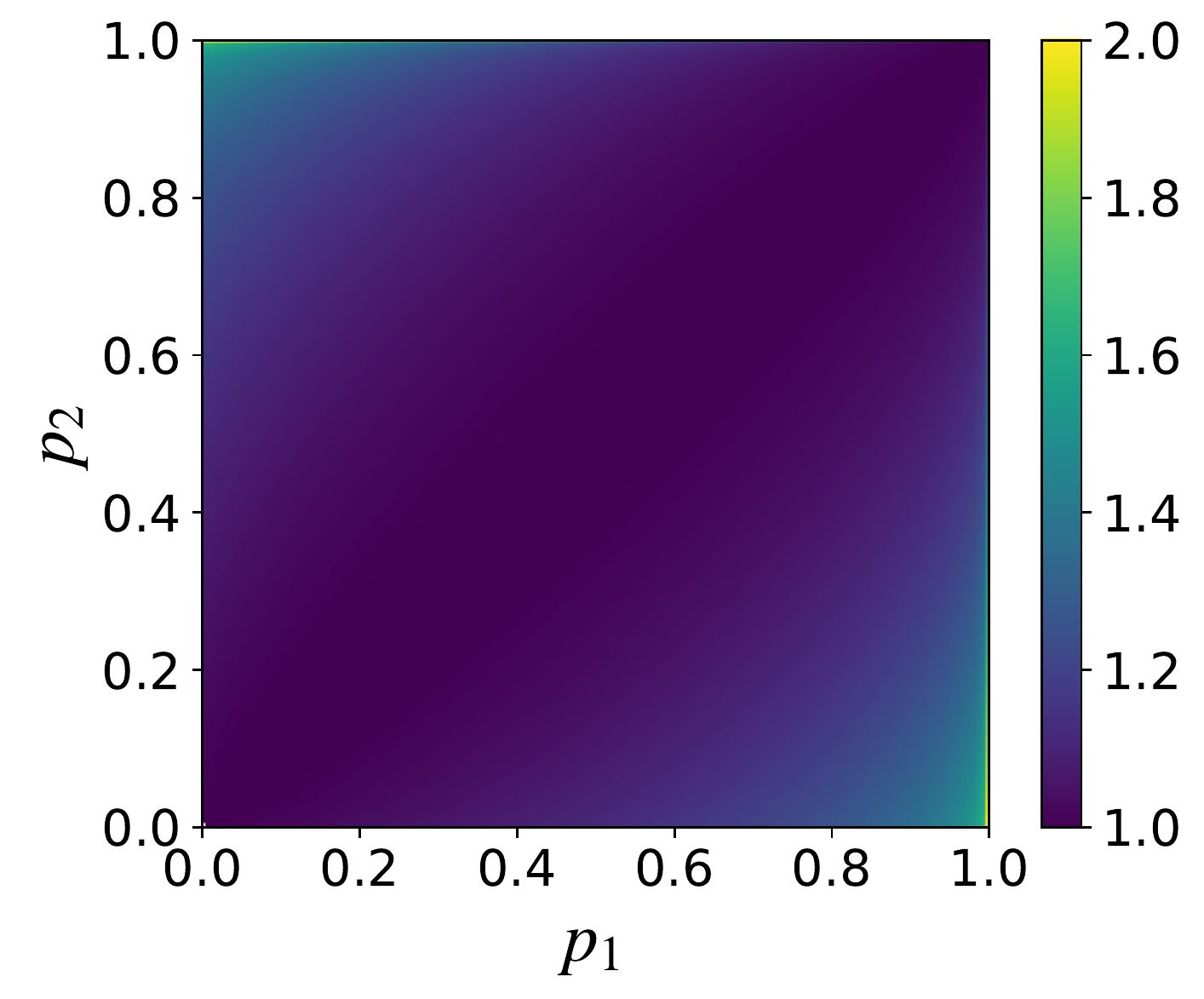}\\
{$\thneym(p_1, p_2)$ (middle)\vskip 1mm} & \includegraphics[height=.14\textheight]{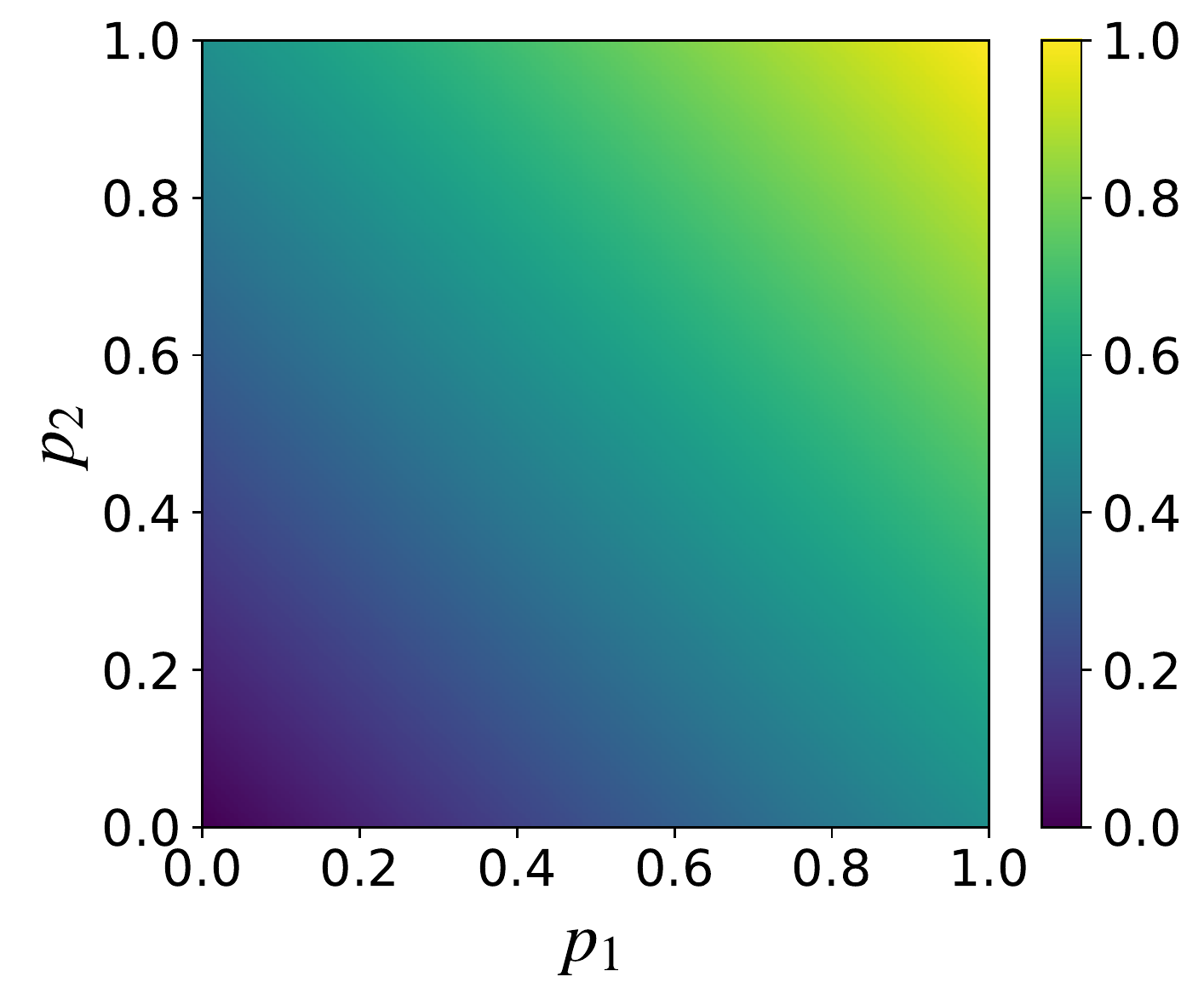} & 
\end{tabular}
\caption{\label{fig:catthresh} Plots of $\thstat(p_1, p_2)$ for the different statistical distances (middle column) arranged in decreasing order of strictness (top to bottom). The right column shows the ratio between $p_\text{stat}^\text{thresh}(p_1, p_2)$ for consecutive distances, in decreasing order of strictness. The ratios being $\geq 1$ demonstrates the correctness of their ordering.}
\end{figure}

This property is written in \eref{eqn:strictorder} as the ``strictness'' order of the statistical distances, with $\tdneym$ and $\tdpear$ being the least and most strict respectively. 
In \fref{fig:catthresh} we plot $\thstat(p_1, p_2)$ for the different statistical distances (middle column) arranged in decreasing order of strictness (top to bottom). \Fref{fig:catthresh} also shows the ratio between $p_\text{stat}^\text{thresh}(p_1, p_2)$ for consecutive distances, in decreasing order of strictness (right panel). The ratios being greater than or equal to one demonstrates the correctness of the ordering in \eref{eqn:strictorder2}.

Finally we want to point out that there is nothing sacred about the six distance measures used in this paper. Any measure of distance that is joint-convex in the distributions $\Hn$ and $\Ha$, and proportional to the Fisher--Rao metric distance between them in the $\Ha \rightarrow \Hn$ limit, can be used for training. The six distances provided in this paper are just a few named distances known in the statistics literature for which we were able to find usable closed form expressions in terms of the number density functions $s_c(\x)$ and $b_c(\x)$. For example, any linear combination of $\tdstat$'s from \eref{eqn:td} can also be used to train categorizers, along with the same linear combination of eventwise contributions $\edstat$ from \eref{eqn:ed}, although this much choice is probably overkill for most applications.

On a related note, we would like to point to following approximation for use in estimation of statistical significance of the deviation of observation from expectation:
\begin{equation}
\left[\sum\limits_i ~2\left[-(\text{obs}_i - \text{exp}_i) + \text{obs}_i \ln\left(\frac{\text{obs}_i}{\text{exp}_i}\right)\right]\right] \approx \left[\frac{1}{3}~\sum\limits_i\frac{(\text{obs}_i - \text{exp}_i)^2}{\text{obs}_i} + \frac{2}{3}~\sum\limits_i\frac{(\text{obs}_i - \text{exp}_i)^2}{\text{exp}_i}\right]
\end{equation}
Here the linear combination of Neyman-$\chi^2$ and Pearson-$\chi^2$ test statistics on the right-hand side, dubbed the ``combined Neyman-Pearson $\chi^2$'' in \cite{Ji:2019cnc}, approximates the left-hand side better than they do individually. In terms of our statistical distances between $\Hn$ and $\Ha$, this approximation can be written as $\tdkl \approx \left(\tdneym + 2~\tdpear\right)/3$.
\end{document}